\documentclass[12pt]{article}

\usepackage{scicite}
\usepackage{epsfig,graphicx}
\usepackage{times}
\usepackage{ccaption}
\usepackage{upgreek}
\usepackage{amsmath}
\usepackage{url}
\usepackage{hyperref}

\def\ra#1#2#3{#1$^{\rm h}$#2$^{\rm m}$#3$^{\rm s}$}
\def\dec#1#2#3{$#1^\circ#2'#3''$}

\newenvironment{sciabstract}{%
\begin{quote} \bf}
{\end{quote}}

\title{A hot and fast ultra-stripped supernova that likely formed a compact neutron star binary} 
\author
{K. De,$^{1\ast}$ M. M. Kasliwal,$^1$ E. O. Ofek,$^2$ T. J. Moriya,$^{3}$ J. Burke,$^{4,5}$ Y. Cao,$^6$\\ S. B. Cenko,$^{7,8}$ G. B. Doran,$^{9}$ G. E. Duggan,$^1$ R. P. Fender,$^{10}$ C. Fransson,$^{11}$\\ A. Gal-Yam,$^{2}$ A. Horesh,$^{12}$ S. R. Kulkarni,$^1$ R. R. Laher,$^{13}$ R. Lunnan,$^{11}$\\ I. Manulis,$^2$ F. Masci,$^{13}$ P. A. Mazzali,$^{14, 15}$ P. E. Nugent,$^{16, 17}$  D. A. Perley,$^{14}$\\ T. Petrushevska,$^{18, 19}$ A. L. Piro,$^{20}$ C. Rumsey,$^{21}$ J. Sollerman,$^{11}$ M. Sullivan,$^{22}$\\ and F. Taddia $^{11}$\\
\\
\normalsize{$^\ast$To whom correspondence should be addressed; E-mail:  kde@astro.caltech.edu}\\
}

\date{}

\begin{document} 
\baselineskip24pt

\maketitle 

\noindent
\normalsize{$^{1}$Cahill Centre for Astrophysics, California Institute of Technology, 1200 East California Boulevard, Pasadena, CA 91125, USA.}\\
\normalsize{$^{2}$Department of Particle Physics and Astrophysics, Faculty of Physics, The Weizmann Institute of Science, Rehovot 76100, Israel.}\\
\normalsize{$^{3}$ Division of Theoretical Astronomy, National Astronomical Observatory of Japan, National Institutes of Natural Sciences, 2-21-1 Osawa, Mitaka, Tokyo 181-8588, Japan.}\\
\normalsize{$^{4}$ Las Cumbres Observatory, 6740 Cortona Drive, Suite 102, Goleta, CA 93117-5575, USA.}\\
\normalsize{$^{5}$ Department of Physics, University of California, Santa Barbara, CA 93106-9530, USA.}\\
\normalsize{$^{6}$Department of Astronomy, University of Washington, Box 351580, Seattle, WA 98195-1580, USA.}\\
\normalsize{$^{7}$Astrophysics Science Division, NASA Goddard Space Flight Center, Mail Code 661, Greenbelt, MD 20771, USA.}\\
\normalsize{$^{8}$ Joint Space-Science Institute, University of Maryland, College Park, MD 20742, USA.}\\
\normalsize{$^{9}$Jet Propulsion Laboratory, California Institute of Technology, Pasadena, CA 91109, USA.}\\
\normalsize{$^{10}$ Department of Physics, Astrophysics, University of Oxford, Denys Wilkinson Building, Oxford OX1 3RH, UK.}\\
\normalsize{$^{11}$Oskar Klein Centre, Department of Astronomy, Stockholm University, 106 91 Stockholm, Sweden.}\\
\normalsize{$^{12}$Racah Institute of Physics, The Hebrew University of Jerusalem, Jerusalem, 91904, Israel.}\\
\normalsize{$^{13}$Infrared Processing and Analysis Center, California Institute of Technology, MS 100-22, Pasadena, CA 91125, USA.}\\
\normalsize{$^{14}$Astrophysics Research Institute, Liverpool John Moores University, Liverpool L3 5RF, UK.}\\
\normalsize{$^{15}$Max-Planck-Institut f{\"u}r Astrophysik, Karl-Schwarzschild-Str. 1, D-85748 Garching bei M{\"u}nchen, Germany.}\\
\normalsize{$^{16}$Lawrence Berkeley National Laboratory, Berkeley, California 94720, USA.}\\
\normalsize{$^{17}$Department of Astronomy, University of California, Berkeley, CA, 94720-3411, USA.}\\
\normalsize{$^{18}$Oskar Klein Centre, Department of Physics, Stockholm University, 106 91 Stockholm, Sweden.}\\
\normalsize{$^{19}$Centre for Astrophysics and Cosmology, University of Nova Gorica, Vipavska 11c, 5270 Ajdov\v{s}\u{c}ina, Slovenia.}\\
\normalsize{$^{20}$The Observatories of the Carnegie Institution for Science, 813 Santa Barbara Street, Pasadena, CA 91101, USA.}\\
\normalsize{$^{21}$ Astrophysics Group, Cavendish Laboratory, 19 J J Thomson Avenue, Cambridge CB3 0HE, UK.}\\
\normalsize{$^{22}$Department of Physics and Astronomy, University of Southampton, Southampton, SO17 1BJ, UK.}\\

\pagebreak

\baselineskip24pt 

\begin{sciabstract}
Compact neutron star binary systems are produced from binary massive stars through stellar evolution involving up to two supernova explosions. The final stages in the formation of these systems have not been directly observed. We report the discovery of iPTF\,14gqr (SN 2014ft), a Type Ic supernova with a fast evolving light curve indicating an extremely low ejecta mass ($\approx 0.2$  solar masses) and low kinetic energy ($\approx 2 \times 10^{50}$ ergs). Early photometry and spectroscopy reveal evidence of shock cooling of an extended He-rich envelope, likely ejected in an intense pre-explosion mass loss episode of the progenitor. Taken together, we interpret iPTF\,14gqr as evidence for ultra-stripped supernovae that form neutron stars in compact binary systems.

\end{sciabstract}

Core-collapse supernovae (SNe) are the violent deaths of massive stars when they run out of nuclear fuel in their cores and collapse, forming a neutron star (NS) or black hole (BH) \cite{Woosley2002}. For massive stars that have lost some or all of their outer hydrogen (H) and helium (He) envelope, the resulting collapse produces a stripped envelope SN \cite{Smartt2009}. The amount of material stripped from the star is a sensitive function of the initial mass of the star and its environment; if the star was born in a binary system, it also depends on the orbital properties of the system and the nature of the companion \cite{Smartt2009,Langer2012}.\\

Because most massive stars are born in close binary systems \cite{Sana2012}, stripping via binary interactions likely plays a large role in producing the observed diversity of stripped envelope SNe \cite{Yoon2010,Smith2011}. For the most compact companions in close orbits, the stripping of massive stars may be large enough to completely remove its outer layers, leaving behind a naked metal core close to the minimum mass required for the core to collapse (the Chandrasekhar mass). If massive enough, the highly stripped core eventually collapses to produce a faint and fast evolving SN explosion which ejects a small amount of material \cite{Tauris2013,Tauris2015}. Although it has been difficult to securely identify these explosions, such `ultra-stripped' SNe have been suggested to lead to the formation of a variety of compact NS binary systems (i.e. a NS in orbit around another NS, white dwarf (WD) or BH) \cite{Tauris2013,Tauris2017}.\\

\section*{Discovery and follow-up of iPTF\,14gqr}

iPTF\,14gqr was discovered by the intermediate Palomar Transient Factory (iPTF; \cite{Cao2016, Masci2017}) on 2014 October 14.18 UT (Coordinated Universal Time) at a $g$-band optical magnitude of $\approx 20.2$ mag. The source was not detected in the previous observation on 2014 October 13.32 (0.86 days before discovery), with a limiting magnitude of $g\geq$ 21.5\,mag. The transient was found in the outskirts (at a projected offset of $\approx 29$ kpc from the center) of a tidally interacting spiral galaxy (IV Zw 155) at a redshift $z = 0.063$ and luminosity distance $D = 284.5$ Megaparsecs (Figure 1). We obtained rapid ultraviolet (UV), optical and near-infrared (NIR) follow-up observations of the source, including a sequence of four spectra within 24 hours from the first detection \cite{SuppMat}. \\

We also obtained multi-epoch X-ray and radio observations and found that the source remained undetected at these wavelengths \cite{SuppMat}. These upper limits rule out luminous non-thermal emission, such as typically seen in relativistic and gamma-ray burst (GRB) associated SNe, but are not stringent enough to constrain the environment of the progenitor (Figure \ref{fig:14gqr_radioLCcompare}, Figure \ref{fig:14gqr_radioLCModel}).\\

Our photometric follow-up indicated that the source rapidly faded within a day of detection, followed by re-brightening to a second peak on a longer timescale (rising over $\approx 7$ days; \cite{SuppMat}) (Figure 2). The early decline was detected in all optical and UV photometric bands, and characterized by a blackbody spectrum which cooled rapidly from a temperature $T > 32000$ K near first detection to $T \sim 10000$ K at one day after discovery (Figure 3; Figure 4). Our early spectra also exhibit blackbody continua with temperatures consistent with those inferred from the photometry, superimposed with intermediate width emission lines of He~\textsc{ii}, C~\textsc{iii} and C~\textsc{iv}. Such high ionization lines, which are typically associated with elevated pre-explosion mass loss episodes in massive stars, have not been seen in early spectra of previously observed hydrogen-poor SNe. Although similar features are present in the early spectra of some hydrogen-rich core-collapse SNe \cite{GalYam2014,  Khazov2016, Yaron2017} (Figure \ref{fig:14gqr_compareFlashSpec}), the relatively large widths of the lines [Full Width at Half Maximum (FWHM) $\sim$ 2000 - 4000 km s$^{-1}$] as well as the rapid evolution of the 4686~\AA\, emission feature (Figure 3) are not.\\

Spectra obtained near the second peak are dominated by emission from the expanding photosphere and exhibit relatively blue continua, with broad absorption features reminiscent of normal stripped envelope SNe of Type Ic, that do not exhibit absorption lines of H or He in the spectra \cite{Gal-Yam2017} (Figure \ref{fig:14gqr_comparePeakSpec}). We find associated absorption velocities of $\sim$ 10,000 km s$^{-1}$ \cite{SuppMat}. The photometric properties of the second peak are broadly consistent with a number of previously observed fast Type Ic events (Figure \ref{fig:14gqr_compareLC}, Figure \ref{fig:14gqr_colorEvol}), but the rapidly declining first peak and the fast rise time to the second peak are unlike previously observed events. The source quickly faded after the second peak, declining at a rate of 0.21 mag day$^{-1}$ in the $g$ band \cite{SuppMat}. Our final spectrum taken at $\approx 34$ days after explosion shows that the source exhibited an early transition to the nebular phase, and on a timescale faster than previously observed core-collapse SNe. The nebular phase spectrum exhibits prominent [Ca~\textsc{ii}] emission similar to several other Type Ic SNe (Figure \ref{fig:14gqr_compareNebSpec}).\\

Multi-color photometry at multiple epochs allow us to trace the evolution of the optical / UV Spectral Energy Distribution (SED), which we use to construct bolometric light curves that contain flux integrated over all wavelengths (Figure 3; Figure 4; \cite{SuppMat}). We fit the pseudo-bolometric light curve of iPTF\,14gqr with a simple Arnett model \cite{Arnett1982} to estimate the explosion parameters. Allowing the explosion time to vary as a free parameter, we estimate an ejecta mass $M_{\textrm{ej}} \approx$ 0.15 -- 0.30 solar masses (M$_{\odot}$), an explosion kinetic energy $E_K \approx$ (1.0 -- 1.9) $\times 10^{50}$ ergs and synthesized Ni mass $M_{\textrm{Ni}} \approx 0.05$ M$_{\odot}$ \cite{SuppMat} (Figure 4, Figure \ref{fig:14gqr_arnettCorner}). The inferred ejecta mass is lower than known core-collapse Type Ic SNe \cite{Drout2011, Lyman2016, Taddia2017}, which have ejecta masses in the higher range of $\sim 0.7 - 15$ M$_{\odot}$, and with a mean of $2 - 3$ M$_{\odot}$ over a sample of $\approx$ 20 SNe. However, the parameters of iPTF\,14gqr are similar to those inferred for the rapidly evolving Type I SNe SN 2005ek \cite{Drout2013} and 2010X \cite{Kasliwal2010}, whose physical origins remain a matter of debate.\\

The rapid decline of the first peak observed in iPTF\,14gqr is reminiscent of shock cooling emission from the outer layers of a progenitor after the core-collapse SN shock breaks out \cite{Nakar2014,Sapir2017} (Figure \ref{fig:14gqr_firstPeak}). We consider alternative explanations \cite{SuppMat} and find them to be inconsistent with the data. In particular, the observed double-peaked light curve in the redder optical bands requires the presence of an extended low mass envelope around the progenitor \cite{Piro2015, Sapir2017}. To constrain the properties of such an envelope, we use models \cite{Piro2015} to construct multi-color light curves for a range of masses and radii of the envelope ($M_e$ and $R_e$ respectively). We find a best-fitting model of $M_e \sim 8 \times 10^{-3}$ M$_{\odot}$ and $R_e \sim 3 \times 10^{13}$ cm ($\sim 450$ solar radii (R$_{\odot}$)) \cite{SuppMat} (Figure 2 \& Figure \ref{fig:14gqr_shockCorner}). Even though the model considered here is simplified (e.g. it ignores the density structure of the envelope), we expect the estimated parameters to be accurate within an order of magnitude \cite{Piro2017}, leading us to conclude that the progenitor was surrounded by an extended envelope with a mass of $\sim 0.01$ M$_{\odot}$ at a radius of $\sim 500$ R$_{\odot}$. \\

We constrain the composition of the outer envelope using the early spectra. The emission lines observed in the early spectra of iPTF\,14gqr can be understood as arising from recombination in the outer regions of the extended circumstellar material (CSM), which was ionized by the high energy radiation produced in the shock breakout (e.g. \cite{GalYam2014, Yaron2017}; Figure \ref{fig:14gqr_compareFlashSpec}). We estimate the location and mass of the emitting He~\textsc{ii} from the luminosity of the early 4686~\AA\, line, and assuming a CSM density profile that varies with radius $r$ as $\propto r^{-2}$ \cite{Yaron2017}. We find the emitting region to be located at $r \sim 6 \times 10^{14} \tau^{-2}$ cm, and contain a helium mass $M_{\textrm{He}} \sim 0.01 \tau^{-3}$ M$_{\odot}$, where $\tau$ is the optical depth of the region \cite{SuppMat}. The absence of prominent Lorentzian scattering profiles in the lines suggest that the optical depth is small and assuming $\tau \approx 1$, we find $r \sim 6 \times 10^{14}$~cm ($8.5 \times 10^{3}$ R$_{\odot}$) and $M_{\textrm{He}} \sim 0.01$ M$_{\odot}$. Because our calculations are based on fitting a simple two-component Gaussian profile to the 4686~\AA\,emission line (to estimate the unknown contamination of C III at 4650~\AA), these estimates are uncertain by a factor of a few. \\

Using the C IV 5801 \AA\,lines and similar methods as above, we estimate a CSM carbon mass of $\sim 4 \times 10^{-3}$ M$_{\odot}$, while the hydrogen mass is constrained to be $< 10^{-3}$ M$_{\odot}$. Additional constraints based on light travel time arguments also suggest that the envelope was located at $r \leq 6 \times 10^{15}$ cm from the progenitor \cite{SuppMat}. The flash-ionized emission lines exhibit complex asymmetric profiles (Figure 3) that we attribute to light travel time effects, given the large size of the envelope and the high inferred wind velocities \cite{Grafener2016,SuppMat}.\\ 

\section*{An ultra-stripped progenitor}

The low ejecta mass and explosion energy, as well as the presence of an extended He-rich envelope, indicate an unusual progenitor channel for iPTF\,14gqr. The detection of the early shock cooling emission indicates a core-collapse origin of the explosion, while the bright radioactivity powered emission suggests that this explosion is associated with the class of iron core-collapse explosions. The low ejecta mass together with the small remaining amount of He in the progenitor rule out models of single star evolution as well as a non-degenerate massive star companion for the progenitor of iPTF\,14gqr \cite{SuppMat}, leaving only the most compact companions (such as a NS, WD or BH) as possible explanations of the highly stripped (or `ultra-stripped') progenitor. \\

Ultra-stripped explosions have been modeled in the case of He star - NS binaries, where stripping of the He star by a NS in a close orbit leads to the subsequent collapse of an ultra-stripped He star \cite{Tauris2013, Tauris2015, Moriya2017}. Hence, we compare theoretical bolometric light curves for ultra-stripped explosions \cite{Moriya2017} to those of iPTF\,14gqr in Figure 5, for a model with $M_{\textrm{ej}} = 0.2$ M$_{\odot}$, $M_{\textrm{Ni}} = 0.05$ M$_{\odot}$ and $E_K = 2 \times 10^{50}$ ergs. To account for the early declining emission, we also add a component corresponding to shock cooling of an extended envelope, for $M_e = 0.01$ M$_{\odot}$ and $R_e = 6 \times 10^{13}$ cm. The two component light curve matches the light curve data. We also compare the spectroscopic properties of iPTF\,14gqr to those of ultra-stripped SN models in Figure 5. The models \cite{Moriya2017} assumed fully mixed ejecta that led to the production of strong line blanketing features below $4000$ \AA\,, unlike this source. Thus, we re-calculated the models for ejecta with no mixing (as with the light curve calculations), and were able to match to the spectra of iPTF\,14gqr near the second peak (Figure 5, Figure \ref{fig:14gqr_simSpecModels}).\\

Our observations indicate the presence of an extended He-rich envelope around the progenitor at the time of collapse, thus providing insight into the terminal evolution of the progenitors of ultra-stripped SNe, and more broadly, the lowest mass progenitors of core-collapse SNe. Using the line widths in our early spectra, we estimate that the emitting envelope was expanding with a velocity of $\sim 1000 - 2000$ km s$^{-1}$ at the time of collapse, consistent with the escape velocity from a compact He star \cite{SuppMat}. When considered with the inferred size of the envelope (at least $\sim$ 500 R$_{\odot}$), the velocities suggest that the envelope was ejected $\sim 8 - 20$ days prior to the explosion. \\

The temporal coincidence of the ejection with the final SN suggests that the envelope was likely associated with an intense pre-SN mass loss episode of the progenitor \cite{SuppMat}. Despite the close stripping, ultra-stripped progenitors are expected to retain a small amount of He ($\sim 0.01$ M$_{\odot}$) in their outer layers. The prominent He and C lines in the early spectra are consistent with eruptive mass loss when considering the expected surface compositions of ultra-stripped progenitors \cite{Tauris2015}. The timescale of the ejection is similar to that expected for silicon flashes ($\sim 2$ weeks before explosion) in the terminal evolution of low mass metal cores \cite{Woosley2015}, that have been suggested to lead to elevated mass loss episodes prior to the explosion. Such mass loss episodes are relevant to ultra-stripped progenitors as well \cite{Woosley2015,Moriya2017,Muller2018}. \\

iPTF\,14gqr exhibits a projected offset of $\sim 15$ kpc from the nearest spiral arms of its star forming host galaxy \cite{SuppMat}, which is puzzling when compared to the expected locations of ultra-stripped SNe \cite{Tauris2015}. While we do not find evidence of an underlying stellar association or of galaxy emission features in late-time imaging and spectroscopy, the limits are not sensitive enough to rule out the presence of a dwarf galaxy or a star forming H-II region (characterized by its H$\alpha$ emission) at or near the transient location \cite{SuppMat}. Nonetheless, the tidally interacting environment of the host galaxy suggests that outlying star formation in collisional debris is likely in this system \cite{Boquien2009,SuppMat}, which could harbor young stellar systems (with ages of $\sim$ 5 - 100 Myrs) in the faint tidal tails (Figure \ref{fig:14gqr_lateDeep}). Hence, the discovery of a core-collapse SN in these outskirts is consistent with our interpretation.\\ 

While a number of previously observed fast Type Ic SNe (e.g. SN\,2005ek \cite{Drout2013} and SN\,2010X \cite{Kasliwal2010}) were suggested to be members of the ultra-stripped SN class, it has been difficult to confirm a core-collapse origin for these explosions because these events were discovered only near maximum of the radioactively powered peak. Specifically, without early photometry and spectroscopy that can reveal the presence of a shock cooling component, these fast transients are also consistent with variants of models involving thermonuclear detonations on white dwarfs (e.g. \cite{Shen2010,Metzger2012,Darbha2010}). The early discovery and prompt follow-up of iPTF\,14gqr establish the presence of a shock cooling emission component that requires an extended progenitor consistent with a core-collapse explosion. In the probable scenario that iPTF\,14gqr formed a NS in the explosion (we find a BH remnant to be unlikely given the observed properties of the SN \cite{SuppMat}), the low ejecta mass in the system suggests that the SN results in the formation of a bound and compact NS binary system \cite{SuppMat}. \\

\section*{Implications for formation of compact NS binaries}

Our interpretation of iPTF\,14gqr as an ultra-stripped SN has implications in the wider context of stellar evolution. Compact NS binary systems evolve from binary massive stars that undergo several phases of mass transfer over their lifetime (Figure 6). The initial phases of such evolution, in which two massive stars evolve into interacting binaries consisting of a compact object in orbit around a massive star (X-ray binaries) have been observed in several systems in the local Universe \cite{Benvenuto2017,Walter2015}. However, the subsequent phases that lead to the formation of compact NS binary systems, have not been observed. This is due to the low occurrence rates of such systems, the short lifetimes ($\sim 10^{6}$ years) of the final stages and observational selection effects disfavoring their detection \cite{Tauris2015,Gotberg2017,Zapartas2017}.\\

Binary evolution models suggest that the subsequent evolution proceeds via a common envelope phase, during which the loss of angular momentum via dynamical friction leads to the formation of a close He star - compact object binary \cite{Bhattacharya1991,Tauris2006,Tauris2017}. An additional phase of close gravitational stripping by the compact companion then leads to the formation of an ultra-stripped SN progenitor \cite{Tauris2017}, with properties which can be inferred from our observations of iPTF\,14gqr. The measured orbital properties of known double NS systems suggest that the second NSs were created in weak and low ejecta mass explosions that impart a small natal kick to the newborn NS \cite{Ferdman2013, Beniamini2016a}.\\

The presence of the extended He-rich envelope in iPTF\,14gqr along with the lack of He in the low mass of ejecta suggest that the progenitor was highly stripped by a compact companion, such that only a thin He layer was retained on its surface. This He layer was then ejected in an intense pre-SN mass loss episode, as shown by the high velocity of the envelope. Taken together, these observations provide evidence of the terminal evolution of a post common envelope He star - compact object binary leading to the formation of a compact NS binary system (Figure 6). \\

While wide binaries containing a NS and another compact object may be formed in non-interacting systems of binary massive stars, ultra-stripped SNe have been suggested to precede the formation of almost all compact NS binary systems \cite{Tauris2015}. Thus, these explosions likely represent the only channel to forming NS-NS and NS-BH systems that are compact enough to merge within the age of the universe and produce observable merger signals for joint gravitational wave (e.g. \cite{Abbott2017a}) and electromagnetic (e.g. \cite{Abbott2017b,Pian2017,Kasen2017}) observations \cite{Voss2003,Tauris2015,DNSDynamical}. Given that only a fraction of the systems produced by these explosions will merge within that time, the rates of ultra-stripped explosions must be higher than the rates of their mergers. \\\\

\noindent

\bibliographystyle{Science.bst}

\clearpage

\section*{Acknowledgements}
We thank the anonymous referees for a careful reading of the manuscript, that helped improve the quality of the paper. We thank C. Steidel, N. Stone, D. Stern, P. Hopkins, S. de Mink, Y. Suwa, A. Heger and T. M. Tauris for valuable discussions. MMK thanks J. Fuller, E. S. Phinney, L. Bildsten and E. Quataert for stimulating discussions at the Skyhouse during a PTF-TN meeting. We thank Tim Staley and Gemma Anderson for help with scheduling of the AMI observations. Additional facility acknowledgments are given in the Supplementary Material.

\section*{Funding}
The Intermediate Palomar Transient Factory project is a scientific collaboration among the California Institute of Technology, Los Alamos National Laboratory, the University of Wisconsin, Milwaukee, the Oskar Klein Center, the Weizmann Institute of Science, the TANGO Program of the University System of Taiwan, and the Kavli Institute for the Physics and Mathematics of the Universe. This work was supported by the GROWTH (Global Relay of Observatories Watching Transients Happen) project funded by the National Science Foundation under PIRE Grant No 1545949. GROWTH is a collaborative project among California Institute of Technology (USA), University of Maryland College Park (USA), University of Wisconsin Milwaukee (USA), Texas Tech University (USA), San Diego State University (USA), Los Alamos National Laboratory (USA), Tokyo Institute of Technology (Japan), National Central University (Taiwan), Indian Institute of Astrophysics (India), Indian Institute of Technology Bombay (India), Weizmann Institute of Science (Israel), The Oskar Klein Centre at Stockholm University (Sweden), Humboldt University (Germany), Liverpool John Moores University (UK).

A.H. acknowledges support by the I-Core Program of the Planning and Budgeting Committee and the Israel Science Foundation. A.G.-Y. is supported by the EU via ERC grant No. 725161, the Quantum Universe I-Core program, the ISF, the BSF Transformative program and by a Kimmel award. E.O.O. is grateful for support by grants from the Willner Family Leadership Institute Ilan Gluzman (Secaucus NJ), Israel Science Foundation, Minerva, BSF, BSF-transformative, and the I-Core program by the Israeli Committee for Planning and Budgeting and the Israel Science Foundation (ISF). F. T. and J. S. gratefully acknowledge the support from the Knut and Alice Wallenberg Foundation. The Oskar Klein Centre is funded by the Swedish Research Council. This research used resources of the National Energy Research Scientific Computing Center, a DOE Office of Science User Facility supported by the Office of Science of the U.S. Department of Energy under Contract No. DE-AC02-05CH11231.  M. S. acknowledges support from EU/FP7 ERC grant no. [615929]. P. E. N. acknowledges support from the DOE through DE-FOA-0001088, Analytical Modeling for Extreme-Scale Computing Environments. T. J. M. is supported by the Grants-in-Aid for Scientific Research of the Japan Society for the Promotion of Science (16H07413, 17H02864). Numerical computations were in part carried out on PC cluster at Center for Computational Astrophysics, National Astronomical Observatory of Japan. Part of this research was carried out at the Jet Propulsion Laboratory, California Institute of Technology, under a contract with the National Aeronautics and Space Administration. 

\section*{Author contributions}
KD and MMK initiated the study, conducted analysis and wrote the manuscript. IM initiated the follow-up of the young transient. DAP, GED and YC conducted Keck and Palomar observations and contributed to data reduction and manuscript preparation. SBC conducted Keck and Swift observations and contributed to data reduction and manuscript preparation. MS conducted the WHT observations and data reduction. FT and JS conducted NOT observations, data analysis and contributed to manuscript preparation. JB conducted the LCO observations and data reduction. TP conducted Gemini observations and data analysis. CR and RPF conducted the AMI observations and data reduction. AH conducted the VLA observations and data reduction. SRK is iPTF PI and contributed to manuscript preparation. TJM and PAM prepared the ultra-stripped SN models presented in the paper. EOO, CF, AGY, RL, PEN and ALP contributed to manuscript preparation. GBD, RRL and FM contributed to the machine learning codes used to search for young transients.

\section*{Competing interests}
The authors declare no competing interests.

\section*{Data and materials availability}
All photometric data used in this paper are provided in the supplementary material (Table S1 and Table S2), while all spectra are available via the WISeREP repository at \url{https://wiserep.weizmann.ac.il/}. The codes used for the ultra-stripped SN modeling are presented in \cite{Mazzali1993}, while the synthetic spectra presented in this paper are available at \url{https://goo.gl/9gkc9M}.\\

\noindent
\textbf{\Large List of Supplementary materials}\\
\noindent
{\large Materials and Methods\\
Supplementary text\\
Figures S1 - S15\\
Tables S1 - S7\\
References (51 - 181)\\
}\\

\newpage


\begin{figure}
\includegraphics[width=\textwidth]{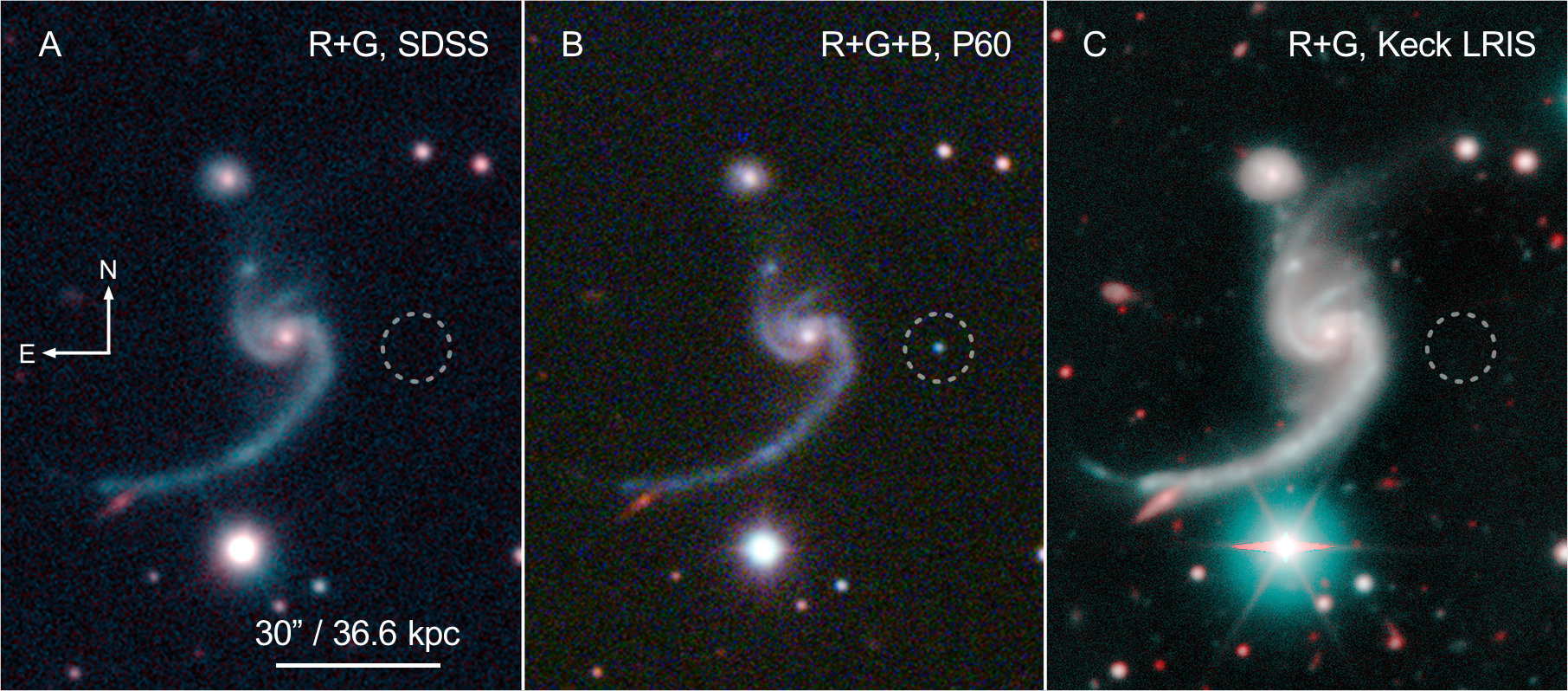}
\caption{\textbf{Discovery field and host galaxy of iPTF\,14gqr}. A. An optical image of the field from the Sloan Digital Sky Survey (SDSS); r and g filter images have been used for red and cyan colors respectively). B. Composite RGB image (r, g and B filter images have been used for red, green and blue colors respectively) of the iPTF 14gqr field from images taken near the second peak (19 October 2014) with the Palomar 60-inch telescope (P60), showing a blue transient inside the white dashed circle at the discovery location. C. Late-time composite R+G image (R and G filter images have been used for red and cyan colors respectively) of the host galaxy taken with the Low Resolution Imaging Spectrograph on the Keck-I telescope. }
\label{fig:14gqr_discovery}
\end{figure}



\begin{figure}
\includegraphics[width=\textwidth]{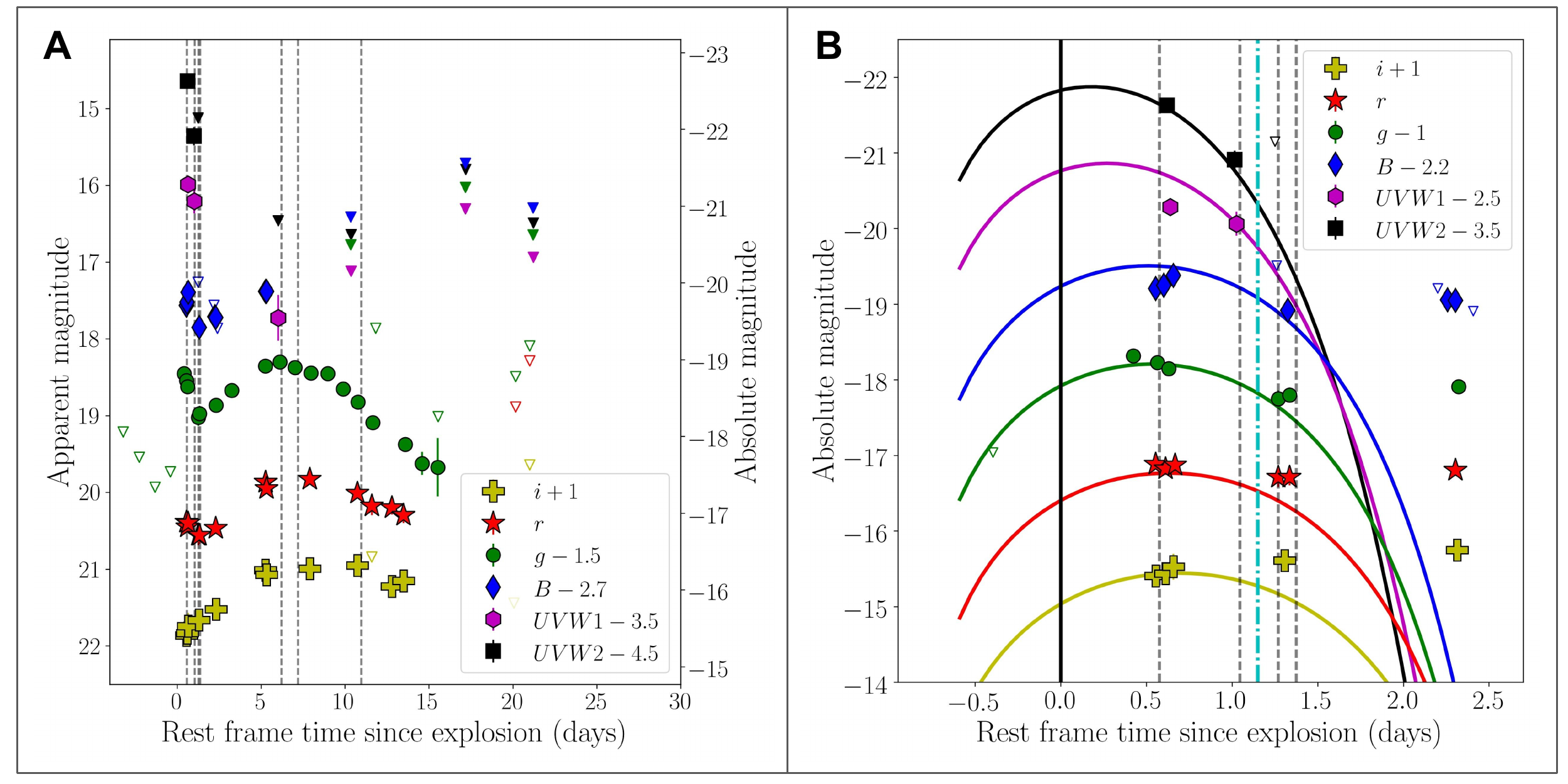}
\caption{\textbf{Multi-color photometric observations of iPTF\,14gqr.} A. Multi-color light curves of iPTF\,14gqr from our photometric follow-up observations (magnitudes are corrected for galactic extinction, and offset vertically as indicated in the legend). Inverted triangles denote 5$\sigma$ upper limits while other symbols denote detections. Hollow inverted triangles are upper limits from P48/P60 imaging and the filled inverted triangles are upper limits from Swift observations (filled green triangles are $V$ band limits from Swift). Epochs when spectra were obtained are marked in both panels by vertical black dashed lines. B. Zoom-in of the early evolution of the light curve. The black solid line shows the assumed explosion epoch. The colored solid lines show the best-fitting shock cooling model for extended progenitors \cite{Piro2015}. Only photometric data before the cyan dot-dashed vertical line were used in the fitting \cite{SuppMat}. }
\label{fig:14gqr_lightCurve}
\end{figure}



\begin{figure}
\includegraphics[width=\textwidth]{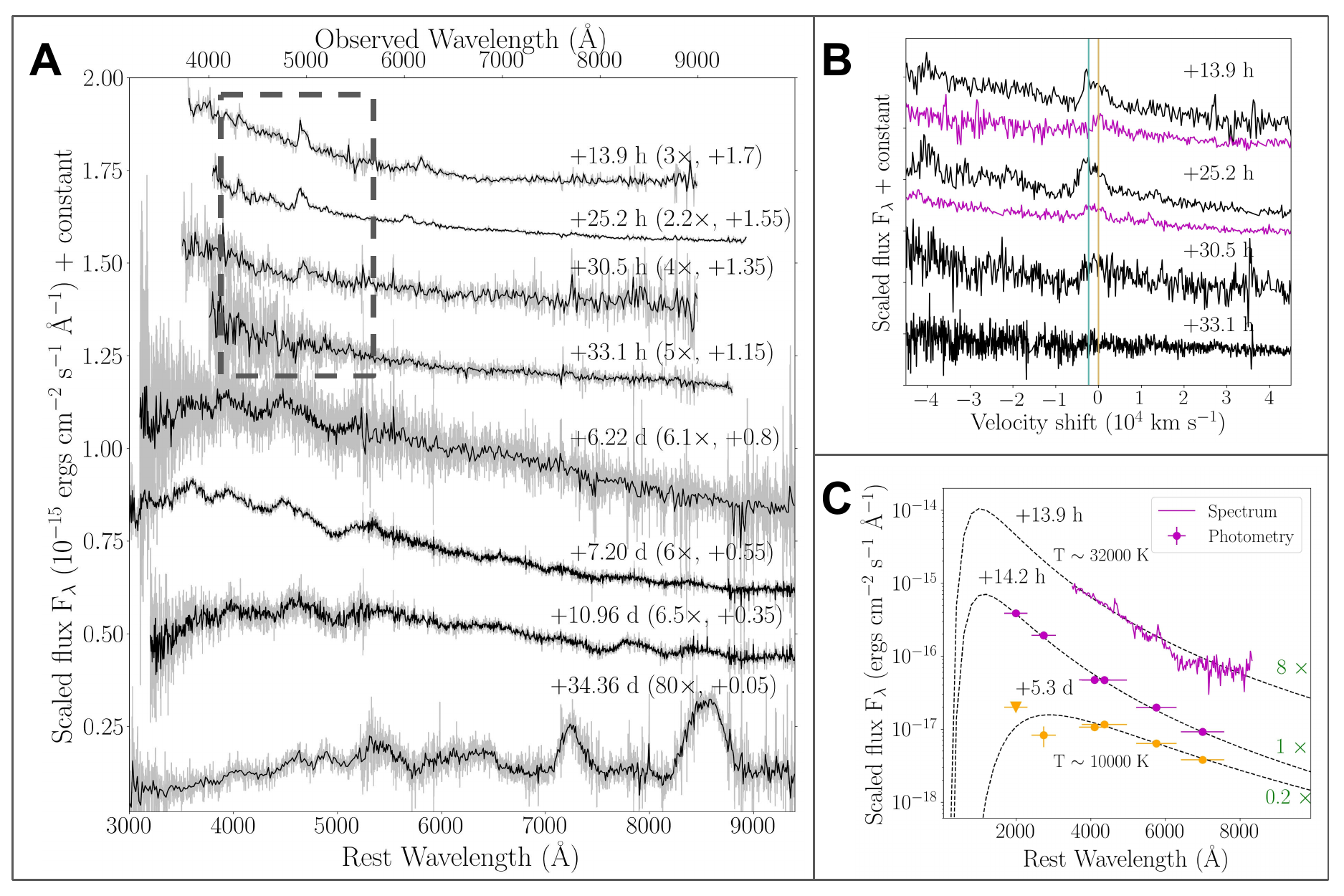}
\caption{\textbf{Spectroscopic evolution of iPTF\,14gqr.} A. Observed spectra before (gray) and after (black) binning. The epochs of the spectra along with the scaling and vertical shifts used are indicated next to each spectrum. B. Zoom-in of the early spectra, indicated by the black dashed box in (A), showing rapid evolution of the $\lambda$4686 feature within 24 hours of discovery. The x-axis indicates the velocity shift from the He II $\lambda$4686 line. The orange and cyan lines mark the locations of the $\lambda$4686 line and the C III $\lambda$4650 line respectively. For the +13.9 h and +25.2 h spectra, additional magenta lines show the profiles of the C IV $\lambda$5801 and the C III $\lambda$5696 features respectively, at the same epochs. C. Scaled optical / UV SEDs of the photometry and spectra obtained within the first light curve peak (see Figure 2) in magenta, along with photometry near the second peak in orange. The circles indicate observed photometric fluxes, while the triangle is a 5$\sigma$ upper limit. The dashed black lines indicate the best fitting blackbody SEDs including all optical / UV data points for the first peak and including only the optical data points for the second peak \cite{SuppMat}.}
\label{fig:14gqr_specEvolution}
\end{figure}



\begin{figure}
\includegraphics[width=\textwidth]{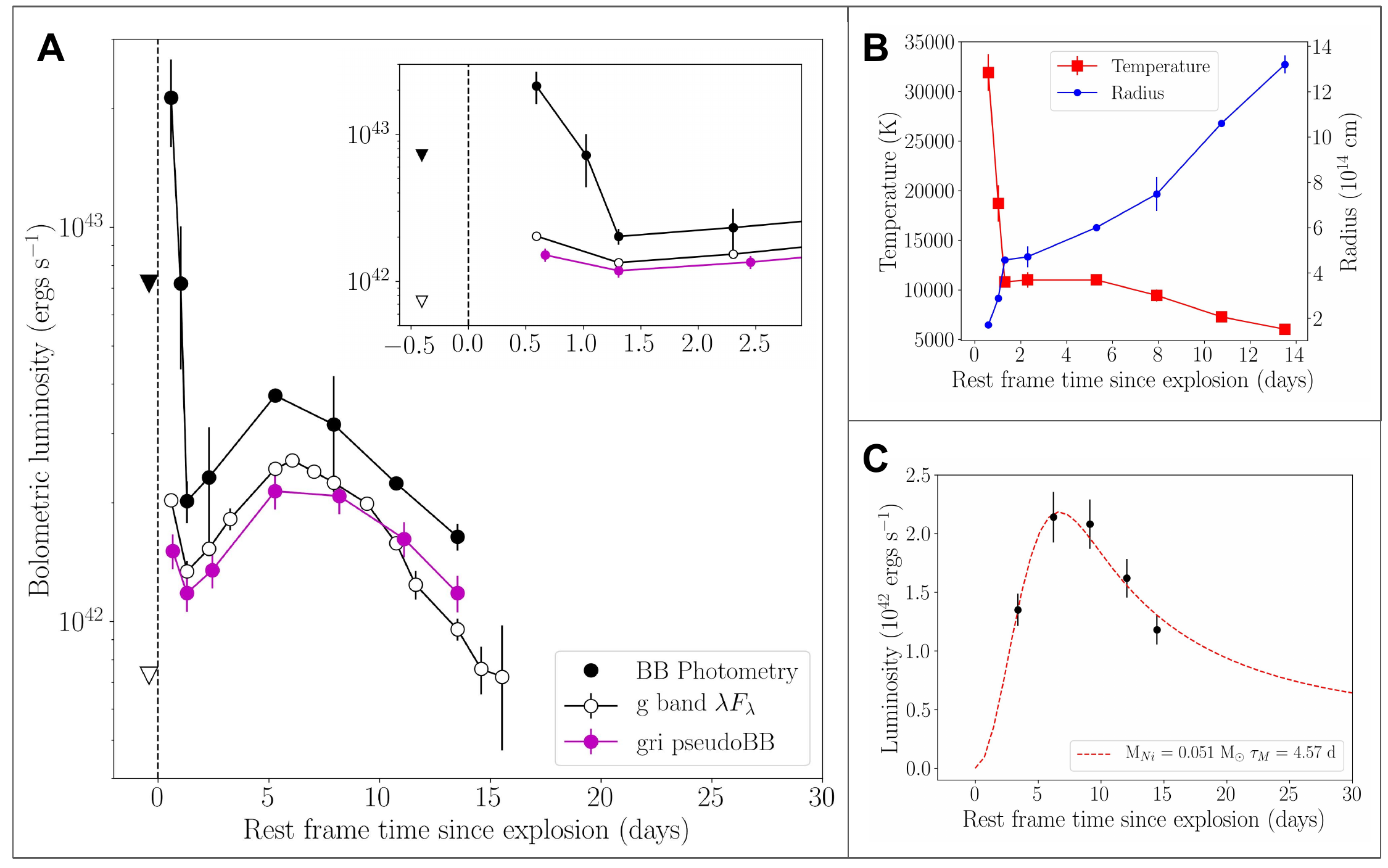}
\caption{\textbf{Bolometric light curve and Arnett modeling of iPTF\,14gqr.} A. Bolometric light curve of iPTF\,14gqr. Filled black points indicate blackbody luminosities obtained from fitting multi-color photometry while the magenta points correspond to pseudo-bolometric luminosities \cite{SuppMat}. The empty black circles indicate $g$-band luminosities obtained by multiplying the $g$-band flux F$_\lambda$ with the wavelength $\lambda$ of the filter. The inverted triangles denote estimated pre-detection 5$\sigma$ upper limits on the respective luminosities \cite{SuppMat}. The inset shows the bolometric light curves zoomed into the region of the first peak. B. The radius and temperature evolution of the fitted blackbody functions. C. Best-fitting Arnett model of the pseduo-bolometric light curve of the main (second) peak of iPTF\,14gqr. The $^{56}$Ni mass $M_{\textrm{Ni}}$ and diffusion timescale $\tau_M$ corresponding to the model are indicated in the legend \cite{SuppMat}.}
\label{fig:14gqr_boloCurve}
\end{figure}



\begin{figure}
\includegraphics[width=\textwidth]{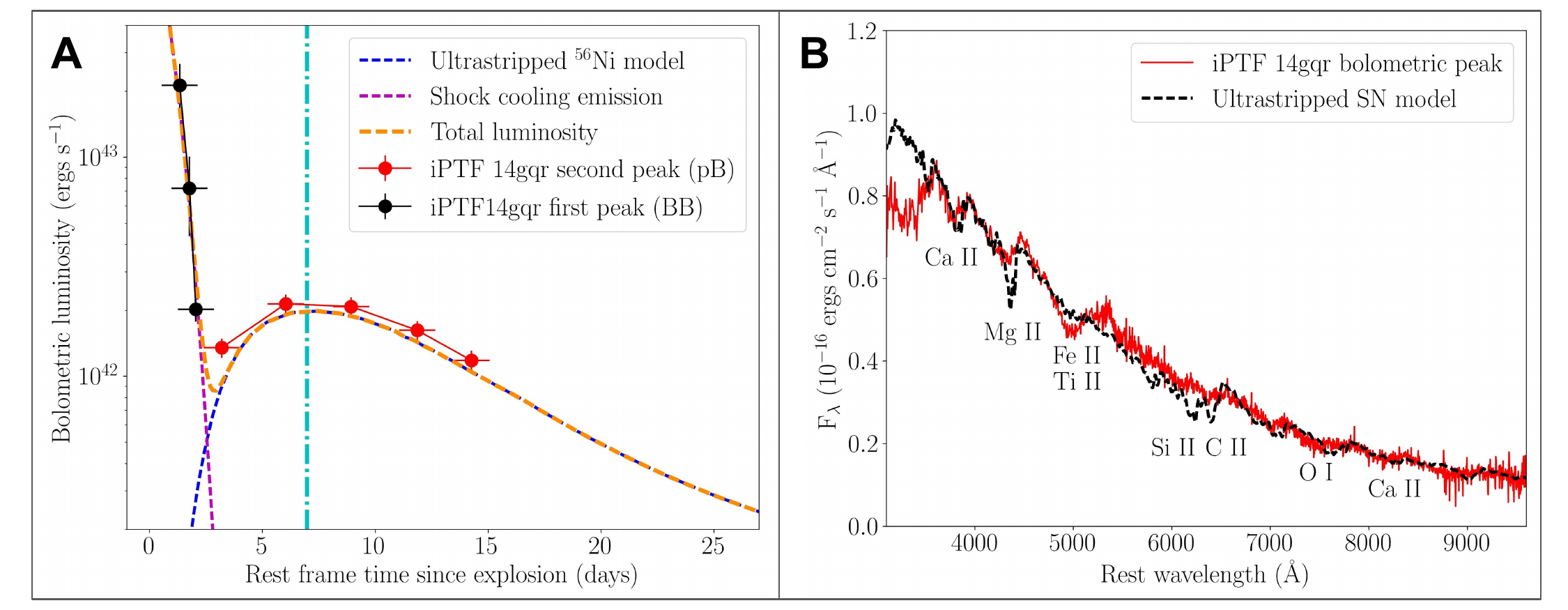}
\caption{\textbf{Comparison of iPTF\,14gqr to theoretical models of ultra-stripped SNe.} A. The bolometric light curve of iPTF\,14gqr shown with a composite light curve consisting of ultra-stripped Type Ic SN models \cite{Moriya2017} and early shock cooling emission \cite{Piro2015}. The blue dashed line corresponds to the $^{56}$Ni powered peak in the ultra-stripped SN models for $M_{\textrm{ej}} = 0.2$ M$_{\odot}$, $M_{\textrm{Ni}} = 0.05$ M$_{\odot}$ and $E_K = 2 \times 10^{50}$ ergs, the magenta line corresponds to the early shock cooling emission and the orange line is the total luminosity from summing the two components. We use the blackbody (BB) luminosities to represent the early emission, while we use the pseudo-bolometric (pB) luminosities for the second peak \cite{SuppMat}. B. Comparison of the peak photospheric spectra of iPTF\,14gqr (the epoch is indicated by the cyan dashed line in (A)) to that of the model in (A). The overall continuum shape, as well as absorption features of O I, Ca II, Fe II and Mg II are reproduced \cite{SuppMat}.}
\label{fig:14gqr_USmodel}
\end{figure}



\begin{figure}
\includegraphics[width=\textwidth]{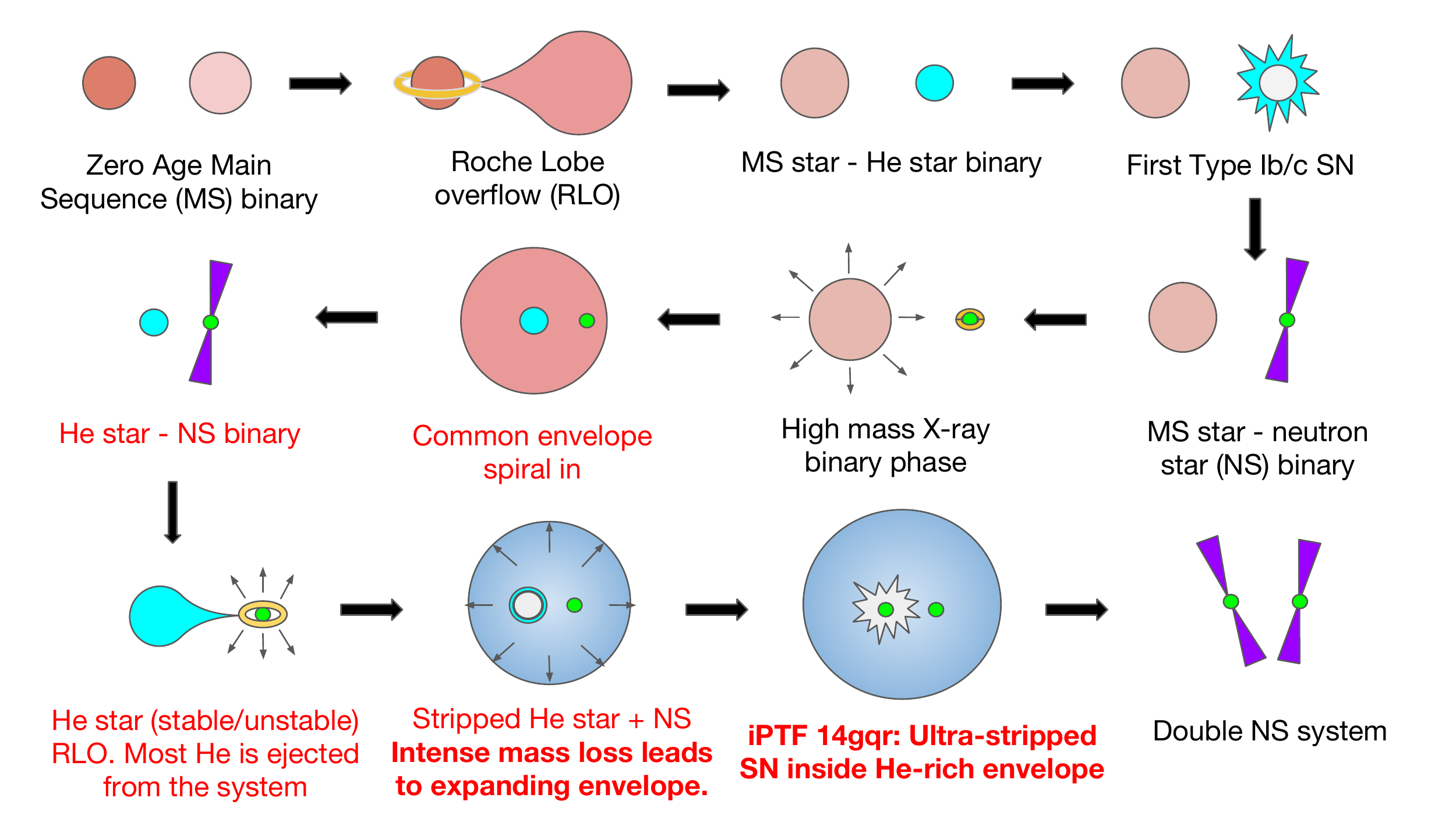}
\caption{\textbf{Stellar evolutionary sequence leading from a binary system of massive stars (starting from the top left) to a NS-NS system, adapted from \cite{Tauris2017}}. NS-BH systems are expected to arise from binaries where the first formed compact object is a BH. NS-WD systems follow a similar evolutionary sequence starting from the HMXB stage (where the NS is replaced by the WD), but require additional mass transfer in the earlier stages \cite{Tauris2000}. The material composition of the stars is indicated by their colors -- red indicates H-rich material, cyan / blue indicate He-rich material, grey indicates CO-rich material and green indicates degenerate matter (in NS). The specific phase of the evolution is indicated by the text next to the systems, with black text indicating phases that have been observed previously, while red text indicates phases that have not been previously observed, and bold red text phases we observed in this work.}
\label{fig:14gqr_stelEvolution}
\end{figure}

\clearpage

\setcounter{figure}{0}    

\renewcommand{\thefigure}{S\arabic{figure}}
\renewcommand{\thetable}{S\arabic{table}}
\renewcommand{\theequation}{S\arabic{equation}}

\noindent
\section*{Materials and Methods}
\subsection*{Observations}
\subsubsection*{iPTF\,Discovery} 
iPTF\,14gqr (SN 2014ft) was discovered by the intermediate Palomar Transient Factory (iPTF; \cite{Law2009, Masci2017}) in data taken with the CFH12K 96-Megapixel camera \cite{Rahmer2008, Law2010} mounted on the 48 inch Samuel Oschin Telescope at Palomar Observatory (P48), on 2014 October 14.18 [Modified Julian Date (MJD) 56944.18; Coordinated Universal Times are used throughout this paper]. The source was discovered at coordinates right ascension ($\alpha$) $=$ \ra{23}{33}{27.95}, declination ($\delta$) $=$ \dec{33}{38}{46.1} (J2000 equinox), while it was not detected on 2014 October 13.32 (MJD 56943.32; 0.86 days before discovery) up to a limiting magnitude of $g\geq$ 21.5. We nominally adopt the average MJD 56943.75 $\pm$ 0.43  as the explosion date, and calculate all phases with reference to this epoch. However, the actual explosion could have taken place before the last non-detection depending on the (unknown) behavior of the early emission. Hence, we allow the explosion time to vary as a free parameter in our modeling, and discuss the last non-detection individually in the context of the physical models. 

\subsubsection*{Optical light curves}
We obtained $g$ band photometry of iPTF\,14gqr with the P48 CFH12K camera, along with additional follow-up photometry in the $Bgri$ bands with the automated 60-inch telescope at Palomar Observatory (P60; \cite{Cenko2006}). Point spread function (PSF) photometry was performed on the P48 images using the Palomar Transient Factory Image Differencing and Extraction (PTFIDE) pipeline \cite{Masci2017}, while the P60 images were reduced using an automated pipeline \cite{Fremling2016}. The photometric evolution of the source is shown in Figure 2 and the data are presented in Table \ref{tab:lc}. The data have been corrected for Galactic extinction \cite{Schlafly2011} for $E(B-V) = 0.082$ mag and $A_{\textrm{V}} = 0.255$ mag. We do not expect any additional host extinction owing to the remote location of the transient, which is consistent with the absence of Na I D absorption lines in all our spectra. \\

Additional follow-up photometry  in $BVgri$ bands was obtained with the Las Cumbres Observatory (LCO) 1-meter telescope located at the McDonald Observatory \cite{Brown2013}. These data were processed using the tools available in the \texttt{lcogtsnpipe} package followed by PSF photometry \cite{Valenti2016}. Owing to the faint peak magnitude of the source (close to the sensitivity limit of LCO), the photometry obtained from these observations are relatively noisy due to their low signal-to-noise ratio. Since P60 obtained contemporaneous observations with LCO with much higher signal to noise ratio, we chose not to include the LCO data in our analysis. Nonetheless, the contemporaneous photometry from LCO and P60 are completely consistent with each other, and we present the LCO photometry in Table \ref{tab:lc} for completeness.

\subsubsection*{\textit{Swift} UV / X-ray observations}
We triggered \textit{Swift} follow-up of the source  in the $V$, $B$, $UVW1$ and $UVW2$ bands with the \textit{Swift} Ultra-Violet/Optical Telescope (UVOT; \cite{Roming2005}) and X-ray follow-up with the \textit{Swift} X-ray telescope (XRT; \cite{Burrows2005}).  We processed the data with the {\tt HEAsoft} package \cite{heasoft} and detected the transient in the $UVW1$ and $UVW2$ bands in the first three and two epochs of observation respectively, while only non-detections were obtained at subsequent epochs. The UV photometric evolution is shown in Figure 2 along with the optical light curves, and the data are presented in Table \ref{tab:lc}, where we used empirical fits \cite{Cardelli1989} to compute extinction coefficients for the $UVW1$ and $UVW2$ bands. Only upper limits were obtained in all epochs of \textit{Swift} XRT observations. The corresponding flux limits are summarized in Table \ref{tab:14gqr_xrt}.

We also stacked all of the cleaned event files obtained over a period of 21.9 days, amounting to a total exposure time of 12.2 ks. We obtain a 5$\sigma$ flux upper limit of 1.24 $\times 10^{-3}$ counts s$^{-1}$ corresponding to a 0.3 - 10 keV unabsorbed flux limit of 3.5 $\times$ 10$^{-14}$ ergs cm$^{-2}$ s$^{-1}$ (assuming a photon index $\Gamma$ = 2). The Galactic neutral hydrogen column density along this line of sight is 5.7 $\times$ 10$^{20}$ cm$^{-2}$ \cite{Winkel2016}, yielding a corresponding X-ray luminosity limit of 3.4 $\times$ 10$^{41}$ ergs s$^{-1}$ for the source at a distance of $D$ = 284.5 Mpc.

\subsubsection*{Optical spectroscopy}
We obtained a sequence of spectroscopic observations of the source starting from 4.3 hours after first detection to +59 days after $r$ band peak using the Dual Imaging Spectrograph (DIS) mounted on the 3.5 m Astrophysical Research Consortium telescope at Apache Point Observatory (APO), the auxiliary port camera (ACAM; \cite{Benn2008}) on the 4.2 m William Herschel Telescope (WHT), Andalusia Faint Object Spectrograph and Camera (ALFOSC) on the Nordic Optical Telescope (NOT), the Gemini Multi-Object Spectrograph (GMOS; \cite{Hook2004}) on the Gemini North (N) telescope, the Low Resolution Imaging Spectrograph (LRIS; \cite{Oke1995}) on the Keck I telescope and the DEep Imaging Multi-Object Spectrograph (DEIMOS; \cite{Faber2003}) on the Keck II telescope. All spectra were reduced using standard tasks in IRAF and IDL, including wavelength calibration using arc lamps and flux calibration using standard stars.

\begin{figure}[h]
\centering
\includegraphics[width=0.7\textwidth]{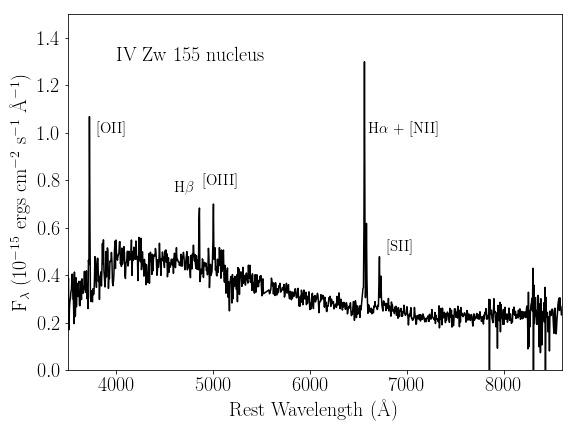}
\caption{\textbf{Spectrum of the nucleus of the apparent host galaxy of iPTF\,14gqr.} Prominent emission lines from the galaxy are marked in the spectrum.}
\label{fig:hostNucSpec}
\end{figure}

The sequence of spectra obtained are shown in Figure 3, and the times of the spectra are shown as dashed vertical lines in Figure 2. The spectroscopic observations are summarized in Table \ref{tab:spectra}. We were unable to obtain a high signal-to-noise ratio (SNR) spectrum of the transient at epochs beyond $\approx$ 30 days from light curve peak.  We also obtained a spectrum of the apparent host galaxy nucleus with APO DIS on 2014 October 14 (shown in Figure \ref{fig:hostNucSpec}) which was found to exhibit narrow emission lines of H$\alpha$, H$\beta$, [SII], [NII], [OII] and [OIII]. Additionally, we obtained one spectrum of the transient location $\sim$ 800 days after the explosion as a part of a spectroscopic mask observation and did not detect any nebular emission features at the source location. 
All spectra are available via the WISeREP repository \cite{Yaron2012}.

\subsubsection*{NIR imaging}
We observed the field of iPTF\,14gqr using the Wide Field Infrared Camera (WIRC; Wilson et al. 2003) on the Palomar 200-inch telescope on the night of 2014 Oct 19 (UT).  We obtained 21 images (1$\times$60 s each) using the $J$ filter, 25 images (2$\times$ 20\,s each) using the $K_s$ filter, and 24 images (2$\times$20\,s each) using the $H$ filter, for a total integration time of 21, 16.7, and 16 minutes respectively in each band.

The data were processed using a custom reduction pipeline including
flat-fielding and sky subtraction as well as special filtering steps to remove artifacts associated with the replacement detector in use at the time.  The source is well-detected in all three filters in the final stacks.  We performed aperture photometry within IDL and obtain magnitudes of $J = 19.76 \pm 0.08$, $H  = 19.58 \pm 0.12$, and $K_s = 19.05 \pm 0.15$.

\subsubsection*{Radio observations}
We observed iPTF\,14gqr with the Very Large Array (VLA) radio telescope on both 2014 October 15.4 and 2014 October 25.0. Each observation was performed using C-band (centered at 6.1\,GHz) and K-band (centered at 22\,GHz) in the C configuration. The Wideband Interferometric Digital Architecture (WIDAR) correlator was used in continuum mode with a 2 GHz bandwidth in C-band and a 8 GHz bandwidth in K-band. We analyzed the data with standard AIPS routines, using 3C\,48 as the flux calibrator and NVSS J234029+264157 as the phase calibrator. Our observations resulted in null detections in both bands at each epoch. The observational limits are $11.6$ microJanskys ($\upmu$Jy) and $11.7\,\upmu$Jy at C-band and K-band [measured as the $1\,\sigma$ root-mean-squared (RMS) noise of the reduced image], respectively, on 2014 October 15. On 2014 October 25, the observational limits are $13.0\,\upmu$Jy and $15.0\,\upmu$Jy, respectively. An additional limit was obtained using the Arcminute Microkelvin Imager (AMI; \cite{Zwart2008}) Large Array telescope. The AMI observation was undertaken on 2014 October 14.7 at a central frequency of 15\,GHz. The reduction of the AMI observation was conducted using the fully-automated calibration and imaging pipeline AMIsurvey \cite{Staley2015} and resulted in null detection with a $1\,\sigma$ RMS of $58\,\upmu$Jy.

\subsection*{Host imaging and spectroscopy}
\subsubsection*{Late-time imaging}

We undertook deep imaging of the source region in the $g$-band and $R$-band filters with LRIS on 2015 June 13 (MJD 57186.5) for a total exposure time of 960 s and 840 s respectively. The data were reduced and processed with standard image reduction procedures in \texttt{lpipe} \cite{lpipe}.  No source was detected at the transient location down to a 3$\sigma$ AB magnitude of $R$ $>$ 26.1  mag and $g$ $>$ 26.5 mag (without extinction correction). This constrains the presence of any stellar association at the location of the transient to $M_R >$ $-11.4$ mag and $M_g >$ $-11.1$ mag. Late-time images of the host galaxy region are shown in Figure \ref{fig:14gqr_lateDeep}.

\subsubsection*{Host environment spectroscopy}
\label{sec:specMask}

iPTF\,14gqr was discovered in the outskirts of an extended spiral galaxy showing clear signs of tidal interactions with nearby companions. Since none of the apparent extended sources near the transient region had measured redshifts in the NASA/IPAC Extragalactic Database (NED), we undertook one spectroscopic mask observation on 2016 November 28 ($\approx$ 800 days after discovery) of the region with LRIS on Keck I in order to confirm the interaction scenario. Additionally, this would allow us to ascertain whether any of the other nearby fainter galaxies could have potentially hosted the transient (i.e. was at a similar redshift) and if the spiral host galaxy itself was a part of a galaxy group or cluster. 

We selected a total of 32 extended sources classified as galaxies (including the apparent spiral host) in the Sloan Digital Sky Survey (SDSS) within 5.4$'$ of the transient location (out of a total of 254 objects) to place the slits on the spectroscopic mask, along with one slit at the location of the transient. The selection of sources for the slit mask was prioritized based on the projected distance of the galaxy from the transient. The spectra were reduced with standard routines in IRAF. Details of the spectroscopic mask observation are given in Table \ref{tab:spectra}, while Table \ref{tab:14gqr_hostSpec} lists the redshifts of the galaxies identified from the mask observation. The positions of the galaxies whose redshifts could be determined are shown in Figure \ref{fig:lateLRISMask_14gqr} as circles while all other cataloged SDSS galaxies are marked by crosses. 

The faintest source placed in the mask had a SDSS magnitude of $r$ $\approx$ 22.51 mag, while the faintest mask source within 100 kpc ($\approx$ 81$''$) of the transient had $r$ $\approx$ 21.89 mag. Amongst the galaxies whose redshifts could be determined, the faintest source had SDSS magnitude of $r$ $\approx$ 22.11 mag, while the same for galaxies within 100 kpc of the transient was $r$ $\approx$ 21.60 mag. Our redshift identification was complete for all sources brighter than $r$ $\approx$ 20.29 mag within 100 kpc of the transient, corresponding to an absolute magnitude of $M_r \approx -17.0$ at the source redshift. 

\subsection*{Photometric evolution}
\subsubsection*{Basic properties}
We summarize the basic photometric properties of the light curve in the $g$, $r$ and $i$ bands (where we had sufficient coverage) in Table \ref{tab:photoTable}. We find the peak magnitudes, time of maximum and corresponding rise time (between assumed explosion time and peak of light curve) in each filter by fitting a low order polynomial to the photometry near peak. We characterize the post-peak light curve in terms of the light curve decline rate (magnitudes per day). The absence of photometric data points beyond $\sim$ 10 days after peak does not allow us to estimate the more commonly used quantity $\Delta m_{15}$, the drop in magnitude in 15 days after light curve peak. The uncertainties on these parameters were estimated by Monte Carlo sampling of 1000 realizations of the photometric data points using their associated uncertainties. 

The observed peak absolute magnitudes are on the low end of the distribution of SN Ic peak magnitudes reported in \cite{Drout2011}, where SNe Ic were found to have peak $\langle M_R \rangle$ = $-18.3 \pm 0.6$ (uncertainties are 1$\sigma$), while they are within the typical peak magnitudes found in the sample of \cite{Taddia2017} (who find $\langle M_r \rangle$ = $-17.66 \pm 0.21$ and $\langle M_g \rangle$ = $-17.28 \pm 0.24$). The rise times of the light curves are shorter in the bluer bands as typically observed in Type Ib/c SNe \cite{Taddia2015}. When compared to other Type Ic SNe, the $r$ band light curve rise time for iPTF\,14gqr falls on the extreme low end of the distribution observed in Type Ic and Ic-BL SNe. For example, \cite{Taddia2015} find $\langle t_{rise} \rangle$ of 11.5 $\pm$ 0.5 days and 14.7 $\pm$ 0.2 days for Type Ic and Ic-BL SNe respectively based on light curve from the SDSS-II supernova survey. If the decline rates are converted to an equivalent $\Delta m_{15}$, we find the estimated $\Delta m_{15}$ ($\approx$ 2.5 mag in $g$ band and $\approx$ 1.3 mag in $r$ band) to be much higher than that observed for all normal Type Ib/c SNe \cite{Drout2011,Taddia2017}, but similar to those observed in the fast Type I events SNe\,2002bj \cite{Poznanski2010}, 2010X \cite{Kasliwal2010} and 2005ek \cite{Drout2013}.

\subsubsection*{Comparison with other sources}
We compare the multi-color light curves of iPTF\,14gqr to other relatively faint and fast evolving Type I SNe from the literature in Figure \ref{fig:14gqr_compareLC}. These include the core-collapse Type Ic SN\,1994I \cite{Richmond1996}, the Ca-rich transient PTF\,10iuv \cite{Kasliwal2012} and the rapidly evolving transients SNe\,2002bj \cite{Poznanski2010}, 2010X \cite{Kasliwal2010} and 2005ek \cite{Drout2013}. We plot the photometric evolution normalized to peak magnitude in the upper panel and on an absolute scale in the lower panel. Owing to the lack of observations in identical filters, we are constrained to compare the light curve evolution of these transients in corresponding pairs of $R$/$r$, $V$/$g$ and $I$/$i$ bands (which in general we refer to as $r$, $g$ and $i$ bands respectively). We first focus on a comparison of the second peak of the light curve of iPTF\,14gqr to that of these events.

Figure \ref{fig:14gqr_compareLC} shows that the light curve shape and timescales (for the second peak) of iPTF\,14gqr are broadly similar to several of the events in this sample. In general, the light curves are faster than those of PTF\,10iuv but slower than the very fast decay exhibited by SN\,2005ek. Overall, SN\,1994I exhibits light curves most similar to that of iPTF\,14gqr near peak light, although the rise time to peak for iPTF\,14gqr is shorter than that of SN\,1994I. For SN\,2005ek, the best upper limit for the rise time was at $\approx$ 9 days before $R$ band maximum while the same for SN\,2002bj was 7 days. On the other hand, iPTF\,14gqr attained a peak absolute magnitude (see Table \ref{tab:photoTable}) fainter than that of SN\,2002bj and SN\,1994I, but similar to that of SN\,2005ek and SN\,2010X. 

We compare the color curves of these transients to that of iPTF\,14gqr in Figure \ref{fig:14gqr_colorEvol}, in corresponding pairs of $V$/$g - R$/$r$ and $R$/$r - I$/$i$ colors. The sharp color jump after the first data point arises from the rapid blue first peak. Subsequently, iPTF\,14gqr has one of the fastest color $g - r$ color evolution among this set of transients, displaying a progression from a very blue transient ($g - r$ $\approx -0.4$ mag) at early times to a relatively red transient ($g - r$ $\approx 0.5$) within 10 days of explosion. This fast reddening is indicative of rapid cooling of the ejecta since the spectra at these phases are broadly consistent with featureless continua with weak broad features. We conclude that the multi-color light curves of the main peak of iPTF\,14gqr exhibit several similarities (light curve shape and timescales) as well as unique differences (short rise time) in this sample of transients.

The rapid first peak of the light curve is perhaps the most distinguishing feature of iPTF\,14gqr when compared to this sample of transients, and hence we compare this first peak to that of other known SNe exhibiting double peaked light curves in Figure \ref{fig:14gqr_firstPeak}. These include the Type Ic iPTF\,15dtg \cite{Taddia2016}, the GRB associated broad-lined Type Ic SN\,2006aj \cite{Brown2009}, the Type Ib SN\,2008D \cite{Bianco2014} and the Type IIb SN\,2011dh \cite{Arcavi2011}. We show the iPTF\,14gqr light curve in $g$ band since it has the best coverage. As shown, iPTF\,14gqr has one of the fastest rise times ($\leq 0.5$ days, as constrained ) and decay rates of the first peak in this sample. The width of the first peak of iPTF\,14gqr is most similar to that of the Type Ic-BL SN\,2006aj, while the peak absolute magnitude is similar to that that of iPTF\,15dtg.

\subsubsection*{Optical / UV SEDs}
We performed blackbody fitting of the multi-color photometry at all epochs for which we had photometric detections in 3 or more filters. In particular, we have two epochs with photometric data from all optical / UV bands, and the resulting blackbody SEDs are shown in Figure 3. The first epoch was within the first peak of the light curve (at $\approx$ 14 hours after explosion), where the UV / optical photometry is consistent with a blackbody of temperature $>$ 30,000 K. For comparison, we plot a blackbody fit of a spectrum obtained within an hour before this epoch of photometry. This spectrum is also well described by a blackbody consistent with the photometric fit within the uncertainties. We also plot a blackbody fit to the second epoch where we had complete multi-color coverage, which was near the main peak of the light curve (at $\approx$ 5.3 d after explosion). In this case, the black dashed line represents a blackbody fit to the optical photometry only ($\lambda_{\textrm{eff}} >$ 4000 \AA). The UV photometric points are found to be significantly fainter than the optical blackbody fit at this epoch (with T $\sim 10000$ K), which is indicative of significant line blanketing at UV wavelengths (as expected from Fe group elements in the ejecta). The NIR photometric magnitudes obtained near this epoch ($\approx 1$ day earlier) are also consistent with the optical blackbody fit.

\subsubsection*{Bolometric light curve}
We construct a bolometric light curve of iPTF\,14gqr using three methods. We first fit a Planck blackbody function to the observed photometry at all epochs where we have detections in 3 or more filters to obtain a best-fitting blackbody and corresponding temperature, radius and luminosity. The relatively featureless optical continua of the source allows us to obtain good blackbody fits at all epochs where multi-color photometry is available. We constrain the blackbody luminosity before the first detection by calculating the fraction of total flux within $g$ band at first detection ($\approx$ 2.6\%), and use it to estimate the luminosity upper limit with the latest pre-discovery limit in $g$ band. The resulting bolometric luminosity curve is shown in Figure 4 as black filled circles. The corresponding radius and temperature evolution is also shown in Figure 4. Although the blackbody approximation is valid within the first peak (as indicated by the best fitting SEDs), the fit overestimates the luminosity during the second peak of the light curve due to line blanketing in the UV.

We therefore compute a pseduo-bolometric light curve of the source by performing trapezoidal integration of the $gri$ photometric fluxes (from the lower wavelength edge of the $g$ band to the higher wavelength edge of the $i$ band) at all available epochs. Our peak photospheric spectra indicate the presence of additional flux at wavelengths $\lambda \leq 4000$ \AA\,(the lower edge of the $g$ band) as well as at $\lambda \geq 8000$ \AA\,(the upper edge of the $i$ band), which are not accounted for by a simple trapezoidal integration. Hence, we estimate the contribution of these fluxes by integrating the optical spectra from 4000 \AA\, to 8000 \AA\, and comparing them to the integrated value over 3000 \AA\, to 10000 \AA\, (the full range of our optical spectra). Performing this procedure on multiple spectra around peak light, we find that the simple trapezoidal integration underestimates the total optical flux by a factor ranging from 1.42 to 1.56 over multiple epochs. Since we do not have simultaneous spectroscopy with all epochs of multi-band photometry, we choose to scale the fluxes obtained from a trapezoidal integration by an average factor of 1.48, while conservatively adding an uncertainty of 10\% to account for the possible errors on this fraction. 

The pseudo-bolometric light curve is shown in Figure 4, which is found to be consistently smaller than the blackbody luminosity as expected. The true bolometric luminosity for the second peak lies in between these two estimates, and is likely to be very close to the pseudo-bolometric luminosity we estimate. Hence, we use the pseudo-bolometric luminosities for modeling the properties of the second light curve peak. We also compute a $g$ band luminosity $\lambda \textrm{F}_{\lambda}$ for comparison, and show it as black empty circles in Figure 4. 

Since our observations did not capture the rise to the first peak, we can only constrain the peak luminosity of this phase to be greater than $\sim 2 \times 10^{43}$ ergs s$^{-1}$ (i.e. the luminosity near first detection), while the corresponding temperature at the first peak to be higher than $\sim 32,000$\,K. The second peak reaches a peak (pseudo-bolometric) luminosity of $\sim 2.2 \times 10^{42}$ ergs s$^{-1}$ with a peak temperature of $\sim$ 10,000 K.  The total integrated energy emitted within the second peak (starting from $\approx$ 2 days to $\approx$ 14 days after explosion) of the pseudo-bolometric light curve is $\sim$ 1.8 $\times 10^{48}$ ergs. The blackbody temperatures exhibit rapid cooling from $> 32,000$ K to $\sim 10,000$ K at initial phases ($<$ 2 days after explosion), followed by a slower cooling phase at later epochs. Similarly, the blackbody radius exhibits an initial fast increase with time, followed by a slower increase at later phases. A linear fit to the photospheric radius evolution after peak light (between 8 and 15 days after explosion) produces an expansion velocity of $\sim 11,400$ km s$^{-1}$, while the same for the early light curve ($<$ 2 days from explosion) gives an expansion velocity of $\sim 33,500$ km s$^{-1}$ .

\subsection*{Spectroscopic evolution}
\subsubsection*{First peak}

The spectroscopic sequence for iPTF\,14gqr is shown in Figure 3. The earliest spectra were obtained within a day from explosion, and exhibit intermediate width emission lines of He~\textsc{ii}, C~\textsc{iii} and C~\textsc{iv} which evolve rapidly with time. In particular, the earliest +13.9 h spectrum exhibits emission lines of He~\textsc{ii} ($\lambda$4686) and C~\textsc{iv} ($\lambda$5801, 5812) with FWHMs of $\sim$ 5000\,km\,s$^{-1}$ and $\sim$ 2000 km s$^{-1}$ respectively, although the He~\textsc{ii} line may be contaminated with emission from a nearby C~\textsc{iii} $\lambda$4650 line. The presence of C~\textsc{iii} is confirmed from the spectrum taken at +25.2 hours where the C~\textsc{iii} $\lambda$5696 line becomes prominent.

The intermediate width He~\textsc{ii} line weakens in the +30.5 h spectrum while exhibiting signatures of an extended red wing, before disappearing in a spectrum taken about 3 hours later. We highlight the rapid evolution of the He~\textsc{ii} line in Figure 3, where we plot F$_{\lambda}$ as function of velocity shift from the He~\textsc{ii} line for all spectra taken within the first day of detection. In Figure \ref{fig:14gqr_compareFlashSpec}, we compare our early spectra to those of the SNe iPTF\,13ast \cite{GalYam2014} and iPTF\,13dqy \cite{Yaron2017}, where early spectroscopy had also revealed significant temporal evolution of the flash ionized spectra. Both these events were of Type II and hence exhibit prominent H emission lines which are absent in iPTF\,14gqr. The He~\textsc{ii} $\lambda$4686 line is a common prominent feature of the flash ionized spectra of these events, and the C~\textsc{iii} $\lambda$4650 and C~\textsc{iv} $\lambda$5801 lines were also observed in iPTF\,13ast. Note that the flash ionized spectra of these events exhibit significantly narrower emission lines (with central FWHM of $\sim$ 100 km s$^{-1}$ superimposed on broad Lorentzian wings) compared to those of iPTF\,14gqr.

The spectral evolution of the C high ionization lines in the early spectra is very similar to that seen in the WC sub-type evolution of galactic Wolf-Rayet stars \cite{Crowther2007, Sander2012}. In particular, the C~\textsc{iii} $\lambda$5696 / C~\textsc{iv} $\lambda$5801 ratio increases in the later and cooler sub-types (WC7 - WC9) of this class, consistent with the increasing ratio observed in this source with decreasing photospheric temperature.  On the other hand, the C~\textsc{iii} $\lambda$4650 line  decreases in strength with decreasing temperature in the later WC stars as the C~\textsc{iii} $\lambda$5696 line becomes stronger \cite{Sander2012}. We can confirm the presence of He~\textsc{ii} in the spectra by noting that the $\lambda$4686 emission feature is continuously present from +13.9 h to +30.5 h, even after the C~\textsc{iii} $\lambda$5696 disappears in the +30.5 h spectrum (this C~\textsc{iii} feature is expected to become stronger than the $\lambda$4650 feature at lower temperatures \cite{Sander2012,Crowther2007}), suggesting that this emission feature has a dominant contribution from a species different from C~\textsc{iii} (i.e. He~\textsc{ii}). Nonetheless, the the $\lambda$4686 feature in the earlier spectra (at and before 25 h) may have a contribution from C~\textsc{iii} $\lambda$4650, as is observed in the earlier sub-types of WC stars\cite{Sander2012}.

\subsubsection*{Photospheric phase}
Spectra taken about a week after explosion show characteristic absorption features of Type Ic SNe, including lines of Fe~\textsc{ii}, Ca~\textsc{ii}, O~\textsc{i} and Ti~\textsc{ii}. We compare the photospheric phase spectra of iPTF\,14gqr to those of other fast and normal Type Ic SNe in Figure \ref{fig:14gqr_comparePeakSpec}. The comparison candidates include the fast Type Ic events SN\,2010X \cite{Kasliwal2010} and SN\,2005ek \cite{Drout2013}, as well as the spectroscopically normal events SN\,1994I \cite{Filippenko1995} and SN\,2004aw \cite{Taubenberger2006}. The phases of the spectra indicated in this section are relative to $r$/$R$-band peak because the explosion times for the literature events are not well constrained. The comparison clearly shows that the photospheric spectra of iPTF\,14gqr remain relatively blue and featureless compared to those of the normal Type Ic SNe (SN\,1994I and SN\,2004aw) at similar phases. 

The only prominent features in the $-1$ d and $+0$ d spectra are those of Fe~\textsc{ii} and Ca~\textsc{ii}, which are also seen in the spectra of the other events. Also apparent are weaker features of C~\textsc{ii} at 6300\,\AA $ $ and 6900\,\AA. The spectrum obtained at +4 days shows a progressively cooling continuum as P-Cygni absorption features of O~\textsc{i} and Ca~\textsc{ii} become prominent in the red part of the spectrum. A possible weak absorption feature of Si~\textsc{ii} also appears near 6200 \AA.  The continuum shape as well as several spectral features (Fe~\textsc{ii} and Ca~\textsc{ii} on the blue side in particular) at this phase are best matched to that of SN\,2005ek near peak light. However, SN\,2005ek also displays prominent absorption features of Si~\textsc{ii} and C~\textsc{ii} near 6500 \AA, which are relatively weaker in iPTF\,14gqr.

We measure the photospheric velocities from the spectra obtained at $+0$ d and $+4$ d from $r$ band maximum. We are restricted to measuring only the Fe~\textsc{ii} $\lambda$5169 velocity since we do not see a prominent Si~\textsc{ii} $\lambda$ 6355 line in any of our photospheric spectra. We measure the velocity of the Fe~\textsc{ii} line by fitting a parabola to the absorption minimum. The resulting fits give a velocity of $\approx 10,500$ km s$^{-1}$ and $9,600$ km s$^{-1}$ for the $+0$ d and $+4$ d spectra respectively. A similar fitting procedure performed on the O~\textsc{i} $\lambda$7773 P-Cygni profile in our +4 d spectrum gives a velocity of $\approx 9,100$ km s$^{-1}$.

Type Ic SNe generally also exhibit the nearby Fe~\textsc{ii} lines of $\lambda$5018 and $\lambda$4924, which are blended into a single blue-shifted feature with respect to the $\lambda$5169 line (see for example, the spectra of SN\,1994I and SN\,2004aw in Figure \ref{fig:14gqr_comparePeakSpec}). The three features are blended into a single broad absorption component in the case of the high velocity Type Ic-BL SNe, and can potentially cause errors in a velocity measurement if this effect is not taken into account \cite{Modjaz2016}. Although we do not separately detect these features in our spectra, we do see a broad Fe II absorption feature, so followed the methods given in \cite{Modjaz2016} to ascertain if the absence of these features could be explained by velocity broadening. The best fitting models produce absorption features that have markedly different shapes, in particular, that are symmetric with respect to the minimum, unlike the shape observed here, and hence, a high velocity broadening is unlikely to be present. Thus, we conclude that the $\lambda$5018 and $\lambda$4924 features are present (owing to the presence of an asymmetric Fe feature) but not prominent enough to create a separate absorption feature in these spectra.

\subsubsection*{Early nebular phase}

Our final spectrum (with good SNR) was obtained $\approx$ 34 days after explosion, and show that the source was transitioning very early into the nebular phase. Prominent features in this spectrum include the Ca~\textsc{ii} IR triplet, [Ca~\textsc{ii}] $\lambda \lambda$7291, 7324, and a weak feature of [O~\textsc{i}] $\lambda \lambda$ 6300, 6364. In particular, the early appearance of [Ca~\textsc{ii}] $\lambda \lambda$7291, 7324 and the apparent high ratio of [Ca~\textsc{ii}]/[O~\textsc{i}] is similar to that seen in the class of Ca-rich gap transients \cite{Kasliwal2012, Lunnan2017}. We compare the only nebular spectrum of iPTF\,14gqr to the nebular spectra of other Type I SNe which exhibited an early nebular transition at similar phases in Figure \ref{fig:14gqr_compareNebSpec}. This sample includes the Type Ic SNe SN\,1987M \cite{Filippenko1990} and SN\,1994I \cite{Filippenko1995}, the Ca-rich transients PTF\,10iuv \cite{Kasliwal2012} and SN\,2012hn \cite{Valenti2014}, SN\,2010X \cite{Kasliwal2010} and SN\,2005ek \cite{Drout2013}. 
 
Amongst the normal Type Ic SNe, the nebular spectrum of iPTF\,14gqr is perhaps closest to that of SN\,1987M in terms of the prominent nebular lines ([Ca~\textsc{ii}] $\lambda \lambda$7291, 7324 and the Ca~\textsc{ii} near-IR triplet), albeit at a later phase ($\approx$ +60 d). SN\,1994I exhibits a stronger [O~\textsc{i}] feature compared to iPTF\,14gqr. Both SN\,2010X and SN\,2005ek also show a strong Ca~\textsc{ii} near-IR feature but still show prominent P-Cygni profiles of O~\textsc{i} near 7700 \AA. The best spectral match to iPTF\,14gqr in this sample are to that of the Ca-rich gap transients PTF\,10iuv and SN\,2012hn. Both these transients subsequently evolved to exhibit a strong [Ca~\textsc{ii}] 7300 \AA $\,$ feature at later phases (with a high [Ca~\textsc{ii}]/[O~\textsc{i}] ratio) characteristic of Ca-rich transients. We could not obtain a good SNR spectrum of iPTF\,14gqr at later epochs to trace the evolution of the Ca features, although the association of iPTF\,14gqr to the class of Ca-rich transients is unlikely (see supplementary text).

\subsection*{Modeling}
\subsubsection*{Arnett Model for the main peak}

Type I SNe which do not show signs of interaction (such as iPTF\,14gqr) are predominantly powered by energy released in the radioactive decay chain of $^{56}$Ni to $^{56}$Co to $^{56}$Fe. Since the peak photospheric spectra of iPTF\,14gqr display a number of similarities to those of fast and normal SNe Ic, we prefer the scenario where this peak is powered by $^{56}$Ni decay as in normal SNe Ic, and derive explosion parameters using a simple Arnett model \cite{Arnett1982}. We use the analytic light curve expressions for a $^{56}$Ni powered photospheric phase light curve given in \cite{Lyman2016} and \cite{Valenti2008a} to fit the pseudo-bolometric light curve of iPTF\,14gqr. We exclude the data points within the first peak ($<$ 2 days from assumed explosion) for this modeling.

The only parameters of this model are the nickel mass $M_{\textrm{Ni}}$ and the photon diffusion timescale $\tau_M$. The photon diffusion timescale is related to the ejecta mass $M_{ej}$ and explosion kinetic energy $E_{k}$ by equation (3) in \cite{Lyman2016}, or equivalently to the peak photospheric velocity $v_{ph}$ and $M_{ej}$ by their equation (1). We fit the observed bolometric light curve to this model using the Markov Chain Monte Carlo (MCMC) method in the Python \texttt{emcee} package \cite{Foreman-Mackey2013}. Keeping the explosion time $t_0$ as an additional free parameter, we obtain a best-fit model (shown in Figure 4) with $M_{Ni} = 0.051^{+0.002}_{-0.002}$ M$_{\odot}$, $\tau_M$ = $4.57^{+0.77}_{-0.62}$ days and $t_0 = -0.94^{+0.49}_{-0.62}$ days (i.e. the explosion occurred 0.94 days before the assumed explosion time). The uncertainties indicate 68\% confidence intervals estimated from the MCMC simulations although these are likely more conservative because we adopted conservative uncertainties when calculating the bolometric light curve. We caution that the Arnett model has several simplistic assumptions (see below) which likely affect the estimation of the uncertainty intervals.

The best-fitting explosion time is earlier than our last non-detection (0.43 days before assumed explosion), although the predicted flux from the Arnett model would be below our detection threshold. However, this explosion time could still be inconsistent with the pre-discovery limits depending on the exact evolution of the first peak. We show the degeneracies between the various model parameters in Figure \ref{fig:14gqr_arnettCorner}, where the degeneracy between $t_0$ and $\tau_M$ is particularly prominent, since $\tau_M$ controls the width of the light curve. For a photospheric velocity of $10^4$\,km\,s$^{-1}$ and optical opacity of $\kappa_{opt}$ = 0.07 cm$^2$ g$^{-1}$ relevant for stripped envelope SNe \cite{Cano2013,Lyman2016,Taddia2017}, we derive $M_{ej} = 0.23_{-0.06}^{+0.08}$ M$_{\odot}$ and explosion kinetic energy of $E_K$ $= 1.38_{-0.35}^{+0.51} \times 10^{50}$ ergs, where the uncertainty intervals (given at 68\% confidence) may again be affected by the assumptions of the model. The lack of late-time photometric coverage does not allow us to put any constraints on this model at late times.

The several assumptions of the Arnett model may also affect our estimates. These include assumptions of homologous expansion, spherical symmetry, completely centralized location of $^{56}$Ni and of optically thick ejecta (as in the photospheric phase). The rise and early decay of the bolometric light curve depends sensitively on the extent of mixing of $^{56}$Ni in the outer layers of the ejecta. A comparison of parameters derived from the Arnett model to hydrodynamic simulations by \cite{Dessart2016} suggest that the Arnett model can over-estimate the $^{56}$Ni mass by about 50\%. Additionally, they find the assumption of constant opacity to be a major limitation of this model since the derived parameters depend sensitively on the assumed opacity. For instance, if we adopt $\kappa_{opt} = 0.1$ cm$^2$ g$^{-1}$ (relevant for material with one electron per four nucleons, e.g., singly ionized He), our ejecta mass estimate would change to 0.16 M$_{\odot}$.

Nonetheless, we find our estimate of the extremely low ejecta mass (lower than previously known core-collapse SNe) and low explosion energy to be fairly robust. One potential caveat in the estimation of the ejecta mass is if the ejecta contain He that is effectively transparent. \cite{PiroMor2014} discuss the possibility that some stripped envelope SNe may have transparent He in their outer layers that does not contribute to the ejecta opacity, since He is effectively transparent in the continuum at low temperatures. They show that some stripped envelope SNe indeed show low photospheric temperatures ($< 10^4$ K) simultaneously with low photospheric velocities ($< 8000$ km s$^{-1}$), suggesting that some of the outer (and faster) He layers may be transparent. However, the photospheric temperatures in iPTF\,14gqr are higher than $10^4$ K up to peak bolometric light (near the epochs of the peak photospheric spectra) while the velocities are also high ($\sim$ $10,000$ km s$^{-1}$). Thus, the properties of this source are different from the events where He could have been hidden, suggesting that this effect is not prominent in this source.

\subsubsection*{Shock cooling model for the first light curve peak}
The collapse of the core in a SN explosion produces a radiation mediated shock that travels outward through the stellar envelope, accelerating and heating material along the way. When the optical depth to the shock becomes low enough, the shock breaks out while the shock-heated envelope subsequently cools, producing early optical / UV emission. \cite{Nakar2014} show that the shock cooling emission arising in normal progenitors (i.e. progenitors with no extended envelopes) can create double peaked light curves in the blue and UV bands, but not in the redder $R$ and $I$ bands. The presence of a double peaked structure in the redder bands requires the presence of a low mass extended envelope, where the low mass allows for a short photon diffusion time (and hence a rapid peak and decline) while the extended structure prevents large adiabatic losses at initial times. This corresponds to a case where the shock breaks out from the surrounding circumstellar material (also known as a `CSM breakout'; e.g. \cite{Waxman2016,Ofek2013}).

In the case of iPTF\,14gqr, we observe a double-peaked structure in $r$ band as well, and hence we refer to the extended progenitor models of \cite{Nakar2014}. The envelope mass can be derived using the velocity of the external envelope $v_{ext}$, the rise time to peak $t_p$ and the assumed opacity $\kappa$. While early spectroscopy near the first peak should allow a measurement of photospheric velocity at this phase, the relatively featureless continuum of our spectra do not allow us to measure this velocity. Instead, we use equation (13) in \cite{Nakar2014} to estimate the radius of the external envelope $R_{ext}$ from our first temperature measurement in the bolometric light curve (which is lower than the temperature at bolometric peak $T_{obs, peak}$). For $\kappa \approx$ 0.2 cm$^2$ g$^{-1}$ (relevant for electron scattering dominated opacity in a hydrogen free atmosphere), $t_p \leq$ 0.5 days and $T_{obs, peak} \geq 30000$ K, we get $R_{ext} \geq$ 1.5 $\times$ 10$^{12}$ cm. 

\cite{Piro2015} developed analytic light curves for the shock cooling emission from such an extended envelope, and we use them to compare to the early multi-color light curves of iPTF\,14gqr. The model uses a simple one-zone treatment of the extended envelope, where all of the mass $M_e$ is assumed to be concentrated around a radius $R_e$. The only other parameters are the velocity of the extended material $v_e$ and the mass of the core $M_c$, where we set $M_c = 0.23$ M$_{\odot}$ the explosion energy to $E = 1.38 \times 10^{50}$ ergs to derive $v_e$. We also let the explosion time $t_0$ vary such that the model is consistent with the last non-detection. Comparisons between the model and the data are done by performing synthetic photometry on the blackbody spectra predicted by the model. 

We fitted the model parameters with the observed data using the Markov Chain Monte Carlo method in the Python \texttt{emcee} package \cite{Foreman-Mackey2013}. We only use photometric detections obtained within 1.1 days from the assumed explosion date since we find later times to be affected by the rising portion of the $^{56}$Ni light curve. We find a best-fitting model with $M_e = 8.8^{+0.8}_{-0.7} \times 10^{-3}$ M$_{\odot}$, $R_e = 3.0^{+0.3}_{-0.3} \times 10^{13}$ cm ($430^{+43}_{-43}$ R$_{\odot}$), and $t_0 = -0.58^{+0.04}_{-0.04}$ days (i.e. the explosion occurred 0.58 days before the assumed explosion time). The resulting light curves for this set of parameters is shown in Figure 2. The error bars indicate the 68\% confidence interval for the derived parameters, as estimated from the MCMC simulations. 

The shock cooling models for an extended progenitor envelope reproduce the optical light curves (Figure 2), although there are discrepancies in the UV light curve. Such discrepancies were also noted by \cite{Piro2017} who suggest that the acceleration of the shock near the lower density edge of the envelope can cause a stronger temperature evolution than predicted by the one zone model, leading to a poorer match at shorter wavelengths. The optical light curves (particularly in the redder bands) start rising around 1.5 days after explosion, suggesting that the underlying $^{56}$Ni light curve becomes important at this phase. As in the case of the Arnett model, we expect the simple assumptions of this model to produce values which are only approximately correct. We thus conclude that the early shock cooling emission was produced by an extended envelope with a mass of $\sim 0.01$ M$_{\odot}$ and located at a radius of $\sim 500$ R$_{\odot}$.

We show the degeneracies between the model parameters in Figure \ref{fig:14gqr_shockCorner}. Although there are degeneracies between the parameters, the range of values occupy a relatively small phase space around the best-fitting value. We also consider how the assumed ejecta mass and explosion energy affect the derived parameters by considering the range derived from the Arnett modeling. Adopting the lower end of ejecta mass and explosion energy, we find a best-fitting model with  $M_e = 8.2^{+0.8}_{-0.6} \times 10^{-3}$ M$_{\odot}$, $R_e = 3.6^{+0.3}_{-0.3} \times 10^{13}$ cm, and $t_0 = -0.53^{+0.04}_{-0.05}$ days. On the other hand, adopting the higher end of the ejecta mass and explosion energy distribution, we find $M_e = 10^{+1}_{-0.9} \times 10^{-3}$ M$_{\odot}$, $R_e = 2.4^{+0.2}_{-0.2} \times 10^{13}$ cm, and $t_0 = -0.64^{+0.04}_{-0.04}$ days. Thus, our estimates of the envelope properties appear to be fairly insensitive to the adopted ejecta mass and explosion energy.

\subsubsection*{Interaction with a companion}

An alternative explanation of the early excess emission in iPTF\,14gqr might be the interaction of the SN ejecta with a non-degenerate companion. Such interaction signatures have been previously observed in some Type Ia SNe (\cite{Cao2015, Marion2016, Hosseinzadeh2017}; see \cite{Gal-Yam2017} for a review of SN classification) where comparison of the data to theoretical models \cite{Kasen2010} allows the inference of the orbital separation of the binary system. The models show that the presence of a companion can produce a void in the expanding ejecta, producing a reverse shock that powers excess luminosity at early times when viewed close to the direction of the void. In such a scenario, we might expect high ionization emission lines in the spectra arising from recombination of the companion wind, although the relatively large widths of the lines would suggest unusually high wind velocities for the companion. However, despite the high signal-to-noise ratio of our early spectra, we see no evidence for the presence of broad lines, as expected from the reverse shock produced in the ejecta-companion interaction.

We consider the photometric properties of the first peak in a companion interaction scenario. As shown in \cite{Kasen2010}, the early luminosity evolution depends on the viewing angle of the observer, where the excess flux is most prominent along the direction of the companion and relatively weak along directions perpendicular to or oriented away from the companion. In the case of iPTF\,14gqr, we observe a very rapid decline of the bolometric luminosity at early times ($\sim$ a factor of 10 in $<$1 day), ruling out viewing angles away from the companion (\cite{Kasen2010}, their figure 2). For viewing angles along the companion direction, the bolometric luminosity is well approximated by the analytic equations presented in \cite{Kasen2010}.

We attempted to fit the early multi-color photometry with the models presented in \cite{Kasen2010}, but were unable to obtain a good fit to the data. This is primarily because the luminosity from the interaction is relatively long-lived (a few days) when the peak luminosity is large (at least $\sim 2 \times 10^{43}$ ergs s$^{-1}$), while the predicted color evolution is very different. This is readily apparent when comparing the luminosity and temperature evolution of the model with the observations -- the analytical model predicts a luminosity evolution scaling with time as $t^{-1/2}$, which is almost the same as the color temperature evolution, which scales as $t^{-37/72}$. Our early observations clearly show that the luminosity drops by a factor of 10, while the temperature drops by only a factor of 3. Thus, we find the companion interaction scenario to be inconsistent with the properties of the early peak.

\subsubsection*{Analysis of early spectra}

The early spectra of iPTF\,14gqr exhibit prominent emission features of highly ionized He and C that are broader (FWHM $\sim 3000$ km s$^{-1}$) than that typically observed in the flash ionized spectra of other core-collapse events. We consider two possible scenarios for the origin of these features -- one where they arise from recombination of material surrounding the progenitor after ionization during the SN shock breakout, and one where they are powered by interaction of the SN ejecta with a dense CSM. While the intermediate width features seen in the early spectra are somewhat similar in width to those seen in interacting SNe at and before peak light \cite{Smith2016}, the observed recombination of highly ionized species is better explained by shock cooling of extended material. In particular, we observe clear evidence of recombination of C~\textsc{iv} in our earliest spectra followed by C~\textsc{iii} in later spectra, that is more consistent with shock cooling emission rather than continued energy supply from interaction. We thus favor shock cooling emission as the origin of the early excess emission as well as the intermediate width emission lines. 

We analyze the properties of the flash ionized spectra of iPTF\,14gqr to estimate properties of the emitting material surrounding the progenitor, where the optically thin flash ionized lines originate. Following the methods outlined in \cite{Ofek2013} and \cite{Yaron2017} where they estimate the mass loss rate from the luminosity of the H$\alpha$ line at 6563 \AA, we can relate the mass of the doubly ionized helium region generating the He II $\lambda$4686 line to the line luminosity using,
\begin{equation}
L_{\textrm{He II}, \lambda4686} \sim \frac{A}{m_{He}} \int_{r_1}^{r} n_e 4 \pi r^2 \rho (r) dr 
\end{equation}
where $L_{\textrm{He II}, \lambda4686}$ is the luminosity of the $\lambda$4686 line, $A = \frac{4\pi j_{\lambda 4686}}{n_e\,n_{He^{++}}}$, $j_{\lambda 4686}$ is the emission coefficient for the $\lambda$4686 transition, $m_{He}$ is the mass of a He nucleus, $n_{He^{++}}$ is the number density of the doubly ionized He and $n_e$ is the number density of electrons. The order of magnitude approximation holds because there may be unknown collisional excitation and de-excitation processes operating, in addition to the recombination processes considered here.

We use $A \approx 1.3 \times 10^{-24}$ ergs cm$^3$ s$^{-1}$ \cite{Storey1995} for a temperature of $10^4$ K, electron density of $\sim 10^{10}$ cm$^{-3}$ (the value of $A$ is fairly insensitive to the assumed electron density) and Case B recombination.  We consider a density profile with the density $\rho$ varying as $\rho = K r^{-2}$, where $K$ is the mass loading parameter. The total mass in a region between $r_1$ and $r$ is given by \cite{Ofek2013},
\begin{equation}
M_{He} = \int_{r_1}^{r} 4\pi r^2 \rho(r) dr = 4 \pi K \beta r
\end{equation}
where $\beta$ = $(r-r_1)/r$ is assumed to be of the order unity and $M_{He}$ is the total mass of the emitting material. Assuming $n_e \approx 2 n_{He^{++}}$, and using the density profile, we write the $\lambda$4686 luminosity as,
\begin{equation}
L_{\textrm{He II}, \lambda4686} \sim \frac{8 \pi A \beta}{m_{He}^2} \, K^2 \, r^{-1}
\end{equation}\\

We first attempted to measure the luminosity of the $\lambda$4686 line by subtracting the best-fitting blackbody continuum from the spectrum and fitting a Gaussian profile to the $\lambda$4686 line in the +13.9 h spectrum. With the absence of simultaneous photometry for the +25.2 h spectrum, we are unable to accurately measure the luminosity of the line in this spectrum. We thus only use the +13.9 h line luminosity for our calculations. Since the $\lambda$4686 feature likely has a non-negligible contamination from C~\textsc{iii} $\lambda$4650 feature at this epoch, we find that the simple Gaussian fit has a blue-shifted peak (with respect to $\lambda$4686) and a red shoulder excess (consistent with $\lambda$4686).

Using the +25.2 h spectrum which has a higher signal to noise ratio to only better estimate the relative contributions, we are able to get a better fit using a simple two component Gaussian profile, which suggests that the He~\textsc{ii} line has a FWHM of $\sim 4500$ km s$^{-1}$ and contributes 60\% of the flux in the line, while the C~\textsc{iii} feature has a FWHM of $\sim 2000$ km s$^{-1}$ and contributes the remaining 40\% of the flux. The width of the C~\textsc{iii} feature is similar to that measured for the C~\textsc{iv} $\lambda$5801 feature in the first spectrum. We are unable to better constrain the contributions of He~\textsc{ii} $\lambda$4686 and C~\textsc{iii} $\lambda$4650 in the +13.9 h spectrum than adopting the above measured ratio of 3:2 for the contribution of the He~\textsc{ii} and C~\textsc{iii} lines. This suggests that the He~\textsc{ii} line has a luminosity of $\sim 1.1 \times 10^{40}$ ergs. 

The location of the emitting region can be constrained by requiring that the Thompson optical depth in the region must be small for the lines to escape. The Thompson optical depth produced by this ionized material can be written as,
\begin{equation}
\tau \, = \frac{2 \sigma_T}{m_{He}} K \beta r^{-1}
\end{equation}
where $\tau$ must be small for the lines to escape. This gives a lower limit on the location of the line forming region at,
\begin{equation}
r \sim L_{\textrm{He II}, \lambda4686} \frac{\sigma_T^2 \beta}{2 \pi A \tau^2}
\end{equation}

Taking the +13.9 h luminosity measured for the $\lambda$4686 line, we get $r \sim 6 \times 10^{14} \beta \tau^{-2}$ cm, $K \sim 2.9 \times 10^{15} \tau^{-1}$ g cm$^{-1}$ and $M_{He} \sim 0.01 \beta^2 \tau^{-3}$ M$_{\odot}$ where we take $\beta \sim 1$. These estimates can be affected if the CSM was not characterized by a wind-like $r^{-2}$ profile, if the emitting region was confined to a thin shell ($\beta \ll 1$), if the emitting region was not spherically symmetric (e.g. if it was clumped) or if our estimate of the C III $\lambda$4650 contamination in the $\lambda$4686 emission feature was incorrect by a factor of a few. Given these caveats, we take the calculated values to be accurate to an order of an magnitude only. The inferred radius of the optically thin material is larger than the envelope producing the early shock cooling emission as seen in the light curve, and suggests that the highly ionized lines likely arise from a lower density extension of the same envelope.

An additional consistency check for the location of the emitting material can be found by noting that the line emission produced due to a short ionizing burst of radiation would be visible for at least the light crossing time between the part of the shell pointing towards the Earth and the part perpendicular to this line of sight (i.e. $t_{cross} = r/c$). This is because the inferred recombination times at these high densities ($n_e = 2 K r^{-2} / m_{He} \sim 10^{9} - 10^{10}$ cm$^{-3}$) are short ($\sim$ 10 minutes) and hence the emitting material recombines promptly after shock breakout. While there is some uncertainty in the time of explosion, the light curve models for the first peak, as well as the Arnett model, favor earlier explosion times than the one assumed. We thus constrain the first spectrum to have been taken between $\approx$ +14 h (for the assumed explosion time) and $\approx$ +36 h after explosion (for an explosion at 0.92 days before the assumed time). Similar constraints for the last featureless spectrum within the first day of detection suggest that it was taken between $\approx$ +33 h and $\approx$ +55 h after explosion. This constrains the emitting region to $r \leq 6 \times 10^{15}$ cm, noting that this estimate is relatively insensitive (within a factor of $\approx 2$) to the exact time (between +33 h and +55 h after explosion) when the lines disappear.

Given the large inferred size of the emitting envelope ($\sim 6 \times 10^{14}$ cm, corresponding to a light travel time of $\sim$ 5 hours), we now consider potential effects of light travel time (LTT) on the observed spectrum. A study of the flash-ionized spectrum of SN 2013cu \cite{GalYam2014} by \cite{Grafener2016} suggest that including LTT effects decreases the expected luminosity of the spectral lines and affects the observed line profiles depending on the location of the line forming region in the envelope. The lower luminosity has the equivalent effect of increasing the inferred mass of the envelope, as was inferred in the modeling of SN 2013cu \cite{Grafener2016}. While the emitting mass in iPTF\,14gqr may also be higher than our estimates, the rapid rise time of the early peak constrains the mass to be $\sim 0.01$ M$_{\odot}$.

\cite{Grafener2016} also show that lines formed in the outer region of the envelope are substantially blue shifted for high wind velocities, while lines formed in the central regions are less affected. Comparing to our observations, we note that the blue-shifted peak of the $\lambda$4686 feature and C~\textsc{iii} $\lambda$5696 lines in the +13.9 h and +25.2 h spectra can potentially be explained by LTT effects. The blue-shifted profile of the $\lambda$4686 feature may not require contamination by C~\textsc{iii} $\lambda$4650 in such a case. The observed blue-shift corresponds to a velocity of $\sim$ 2200 km s$^{-1}$, similar to the velocity inferred from the widths of the emission lines. 

Assuming that the electron density in the envelope is dominated by He ionization ($n_e \sim 10^{10}$\,cm$^{-3}$), we try to estimate the amount of C in the envelope. Using the C~\textsc{iv} line in the +13.9 spectrum, we measure a C~\textsc{iv} $\lambda$5801, 5812 luminosity of $\sim 6.1 \times 10^{39}$ ergs. Using a C~\textsc{iv} $\lambda$5801, 5812 recombination coefficient of $A = 1.4 \times 10^{-24}$ ergs cm$^3$ s$^{-1}$ \cite{Kingsburgh1995}, we determine a C mass of $\sim 4 \times 10^{-3}$ M$_{\odot}$. We can use our early spectra to place constraints on the amount of hydrogen in the envelope using the non-detection of the H$\alpha$ emission. We use the +13.9 h spectrum to place a 3$\sigma$ upper limit on the H$\alpha$ luminosity of $4.5 \times 10^{39}$ ergs s$^{-1}$. Taking an electron density of $n_e \sim 10^{10}$ cm$^{-3}$ in the emitting region and recombination coefficient of $A = 2.6 \times 10^{-25}$ ergs cm$^3$ s$^{-1}$ , we constrain the hydrogen mass of the envelope to a 3$\sigma$ upper limit of $\sim 10^{-3}$ M$_{\odot}$. 

\subsubsection*{Radio analysis}

Radio emission in SNe arises from synchrotron radiation produced by shock accelerated electrons in the circumstellar medium. As the forward shock propagates outwards, it accelerates electrons producing a power law electron energy distribution, which then gyrate in the post-shock magnetic field producing synchrotron radiation that peaks in the radio bands \cite{Chevalier1982,Chevalier1998}. Thus, radio emission can be used as a tracer of the environment of the explosion, and place constraints on the mass-loss history of the progenitor.

Radio emission from SNe has been previously observed in a number of events, and are often conspicuously associated with relativistic Type Ic-BL SNe (see \cite{Gal-Yam2017} for a description of this subtype), with or without an associated gamma ray burst (GRB) (see for example, \cite{Kulkarni1998,Soderberg2006}). Normal Type Ib/c SNe can also exhibit bright radio emission depending on the environment of the progenitor. We show a comparison of the observed radio light curves of other Type Ib/c SNe to the radio limits on iPTF\,14gqr in Figure \ref{fig:14gqr_radioLCcompare}. Our early radio observations rule out radio emission similar to most relativistic SNe, as well as the bright afterglows associated with GRBs (which are typically brighter than the Ic-BL SNe; see for example, \cite{Perley2014, Corsi2016}). However, we are not sensitive to the radio emission phase space occupied by the majority of the normal Type Ib/c population, primarily due to the large distance of iPTF\,14gqr and the resulting weak luminosity constraints.

In order to place the radio observations of iPTF\,14gqr in physical context, we use the synchrotron self-absorption (SSA) model of \cite{Chevalier1998} to generate analytic radio light curves for various kinds of circumstellar environments. We generate these light curves for two types of environments $-$ one characterized by a constant pre-explosion mass loss ($\rho = K r^{-2}$, where $K = \frac{\dot{M}}{4 \pi v_w}$ with $\dot{M}$ as the mass loss rate and $v_w$ as the wind velocity) and one for a constant density interstellar medium (ISM; $\rho$ = constant). Since our observations were obtained at relatively early times after the explosion (within $\sim$ 10 days of explosion), we assume that the shock velocity is constant, instead of assuming a density structure for the outer envelope and a corresponding velocity evolution. We consider a range of shock velocities for a filling factor of $0.5$ (see \cite{Chevalier1998}). We also assume that the electrons and magnetic field in the post-shock region share constant fractions ($= 0.1$) of the post-shock energy density, as denoted by $\epsilon_e$ and $\epsilon_B$ respectively, and an electron energy power law index of $p = 3$. 

We show model light curves and the radio upper limits in Figure \ref{fig:14gqr_radioLCModel}. The 22 GHz radio upper limits barely intersect the optically thick locus of the models, while the 6 GHz upper limits are well above the optically thick locus. Hence, these upper limits do not constrain the cicrumstellar environment. The models shown in Figure \ref{fig:14gqr_radioLCModel} are for the closest set of parameters to the radio upper limits, and correspond to $K \sim 6 \times 10^{12}$ g cm$^{-1}$ in the wind CSM case, and to $n = 3 \times 10^5$ cm$^{-3}$ in the constant $\rho$ case. The value of K corresponds to a $\dot{M} \approx 10^{-4} \frac{v_w}{1000\,\textrm{km s}^{-1}}$ M$_{\odot}$ yr$^{-1}$. For each circumstellar density model and observing frequency, we show light curves for shock velocities of $2 \times 10^4$ km s$^{-1}$ and $4 \times 10^4$ km s$^{-1}$. We conclude that while CSM parameters similar to those indicated here would have been detected at the 3$\sigma$ level in our 22 GHz observations, they do not constrain the environment any further.\\

\noindent

\subsubsection*{Ultra-stripped SN modeling}

We compare the bolometric light curve and peak photospheric spectra of iPTF\,14gqr to models of ultra-stripped SNe (as described in \cite{Tauris2013, Moriya2017}) in Figure 5. As discussed earlier, the optical / UV SED of the first peak of iPTF\,14gqr is consistent with blackbody emission through the optical / UV wavelengths, while the second peak SEDs show evidence of line blanketing in the UV. Hence, we compare these models to the bolometric luminosity for the first peak, and to  the pseudo-bolometric luminosity for the second peak. 

While the shape of the $^{56}$Ni powered peak of iPTF\,14gqr was found to be well matched to a model with 0.2 M$_{\odot}$ of ejecta and 2 $\times 10^{50}$ ergs of kinetic energy (similar to the explosion parameters estimated in our Arnett modeling), the observed light curve was found to be marginally brighter than that predicted in the model. This is because current models of ultra-stripped SNe synthesize $\approx 0.03 - 0.04$ M$_{\odot}$ of $^{56}$Ni, compared to the higher value ($\approx 0.04 - 0.05$ M$_{\odot}$) required to explain the luminosity of iPTF\,14gqr. Such discrepancies in the $^{56}$Ni mass may originate from small differences in the progenitor structure, core masses and / or explosion energy (note that 2D models synthesize higher $^{56}$Ni masses compared to the 1D models we use here \cite{Suwa2015}). Recent calculations of nucleosynthesis in ultra-stripped SNe \cite{Yoshida2017} suggest that there are likely to be additional power sources for these light curves (increasing the luminosity by up to a factor of $\sim 2$), including those from additional iron-peak isotopes (mainly $^{57}$Ni and $^{66}$Cu) as well as light trans-iron elements. Hence, the observed luminosity difference is not significant, and we explicitly set the $^{56}$Ni mass to 0.05 M$_{\odot}$ in our 1D models to account for the higher observed luminosity.

Since the early luminous peak arising from the He-rich extended envelope was not predicted in these models, and hence there were large discrepancies in the bolometric light curve at early times. Hence, we added a shock cooling emission component based on the models of \cite{Piro2015} for $M_e = 0.01$ M$_{\odot}$ and $R_e = 6 \times 10^{13}$ cm (as found earlier), and found the best explosion time by minimizing the residuals between the total model luminosity and the observed luminosity. Taken together, we were able to find an excellent match between the observed bolometric light curve and this two-component light curve for an explosion occurring $\approx 0.75$ days before the nominally assumed date (Figure 5).

The light curve models presented in \cite{Moriya2017} did not assume any $^{56}$Ni mixing in the ejecta, whereas the original synthetic spectra in \cite{Moriya2017} assumed complete mixing of the ejecta. This assumption leads to strong line blanketing below 4000 \AA, due to the relatively high fraction of Fe group elements in the outer layers of the ejecta (blue curve in Figure \ref{fig:14gqr_simSpecModels}). Hence, we recalculated these synthetic spectra in the same way as in \cite{Moriya2017} with the one-zone Monte Carlo spectral synthesis code developed by \cite{Mazzali1993}. Since we use the one-zone spectral synthesis code, we do not directly put the stratified chemical structure obtained by the explosive nucleosynthesis modeling into the code. We use the chemical composition just above the photosphere as input. This assumption is valid because most of the lines are formed just above the photosphere. Additionally, we generate these spectra for models with synthesized Ni mass of 0.05 M$_{\odot}$ to account for the higher luminosity.

The higher Ni mass in these models broadly lead to bluer spectra at peak bolometric light, as compared to the original models with 0.03 M$_{\odot}$ of Ni (blue curve in Figure \ref{fig:14gqr_simSpecModels}). Nonetheless, models with complete mixing of the ejecta predict deep absorption features (magenta curve in Figure \ref{fig:14gqr_simSpecModels}) and line blanketing at blue wavelengths (below $\sim 4500$ \AA) unlike what we observe. We find a much better match to the observed peak spectra for ejecta with no mixing and 0.05 M$_{\odot}$ of Ni (black curve in Figure \ref{fig:14gqr_simSpecModels}). As shown, the overall continuum shape, as well as all prominent absorption features (of O I, Ca II, Mg II and Fe II) are very well reproduced in this model. We also observe weak absorption features of Si II and C II near 6300 \AA\, as predicted in the model. Although there are some minor discrepancies in the strengths of the Mg II, Si II and C II features, we overall conclude that the spectral properties of iPTF14gqr are in very good agreement with the ultra-stripped SN models.

We show the chemical abundances of the photosphere found in this model in Table \ref{tab:USAbundance}. The specific model used here was directly taken from the binary evolution simulations of \cite{Tauris2015}, and also contains 0.03 M$_{\odot}$ of He. However, \cite{Moriya2017} show that He optical lines are not expected in this model even when including non-thermal excitations due to the low He mass, which is below the 0.06 M$_{\odot}$ limit suggested by \cite{Hachinger2012}. As shown earlier, the low $\gamma$-ray opacity of the ejecta also suggest that the lack of He excitation lines is likely due to a genuine low abundance of He in the ejecta, even in the limit of very low $^{56}$Ni mixing. Modeling of the light curve and spectral properties of the fast Type Ic SN 1994I, which was also suggested to arise from a highly stripped progenitor \cite{Nomoto1994}, also require a strongly centralized distribution of $^{56}$Ni \cite{Hachinger2012}.

\subsubsection*{Absence of He in the peak photospheric spectra}

The classification of iPTF\,14gqr as a Type Ic SN arises from the absence of H, He and prominent Si lines in the observed peak photospheric spectra. However, the absence of He lines in optical spectra can be either due to the genuine lack of He in the ejecta or due to lack of non-thermal He excitation in the He-rich regions of the ejecta \cite{Dessart2011, Dessart2012, Dessart2015}. In particular, it has been suggested that the absence of He lines in Type Ic SNe can also be explained only by the lack of $^{56}$Ni mixing into the outer He-rich layers, that prevents non-thermal excitation of He by fast electrons arising from Compton scattering of $\gamma$-rays produced in $^{56}$Ni decay. It is thus necessary to consider the effect of mixing in hiding He lines in our peak photospheric spectra.

Modeling a continuum of progenitor structures leading from Type Ib to Type Ic SNe (including a certain amount of mixing), \cite{Hachinger2012} show that a He mass of at least 0.06 M$_{\odot}$ is required to produce He lines in the optical spectra. Our modeling of the peak photospheric spectra and light curve of iPTF\,14gqr in the context of ultra-stripped SNe suggest that the $^{56}$Ni needs to be centrally located in order to explain the lack of line blanketing at $<$ 4000 \AA\, near peak optical light. While this could potentially contribute to hiding He lines in this source, we show below that such a scenario is unlikely due to the low ejecta mass of the system, and the resulting low $\gamma$-ray optical depth of the ejecta. 

We first estimate the $\gamma$-ray escape fraction from $^{56}$Ni decay for a centralized $^{56}$Ni distribution (as inferred from the modeling) using the methods in \cite{Valenti2008b}. The $\gamma$-ray escape fraction is given by $e^{-{(F/t)}^2}$, where $F \approx 32 M_{ej, \odot}$ / $\sqrt{E_{51}}$ days, $M_{ej, \odot}$ is the ejecta mass in units of solar masses and $E_{51}$ is the explosion energy in units of $10^{51}$ ergs. For the parameters inferred for iPTF\,14gqr, we get F $\approx 14.3$ days, and hence find the escape fraction at 11 days (the epoch of the last photospheric phase spectrum) to be $\approx$ 20\%. This is in stark contrast to normal Type Ib/c SNe, where the larger ejecta masses lead to negligible $\gamma$-ray escape up to $\sim$ 2 months after the explosion \cite{Dessart2011}. Hence, Ni mixing into the outer He layers has been suggested to be a necessary criterion for producing He optical lines in normal Type Ib/c SNe. In particular, \cite{Dessart2012} show that a Ni mass fraction of at least 0.01 in the outer layers is enough to produce prominent He optical lines in mixed ejecta, for He mass fractions as low as 0.2 - 0.3 in the outer layers. If we consider the outer 0.05 - 0.1 M$_\odot$ ejecta of iPTF\,14gqr, this corresponds to a Ni mass of (0.5 - 1) $\times 10^{-3}$ M$_{\odot}$ in the outer layers. 

Since the role of $^{56}$Ni in the outer ejecta layers is to non-thermally excite He by $\gamma$-ray energy deposition, we can compare the $\gamma$-ray flux required for the production of He optical lines to the actual rate estimated from the escape fraction. Considering the $\gamma$-ray escape fraction from the entire ejecta of 20\%, we find that the $\gamma$ ray flux from the centrally located 0.05 M$_{\odot}$ of $^{56}$Ni is at least an order of magnitude larger than what would be produced from (0.5 - 1) $\times 10^{-3}$ M$_{\odot}$ of $^{56}$Ni in the outer layers. Taken together, the lack of He excitation lines in the presence of a significant $\gamma$-ray flux suggests that the mass fraction of He in the outer layers is indeed low ($X_{He} < 0.2$, as suggested by \cite{Dessart2012}). Although the above estimate is very simplified compared to the detailed radiative transfer involved in the production of He lines, it nevertheless demonstrates that the ejecta of iPTF\,14gqr is deficient in He.

\subsection*{Host galaxy and environment}
\subsubsection*{Host identification}

iPTF\,14gqr was discovered at a large projected offset of $\approx$ 24$'' $ from the nearest apparent host galaxy, a two-tailed spiral galaxy showing signs of tidal interaction with at least three companion galaxies. The association of the transient with the host galaxy was confirmed from the early spectroscopy of the transient, which exhibited flash ionized He~\textsc{ii} lines at the same redshift $z = 0.063$ as that of the galaxy. The offset corresponds to a projected distance of $\approx$ 29 kpc at the redshift of the galaxy. We nominally adopt this galaxy as the host of iPTF\,14gqr, but also discuss several other possible scenarios below.

Although no stellar association was visible in the SDSS and PTF\,reference images at the location of the transient, it was not possible to rule out the presence of a faint dwarf galaxy or globular cluster underneath the transient. We thus undertook deep late-time imaging of the host region and did not find any stellar association down to absolute magnitude limits of $M_r > -11.4$ mag and $M_g > -11.1$ mag. Although these limits rule out a certain luminosity space for a potential host, they are not stringent enough to rule out the entire population of dwarf galaxies and stellar clusters given the observed luminosity function of these systems in the local group \cite{Lyman2014,Mcconnachie2012,Harris1996}. We also do not detect any galaxy emission lines at the location of the transient in our late-time spectra.

\subsubsection*{Host properties}

The host galaxy IV Zw 155 is a spiral of morphology Sa, and has been previously cataloged in the SDSS \cite{Abolfathi2017} and the Two Micron All Sky Survey (2MASS) \cite{Skrutskie2006} as the source SDSS J233329.84+333848.2 and 2MASX J23332981+3338484 respectively. The galaxy was also detected as a UV source in the Galaxy Evolution Explorer (GALEX) source catalogs \cite{Bianchi2014} as the source GALEXASC J233330.19+333840.3. The broadband photometric magnitudes of the galaxy are summarized in Table \ref{tab:lineFlux}.  We use the spectrum of the core of the galaxy to measure the emission line fluxes of H$\alpha$, H$\beta$, [O II], [O III], [N II] and [S II] by fitting Gaussian profiles to the individual lines. The fluxes obtained are given in Table \ref{tab:lineFlux}.

We estimate the metallicity of the host galaxy using the \texttt{pyMCZ} code \cite{Bianco2016} to calculate the host oxygen metallicity (12 + log(O/H)). The code is based on \cite{Kewley2002} with updates from \cite{Kewley2008}, and calculates a number of metallicity calibrators based on strong emission lines along with uncertainty estimates derived from Monte-Carlo sampling. The oxygen abundance estimates we derive are 8.58$^{+0.07}_{-0.08}$ on the O3N2 scale of \cite{Pettini2004} and 8.78$^{+0.13}_{-0.22}$ on the scale of \cite{Mcgaugh1991}. Similarly, using the multi-color integrated magnitudes of the galaxy in SDSS, we find 12 + log(O/H) = 8.77 on the 03N2 scale of \cite{Pettini2004} using the luminosity - color - metallicity relation of \cite{Sanders2013} with a typical scatter of 0.07. Overall, the estimates are roughly consistent with a solar metallicity host galaxy (where [12 + log(O/H)]$_{\textrm{solar}}$ = 8.69 $\pm$ 0.05; \cite{Asplund2009}).

We fit the broadband photometry measurements of the galaxy to estimate parameters for the host stellar population. We use the \texttt{FAST} code \cite{Kriek2009} to fit a galaxy model to the observed photometry using a \cite{Maraston2005} stellar population synthesis model, a Salpeter IMF, a delayed star formation history and a Milky-way (MW) like extinction law. Additionally, we constrain the metallicity to be Z = Z$_{\odot}$ based on our spectroscopic measurements. Using these inputs, we find a best-fitting stellar mass of 3.02$^{+0.36}_{-0.33} \times 10^{10}$ M$_{\odot}$ and a stellar population age of 7.08$^{+0.17}_{-3.69} \times 10^8$ yrs. The GALEX UV photometry measurements allow us to constrain the best-fitting star formation rate to 2.3 - 7.6 M$_{\odot}$ yr$^{-1}$. Overall, the host of iPTF\,14gqr is found to be a solar metallicity, star forming, spiral galaxy.

Type Ib/c SNe generally show a preference for high metallicity environments \cite{Smartt2009,Leloudas2011,Sanders2012}, which is consistent with the requirement of substantial line driven mass loss in the progenitors. Additionally, \cite{Arcavi2010} show that Type Ic SNe also show a preference for giant host galaxies as compared to Type Ib or Type II events. Thus, taken at face value, the host galaxy of iPTF\,14gqr is indeed similar to the typical host galaxies of Type Ic SNe. All of the fast Type I transients in our light curve comparison sample were also found in star forming galaxies, with the exception of PTF\,10iuv, which was found near an old elliptical galaxy.

\subsubsection*{Explosion site properties}

The 29 kpc central offset of the location of iPTF\,14gqr corresponds to a host normalized offset of $\approx$ 4.8 R$_{{\textrm{eff}}}$ (where R$_{{\textrm{eff}}}$ is the half-light radius of the galaxy), placing this source at the extreme high end of the distribution of host offsets found for all SNe (\cite{Prieto2008}; \cite{Kasliwal2012}). For example, iPTF\,14gqr has the second largest host offset in terms of physical distance when compared to the PTF\,sample of core-collapse SNe, with SN\,2010jp (PTF\,10aaxi; \cite{Smith2012}) as the only other core-collapse event with a larger offset ($\approx$ 33 kpc). When considered in terms of host-normalized offset of PTF\,SNe, iPTF\,14gqr occupies a position on the higher end of the distribution of even SN Ia as well as of short GRBs \cite{Lunnan2017, Fong2013}.

We can attempt to place limits on the presence of an underlying H II region based on the non-detection of nebular H$\alpha$ emission in in our late-time spectra. Using our LRIS spectrum obtained at $\approx$ 61 days after explosion, we place a 5$\sigma$ upper limit of the H$\alpha$ flux to be be $\sim$ 10$^{-17}$ ergs cm$^{-2}$ s$^{-1}$, corresponding to H$\alpha$ luminosity of 9.6 $\times$ 10$^{37}$ ergs s$^{-1}$. We thus derive a corresponding upper limit of 7.6 $\times$ 10$^{-4}$ M$_{\odot}$ yr$^{-1}$  on the star formation rate (SFR) \cite{Kennicutt1998} within the 1" slit ($\sim$ 1.2 kpc at the galaxy redshift) used for the spectrum. The H$\alpha$ luminosity upper limit is much lower than the average luminosity observed for H II regions associated with core-collapse SNe \cite{Crowther2013}, although it does not rule out the complete range of H$\alpha$ luminosities associated with Type Ib/c SNe. Nonetheless, the absence of a H II region would potentially suggest that the progenitor was not formed in situ at the location, or was older than the $\sim$ 10\,Myr lifetimes of giant H II regions \cite{Osterbrock2006}. This would be in contrast to the general trend where Type Ib/c SNe are found to be more likely to be associated with H II regions than Type II events, in accordance with the shorter lifetimes of their more massive progenitors \cite{Anderson2012,Crowther2013}.

\subsubsection*{Tidally interacting environment}

\begin{figure}[!ht]
\centering
\includegraphics[width=0.8\textwidth]{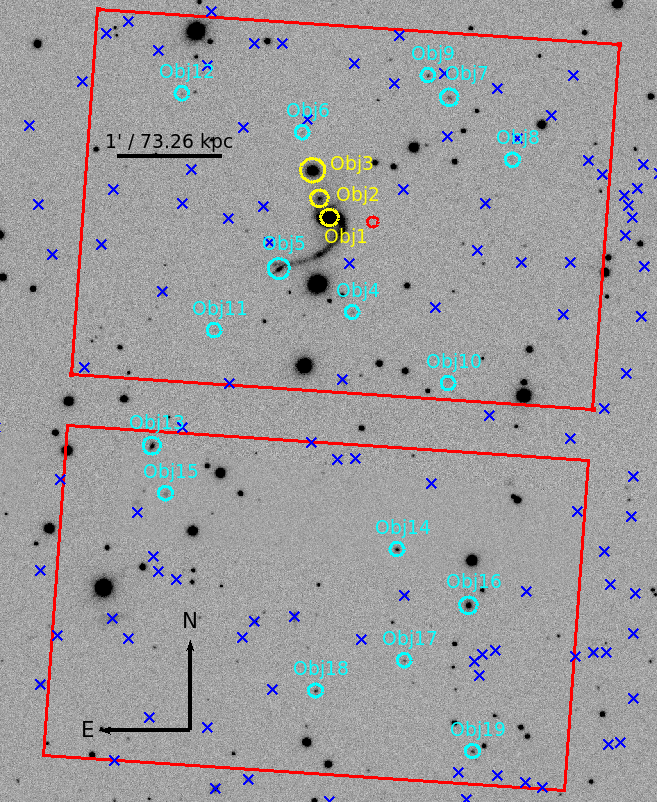}
\caption{\textbf{SDSS $r$ band image of the host environment of iPTF\,14gqr.} The red circle marks the location of the transient, while the yellow circles mark the positions of the galaxies confirmed to be at the same redshift as that of the transient. The cyan circles mark the locations of galaxies identified as background galaxies in the mask. Blue crosses mark the locations of all other objects classified as galaxies in SDSS. The red boxes indicate the field of view of Keck LRIS during the mask observation. Redshifts could not be securely determined for other objects in the mask due to low SNR of the spectra.}
\label{fig:lateLRISMask_14gqr}
\end{figure}

\begin{figure}[!ht]
\centering
\includegraphics[width=\textwidth]{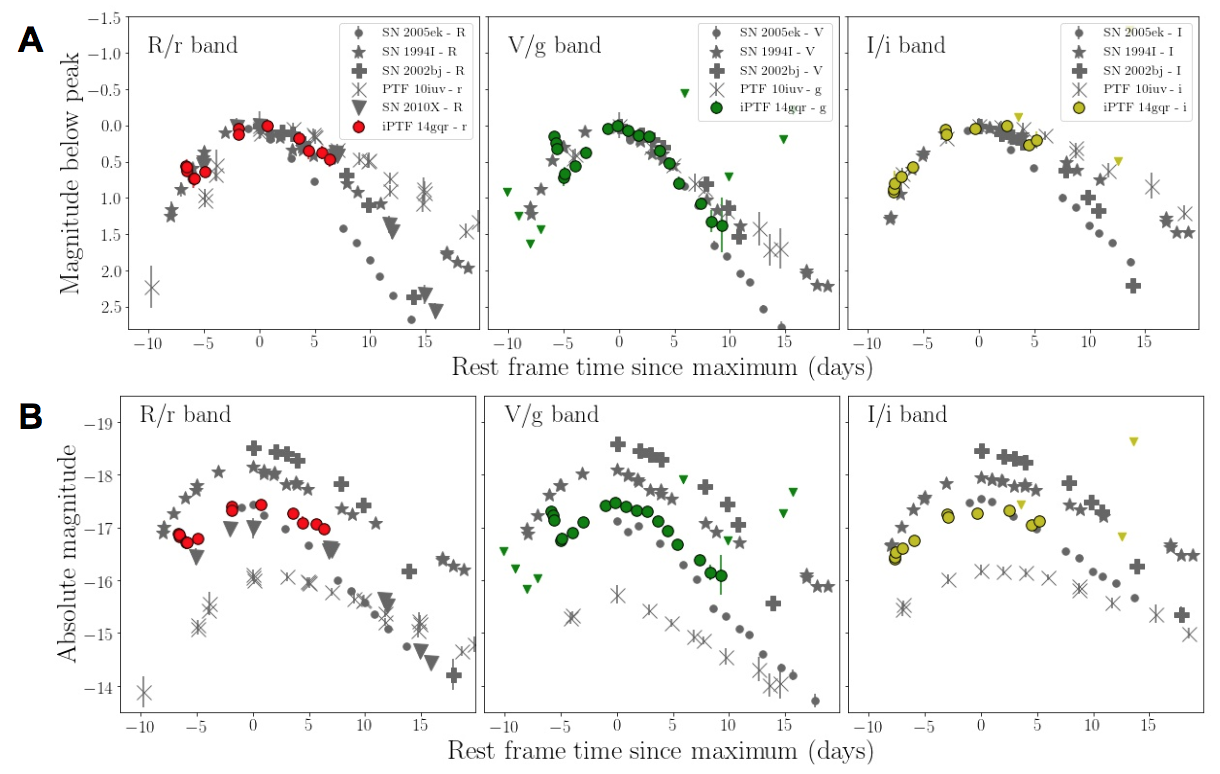}\\
\caption{\textbf{Comparison of the photometric evolution of iPTF\,14gqr to other fast evolving SNe}. These include SN\,1994I \cite{Richmond1996}, SN\,2002bj \cite{Poznanski2010}, SN\,2005ek \cite{Drout2013}, SN\,2010X \cite{Kasliwal2010} and PTF\,10iuv \cite{Kasliwal2012}. Panel A shows the photometric evolution normalized to peak magnitude, while the panel B shows the same on an absolute magnitude scale.}
\label{fig:14gqr_compareLC}
\end{figure}

\begin{figure}[!ht]
\centering
\includegraphics[width=0.49\textwidth]{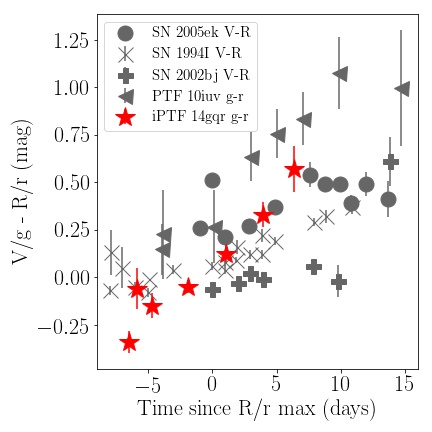}
\includegraphics[width=0.49\textwidth]{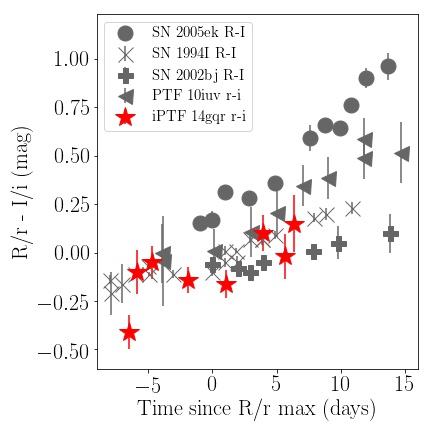}\\
\caption{\textbf{Comparison of the color evolution of iPTF\,14gqr with the other fast SNe shown in Figure \ref{fig:14gqr_compareLC}}. Due to absence of photometry in identical filters, we compare colors in corresponding filter pairs of $V$/$g$, $R$/$r$ and $I$/$i$ as indicated in the legend.}
\label{fig:14gqr_colorEvol}
\end{figure}

\begin{figure}[!ht]
\centering
\includegraphics[width=0.6\textwidth]{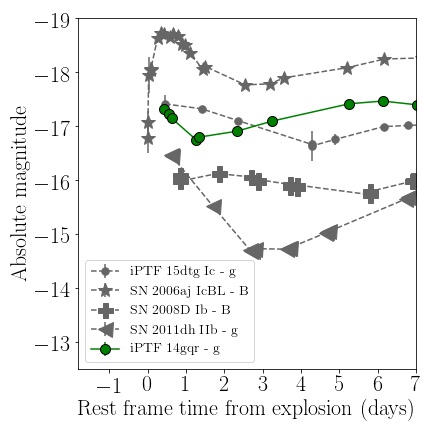}
\caption{\textbf{Comparison of the first peak of the light curve of iPTF\,14gqr with other known SNe exhibiting double peaked light curves.} The comparison sample includes the Type Ic SN iPTF 15dtg \cite{Taddia2016}, the Type Ic-BL SN 2006aj \cite{Brown2009}, the Type Ib SN 2008D \cite{Bianco2014} and the Type IIb SN 2011dh \cite{Arcavi2011}.} 
\label{fig:14gqr_firstPeak}
\end{figure}

\begin{figure}[!ht]
\centering
\includegraphics[width=0.7\textwidth]{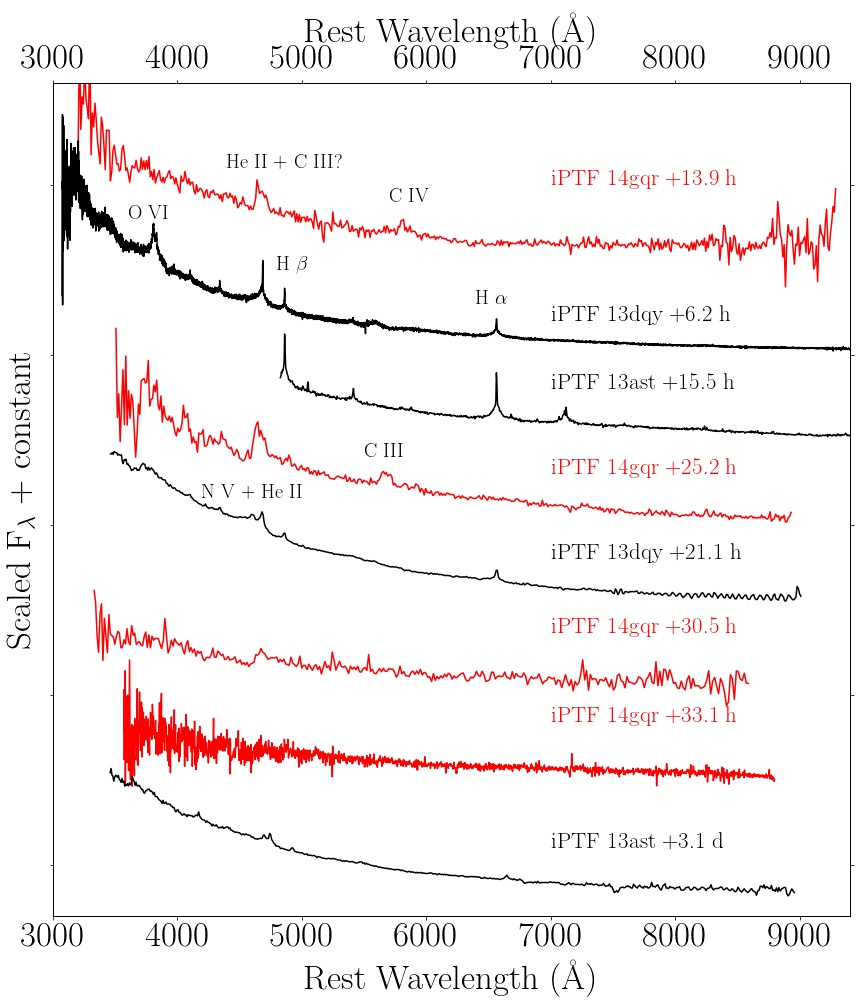}\\
\caption{\textbf{Comparison of early (flash) spectra of iPTF\,14gqr (red) to those of iPTF\,13ast and iPTF\,13dqy (black)}. These SNe also showed rapid temporal evolution in their early spectra. Prominent emission lines are marked along with phases (with respect to explosion time) of the spectra.}
\label{fig:14gqr_compareFlashSpec}
\end{figure}

\begin{figure}[!ht]
\centering
\includegraphics[width=0.7\textwidth]{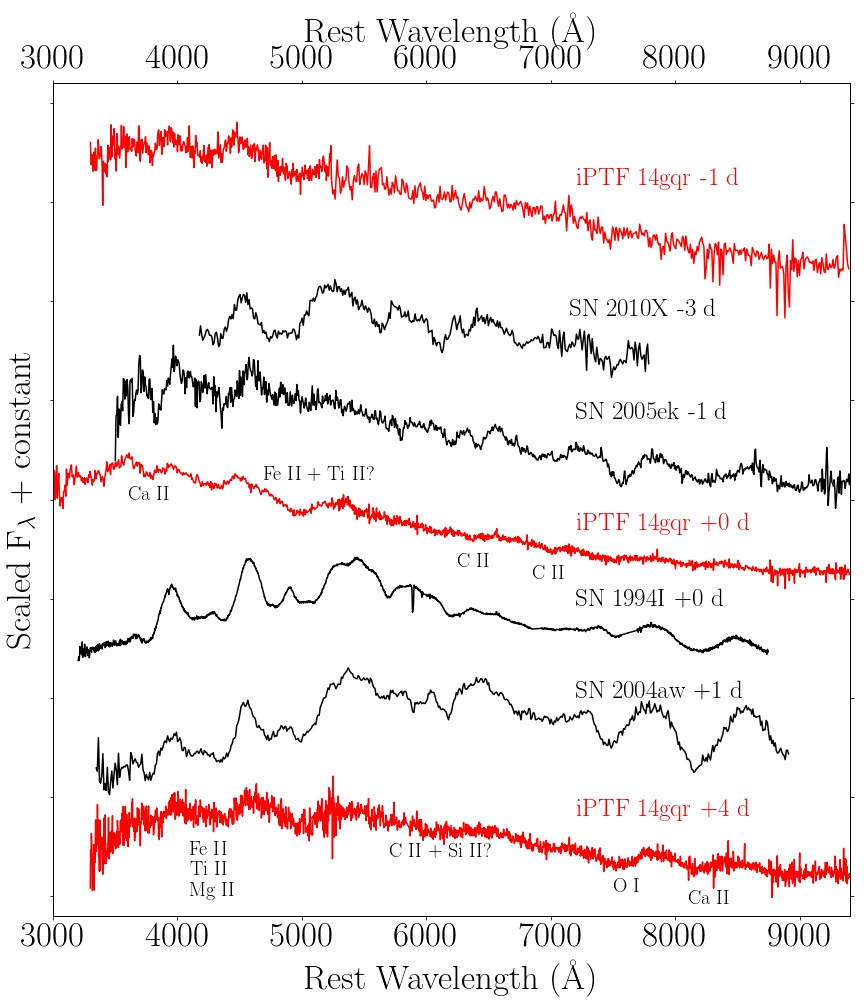}
\caption{\textbf{Comparison of the peak photospheric spectrum of iPTF\,14gqr (red) to some other known Type Ic SNe (black).} Indicated phases are with respect to R/r band maximum since the explosion times are not well constrained for the events from the literature. Prominent absorption features are marked in the spectra.}
\label{fig:14gqr_comparePeakSpec}
\end{figure}

\begin{figure}[!ht]
\centering
\includegraphics[width=0.7\textwidth]{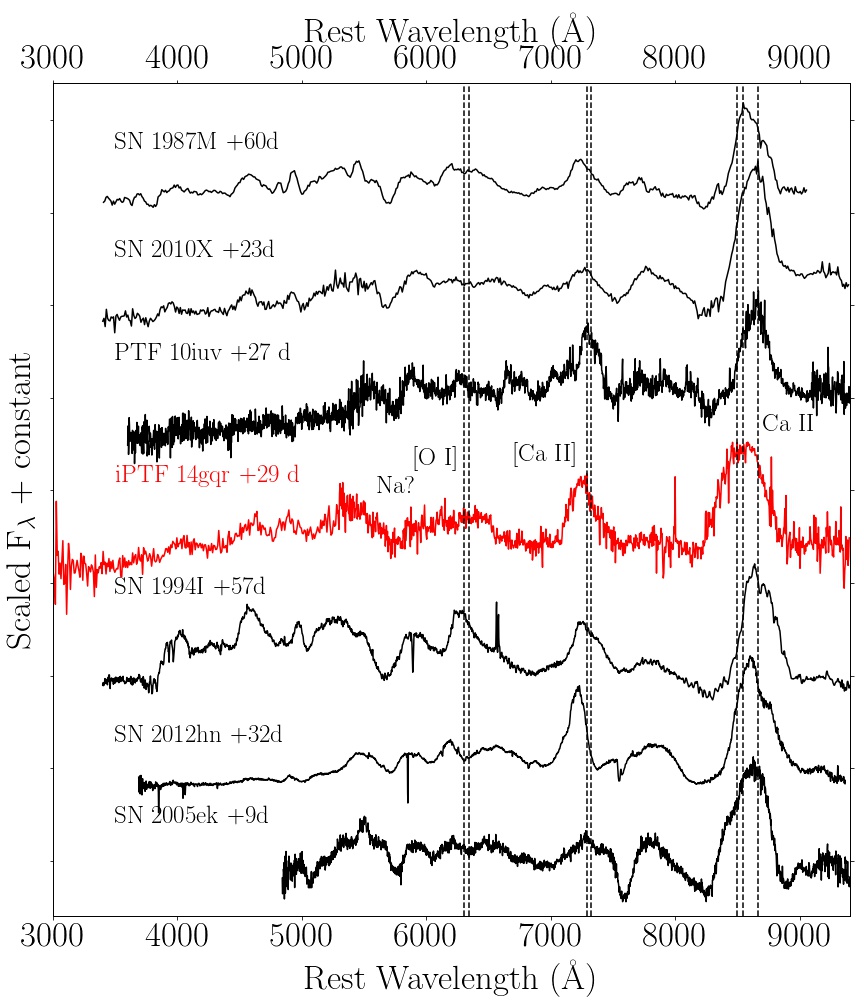}
\caption{\textbf{Comparison of the early nebular spectrum of iPTF\,14gqr (red) to other Type I SNe (black) which exhibited a fast nebular transition.} The phases of the spectra are with respect to r/R band peak. Prominent nebular lines and absorption features are marked. The dashed vertical lines mark the wavelengths of the [O~\textsc{i}] $\lambda \lambda$ 6300, 6364 doublet, the [Ca~\textsc{ii}] $\lambda \lambda$7291, 7324 doublet and the Ca~\textsc{ii} IR triplet. The feature near 5700 \AA\, may be from Na, as was suggested in SN 1987M \cite{Filippenko1990}, or due to Co~\textsc{iii}.}
\label{fig:14gqr_compareNebSpec}
\end{figure}

\begin{figure}
\includegraphics[width=\textwidth]{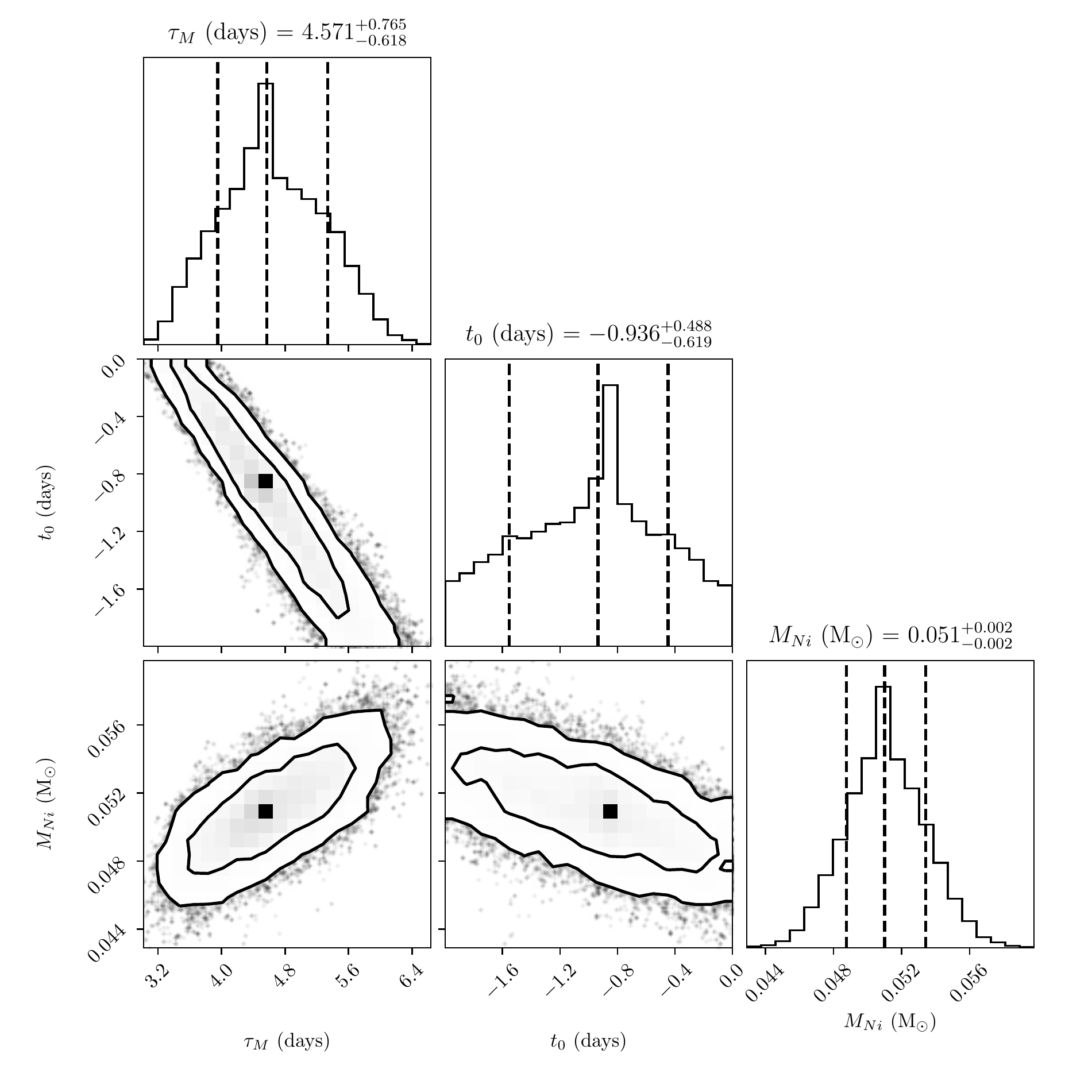}
\caption{\textbf{Confidence intervals of the parameters of the Arnett model that is used to fit the bolometric light curve of iPTF\,14gqr.} The plot has been derived from the MCMC simulations performed with the \texttt{emcee} package. The vertical lines in the histograms indicate the median (best-fitting) value along with the 68\% confidence intervals on the parameters. The contours in the two dimensional scatter plots represent 68\% and 95\% confidence intervals in the respective phase space. The best-fitting values along with their uncertainty intervals are also indicated above the individual histograms.}
\label{fig:14gqr_arnettCorner}
\end{figure}

\begin{figure}
\includegraphics[width=\textwidth]{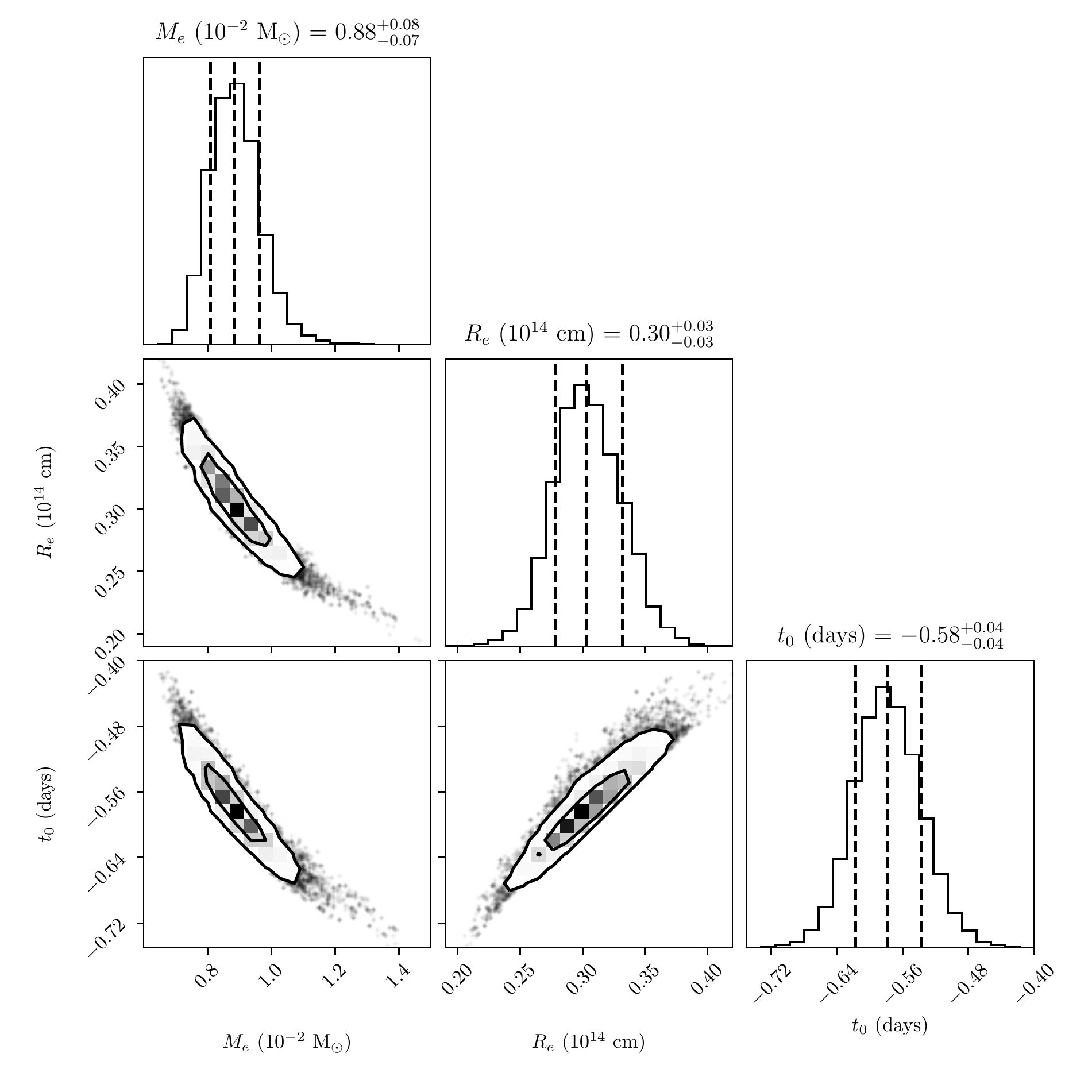}
\caption{\textbf{Confidence intervals of the parameters used to fit the early peak using the shock cooling models for extended progenitors, as presented in \cite{Piro2015}.} The figure shows confidence intervals as in Figure S9, but for the parameters of the \cite{Piro2015} model.}
\label{fig:14gqr_shockCorner}
\end{figure}

\begin{figure*}[!ht]
\centering
\includegraphics[width=0.7\textwidth]{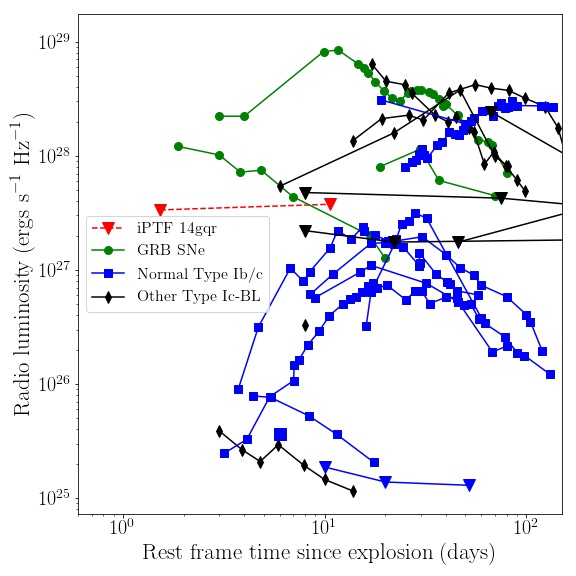}
\caption{\textbf{Comparison of the VLA radio constraints on iPTF\,14gqr (3$\sigma$ limits at 6 GHz) to the observed radio light curves of other Type Ib/c SNe and relativistic SNe at 8.5 GHz.} Events have been color coded into categories of GRB associated Type Ic-BL SNe (green circles), other Type Ic-BL SNe (black diamonds), normal Type Ib/c SNe (blue squares) along with iPTF\,14gqr (red). Inverted triangles denote upper limits. The sources in the GRB SNe sample (and their original published data) are SN\,1998bw \cite{Kulkarni1998}, SN\,2006aj \cite{Soderberg2006} and  SN\,2010bh \cite{Margutti2013}., while the other Type Ic-BL SNe are SN\,2009bb \cite{Soderberg2010a}, SN\,2007bg \cite{Salas2013}, SN\,2002ap \cite{Berger2002}, PTF\,12as, PTF\,13u, PTF\,13ebw, PTF\,14gaq, PTF\,14dby \cite{Corsi2016} and PTF\,12gzk \cite{Horesh2013}. The normal Type Ib/c events include SN\,2013ge \cite{Drout2016}, SN\,2007gr \cite{Soderberg2010b}, SN\,2003L \cite{Soderberg2005}, SN\,2004cc, SN\,2004dk, SN\,2004gq \cite{Wellons2012}, SN\,1994I \cite{Weiler2011}, SN\,2008D \cite{Soderberg2008, vanDerHorst2011} and iPTF\,13bvn \cite{Cao2013}. }
\label{fig:14gqr_radioLCcompare}
\end{figure*}

\begin{figure*}[!ht]
\centering
\includegraphics[width=0.7\textwidth]{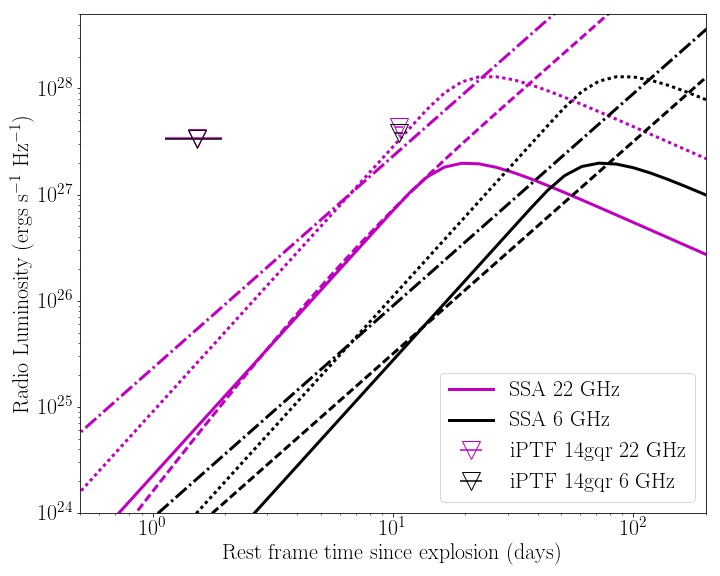}
\caption{\textbf{Radio light curves for the synchrotron self absorption model of \cite{Chevalier1998} plotted along with the radio upper limits of iPTF\,14gqr.} The magenta and black curves correspond to observation frequencies of 22 GHz and 6 GHz respectively, while the upper limits on iPTF\,14gqr are also denoted by identical colors. We assume a constant magnetic field and electron energy fraction with respect to the post-shock energy density of $\epsilon_B = \epsilon_e = 0.1$. Although the radio upper limits lie above the optically thick locus of the model light curves, we show these light curves for the closest set of parameters, corresponding to $K \sim 6 \times 10^{12}$ g cm$^{-1}$ in the wind CSM case, and to $n = 3 \times 10^5$ cm$^{-3}$ in the constant $\rho$ case. The solid and the dotted lines correspond to the wind CSM models (for shock velocities of $2 \times 10^4$ km s$^{-1}$ and $4 \times 10^4$ km s$^{-1}$ respectively), while the dashed and dot-dashed lines indicate the constant density models (for shock velocities of $2 \times 10^4$ km s$^{-1}$ and $4 \times 10^4$ km s$^{-1}$ respectively).}
\label{fig:14gqr_radioLCModel}
\end{figure*}

\begin{figure}[!ht]
\centering
\includegraphics[width=0.75\textwidth]{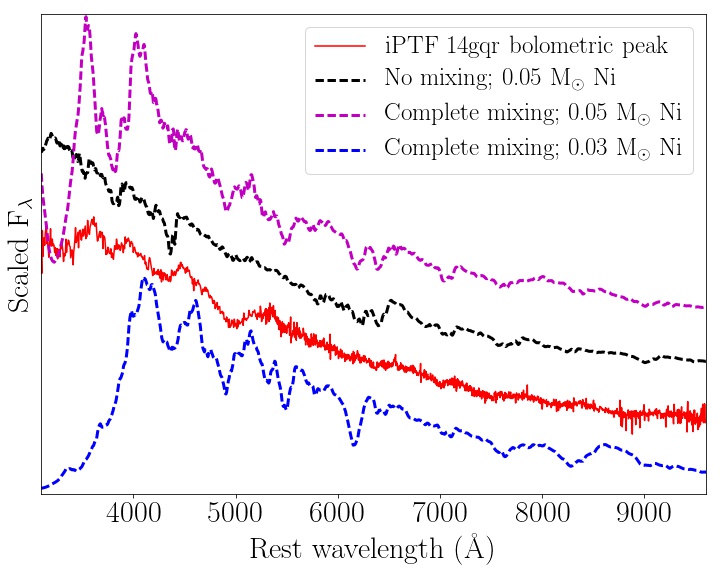}
\caption{\textbf{Comparison of synthetic spectra for ultra-stripped SNe to that of iPTF\,14gqr.} We show the original spectra presented in \cite{Moriya2017} (for $M_{ej} = 0.2$ M$_{\odot}$, $E = 2.5 \times 10^{50}$ ergs and $M_{Ni} = 0.03$ M$_{\odot}$) as the blue dashed line, which exhibits significant line blanketing below 4000 \AA\,due to the assumption of completely mixed ejecta. The magenta line is for a model with complete ejecta mixing and higher Ni mass of 0.05 M$_{\odot}$, and exhibits bluer spectra still affected by deep absorption features at bluer wavelengths. The black line shows a spectrum from a model with the same parameters as the magenta line but with no mixing in the ejecta. This model reproduces both the continuum shape and prominent absorption features.}
\label{fig:14gqr_simSpecModels}
\end{figure}

Since the host galaxy appeared to be part of a tidally interacting galaxy group, we undertook a spectroscopic mask observation of the host region to determine if the spiral host was a part of a galaxy group or a larger cluster in the region. As shown in Figure \ref{fig:lateLRISMask_14gqr}, the galaxies marked \textit{Obj2} and \textit{Obj3} were found to be at a redshift consistent with that of the nominal host \textit{Obj1}. While we did find an unrelated background galaxy cluster at a redshift of $\approx$ 0.19 with at least 8 members within $\pm$ 3000 km s$^{-1}$ of z = 0.19, we did not find evidence of a cluster environment for the host galaxy of the transient.

We conclude that the host of iPTF\,14gqr is part of a galaxy group of at least three tidally interacting members, as evident from the stretched tidal tails of the host (see Figure \ref{fig:14gqr_lateDeep}). All of the three galaxies at the host redshift show prominent signs of star formation via their emission features of H$\alpha$, H$\beta$ and H$\gamma$. However, given that the other two members of this group are at much larger projected offsets from the transient location (Table \ref{tab:14gqr_hostSpec}), it seems relatively unlikely that they could have hosted the progenitor.

\subsubsection*{The remote location in the context of ultra-stripped SN models}

The remote location of iPTF\,14gqr with respect to its host galaxy remains to be  understood in the context of the ultra-stripped SN models discussed earlier. \cite{Tauris2015} suggest that the locations of ultra-stripped SN within their host galaxies should be primarily near star forming regions (within 1 kpc) as indicated by the relatively short lifetimes of the collapsing star ($\sim 7 - 40$ Myr) and the small average systemic velocities ($\sim 10$ km s$^{-1}$) corresponding to average NS kicks of $\sim 50$ km s$^{-1}$ predicted in their calculations. The small systemic velocities arise in particular due to the requirement of wide binary systems that can survive the common envelope (CE) ejection after the High Mass X-ray Binary (HMXB) phase (see \cite{Tauris2006} for a review).

Instead, iPTF\,14gqr displays a $\sim$ 29 kpc projected offset from the center of its host galaxy. If we consider the minimum projected offset of the location from the edge of the visible galaxy, the distance still remains at $\sim 15$ kpc. For the 40 Myr estimated lifetime of the progenitors, the required systemic velocity for the binary system would be $\sim 350$ km s$^{-1}$ (note that the system moves out of its host's gravitational potential, and hence the velocity is unlikely to be constant), which is much larger than the predicted systemic velocities after the first SN. 

There are several possible explanations of this discrepancy. The lifetimes of giant H II regions ($\sim 10$ Myr) are shorter than the expected lifetime of the ultra-stripped progenitors. While we did obtain an upper limit of $\sim 10^{38}$ ergs s$^{-1}$  on the H$\alpha$ luminosity at the location of the transient, lower H$\alpha$ luminosities have been previously observed in H II regions associated with Type Ic SNe \cite{Crowther2013}. Outlying H II regions around galaxies are also not unforeseen, and have been observed out to projected distances larger than the $\sim$ 30 kpc offset observed in iPTF\,14gqr \cite{Werk2010}. Thus, we cannot rule out the presence of either a low luminosity H II region at the transient location or an H II region that has faded away within the progenitor lifetime.

The presence of a tidally interacting galaxy group associated with the host galaxy of iPTF\,14gqr provides another possible clue to its remote location. Tidal tails formed during galaxy mergers are known to host star forming clusters within them \cite{Mullan2011, Knierman2003, Tran2003}, arising from tidally induced periodic density waves similar to those seen in the spiral arms of galaxies \cite{Renaud2009}. Ages inferred from modeling of the broadband photometry as well as spectroscopy of some of these star clusters indicate that they are typically younger than their parent tidal tails, which is strongly suggestive of in-situ formation \cite{Knierman2003, Trancho2007}. Studies of star formation in collisional debris of interacting galaxies in the nearby universe indicate that the instantaneous star formation rate (SFR) in such outlying debris can be as high as 80\% of the total SFR of the entire system, with a median value of 20\% \cite{Boquien2009}.

\begin{figure*}
\includegraphics[width=\textwidth]{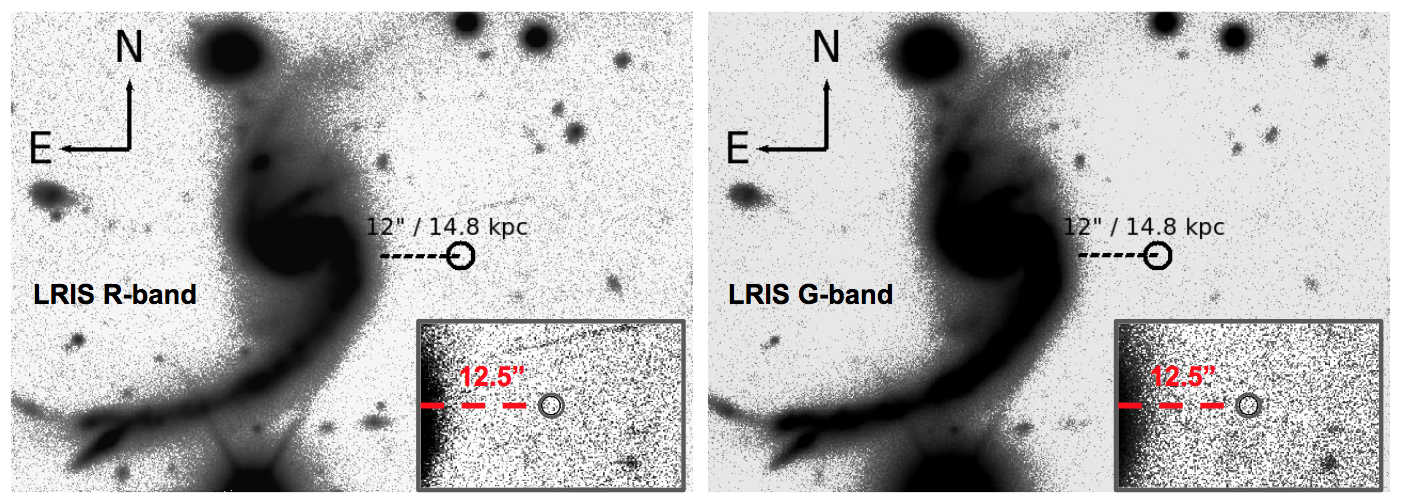}
\caption{\textbf{Late-time $R$-band and $G$-band image of the host galaxy of iPTF\,14gqr.} There are extended tidal tails of the host galaxy arising from tidal interactions of the host with its companions, and an extended halo around the spiral arms. The insets show a zoomed in portion of the image centered on the location of the transient.  Although we are not sensitive to low surface brightness features (spiral arms and tidal tails) at the distance of the galaxy, the brighter tidal tails of the host galaxy extend to much larger distances than the observed offset of iPTF\,14gqr. }
\label{fig:14gqr_lateDeep}
\end{figure*}

Our late-time images of the host galaxy clearly show that the tidal tails of the host galaxy extend to larger projected distances compared to the offset of iPTF\,14gqr, and hence a faint tidal tail at the location of iPTF\,14gqr is not unexpected. However, the large distance to the source and the low surface brightness of tidal tails would not allow us to detect such a feature even if it were present underneath the location of iPTF\,14gqr (note the faint tidal tails in Figure \ref{fig:14gqr_lateDeep}), while our photometric limits are not deep enough to constrain the population of star clusters found in such tidal tails (with typical $M_V \approx -8.5$). There are also known examples of tidal dwarf galaxies forming in tidal tails that can host star formation several $100$ Myrs after the formation of the parent tail \cite{Saviane2004,Knierman2003}.

One potential channel in this context is the delayed core-collapse SNe suggested by \cite{Zapartas2017}, arising from interacting binaries consisting of a massive star and an intermediate mass star, where mass transfer pushes the intermediate mass star above the mass threshold for core-collapse, which can produce a core-collapse SN as late as 200 Myr after the initial star formation episode. In that case, the companion to the progenitor would be a WD while the required systemic velocity would be $\sim$ 70 km s$^{-1}$, if the binary was formed in the spiral arms.

We briefly consider the potential biases against finding faint transients like iPTF\,14gqr on bright galaxy backgrounds, referring to recovery efficiencies of PTF presented in \cite{Frohmaier2017}. From our late time-time imaging,  we estimate that the surface brightness of the host galaxy varies between $< 23$ mag arcsec$^{-2}$ in the spiral arms to $< 21$ mag arcsec$^{-2}$ near the nucleus. Taking the peak magnitude of iPTF\,14gqr and assuming a typical seeing of 1.5 $''$, the recovery efficiency can vary between 80\% in the spiral arms to $<$20\% in the nucleus. On the other hand, the recovery efficiency at the actual location of the transient would be $\approx 95$\% given the negligible background. While this potentially suggests that faint events like iPTF\,14gqr may be missed by PTF on bright galaxy backgrounds, we caution against drawing conclusions about the population of ultra-stripped SNe based on one event, which can be only be resolved with larger samples of these fast transients from wide-field transient surveys \cite{Bellm2017,Tonry2018}.

\noindent
\section*{Supplementary Text}
\subsection*{The nature of the companion star}

Stripping of the outer H and He envelopes in stripped envelope SNe can arise either due to mass loss via strong winds or due to stripping by a binary companion \cite{Smartt2009}. We thus first consider the possibility of explaining the highly stripped progenitor of iPTF\,14gqr from single star evolution. Several theoretical calculations for a wide range of stellar mass loss rates and metallicity show that the minimum progenitor mass from single star evolution of massive stars is $> 5$ M$_{\odot}$ \cite{Meynet2005, Eldridge2006, Georgy2009}, and hence we expect ejecta masses of $> 3$ M$_{\odot}$. Such large pre-explosion masses for single star progenitors are observationally consistent with masses inferred for single Wolf-Rayet (WR) stars in the galaxy. Nonetheless, these estimates are at least an order of magnitude larger than the ejecta mass inferred for iPTF\,14gqr ($\approx 0.2$ M$_{\odot}$), and rules out a single star progenitor for iPTF\,14gqr. 

Next, we consider a scenario where the progenitor of iPTF\,14gqr can be explained from close binary evolution of two non-degenerate massive stars, as calculated in \cite{Yoon2010}. Although \cite{Yoon2010} show that main sequence companions of lower mass stripped-envelope SN progenitors can efficiently strip the progenitor to low core masses at the time of core-collapse, these progenitors end up retaining a significant amount of He in their outer layers in all such cases ($\sim 0.20$ M$_\odot$, where $> 80$\% of the mass in the ejecta would be He). Our estimates of the He mass in this event are $\sim 0.01$ M$_{\odot}$ (in the ejected envelope), with a negligible additional amount in the ejecta. This is again an order of magnitude lower than these predictions. We thus conclude that a binary system consisting of two massive stars cannot reproduce the extreme stripping observed in the progenitor of iPTF\,14gqr. 

We now consider the case for a lower mass main sequence (MS) companions to the progenitor of iPTF\,14gqr. In order to explain the extreme stripping of the He star, one would need a system that is close enough such that the He star fills its Roche lobe during its late evolutionary expansion and begins mass transfer on to the companion. In this context, we refer to \cite{Zapartas2017b}, who presented a binary population synthesis study of the progenitors for stripped envelope SNe for different initial parameters of the binary system. In particular, they find that systems with very small mass ratios initially ($q < 0.2$, as would be required for a compact main sequence companion for iPTF\,14gqr) are very unlikely for progenitor final masses $< 2$ M$_{\odot}$ (\cite{Zapartas2017b}, see their figure 5). Thus, based on the existing literature of stellar evolutionary models, we find that a low mass MS companion does not appear to be a likely channel to explain the extreme stripping.

Several studies in the literature have shown that highly stripped progenitors, as that inferred for iPTF\,14gqr, are expected to arise in the case of compact binary companions such as a NS, WD or BH \cite{Tauris2013, Tauris2015, Zapartas2017b}, where close (and possibly dynamically stable) gravitational stripping by the compact companion is able to strip of most of the outer layers of the initial He star, leaving behind only a thin layer of He before core-collapse. Even in the presence of a NS companion, such extreme stripping appears to be only possible when the initial orbital period (at the start of the He main sequence of the progenitor) is less than 0.1 days \cite{Tauris2015}, and hence an orbital separation of less than $\sim 1$ R$_{\odot}$. This allows only the most compact companions to explain the stripping. Taken together, we conclude that the case for a NS, WD or BH companion is most consistent with the inferred properties of the progenitor of iPTF\,14gqr.

\subsection*{The remnant of the explosion}

The mass of the remnant of iPTF\,14gqr is likely to be dominated by the mass of the iron core at the time of core-collapse (i.e. the Chandrasekhar mass), as is applicable for massive stars with initial masses $< 20$ M$_{\odot}$ that can lead to ultra-stripped progenitors \cite{Tauris2015,Woosley2002,Zapartas2017b}. Taking the ejecta mass of $\approx 0.2$ M$_{\odot}$ observed in iPTF\,14gqr, and assuming negligible fallback of material during the SN explosion, we infer a progenitor mass of $\approx 1.5$ M$_{\odot}$. 

Although some amount of fallback is expected in all SNe, this will be prominent only for stars that were initially more massive than $\sim 30$ M$_{\odot}$ \cite{Fryer1999, Fryer2001, MacFayden2001, Woosley2002}. The binding energy of the progenitor envelope is high in such cases, and leads to infall of the inner layers of the star on to the proto-neutron star, potentially leading to the formation of a black hole. Since most of the $^{56}$Ni is synthesized in the inner layers of the explosion, the resulting SN would be faint with little or no radioactive $^{56}$Ni ejected in the explosion (unless the $^{56}$Ni was well-mixed prior to the fallback). In the context of iPTF\,14gqr, we find that the explosion ejected a relatively large amount of $^{56}$Ni (comprising $\approx 25$\% of the ejecta by mass) that was located in the inner regions of the ejecta, further suggesting that fallback was negligible in this source. We thus conclude that iPTF\,14gqr most likely formed a NS in the explosion.

\subsection*{Origin of the extended He-rich envelope}

Our early spectra indicate that the SN explosion occurred inside a He-rich envelope with a mass of $\sim$ 0.01 M$_{\odot}$, located at a radius of $\sim 500 - 10^4$ R$_{\odot}$. By fitting a two-component Gaussian profile to the $\lambda$4686 feature (to account for the He II and C III features) and a simple Gaussian profile to other C III and C IV lines, we estimate line FWHMs in the range of $\sim 2000 - 4000$ km s$^{-1}$, consistent with an expanding envelope at the time of explosion. Taking the widths of the Gaussian (the standard deviation, or $\sigma$) to be representative of the expansion velocity, we infer the envelope velocity to be $\sim 1000 - 2000$ km s$^{-1}$. This suggests that the envelope was ejected $\sim 8 - 20$ days before the core-collapse when considering that the envelope is located at a radius of at least $\sim 500$ R$_{\odot}$. The high inferred mass loss rate ($\sim 0.05$ M$_{\odot}$ yr$^{-1}$) also suggests that the envelope cannot be associated with a long term wind given the low mass of the progenitor. 

The ejecta traveling at $\sim 10^4$ km s$^{-1}$ would take $\sim 1 - 7$ days to sweep up the extended material lying between 500 and 10$^4$ R$_{\odot}$ and hence, the disappearance of the emission lines within 2 days after explosion favors a recombining envelope scenario. The measured velocity is very similar to the expected escape velocity of $1000$ km s$^{-1}$ for a compact He star (with $M \sim 1.5$ M$_{\odot}$ and $R \sim 0.5$ R$_{\odot}$).

When considering the required presence of a close compact companion to explain the low ejecta mass, a possible explanation of the expanding envelope could be due to a (partial) common envelope ejection preceding the terminal core-collapse. For example, multiple theoretical studies on the evolution of compact He star - NS and He star - WD binaries find evidence of unstable mass transfer between the He star and the compact object as the He star moves off the He burning main sequence \cite{Tauris2000,Dewi2003,Zapartas2017}. The result of such unstable mass transfer is the formation of a common envelope that engulfs the binary. The common envelope phase triggers a rapid inspiral of the compact object, where the common envelope may be ejected on a few dynamical timescales ($\sim$ few days \cite{Ivanova2013}). However, the inspiral may also be slowed down before the envelope is ejected if the binary enters a self-regulated inspiral stage, which may last much longer ($\sim 10^3 - 10^4$ years; \cite{Podsiadlowski2001}). 

Depending on the exact time left until the SN explosion and the ejection of the envelope, the second SN may take place inside a common envelope, or after the compact object has spiraled in to a tight orbit and the common envelope has been ejected \cite{Dewi2003}. Numerical calculations of the evolution of isolated low mass He stars indeed suggest that they undergo rapid expansion at the onset of the He shell burning phase and as they approach core-collapse, reaching radii of the order of $10^2 - 10^3$ R$_{\odot}$ \cite{Woosley1995}. Hence, the formation of a common envelope arising out of such an expansion shortly before core-collapse is indeed plausible. Nonetheless, the exact temporal co-incidence between the ejection and the final core-collapse requires fine-tuning of either the formation of the common envelope (if ejected on a dynamical timescale) or its ejection (if ejected after a self-regulated phase).

Calculations of the evolution of such He stars in He star - NS binaries by \cite{Tauris2015} suggest that although the NS is able to strip off almost the entire He layer of the exploding star, a small amount of He (with a mass of $\sim 0.01$ M$_{\odot}$) remains unstripped in its outer layers at the time of explosion. Given the observed temporal co-incidence between the expansion of the envelope and the final core-collapse, as well as the very similar estimated mass of the He envelope, we favor a scenario where the observed envelope is likely associated with the rapid expansion of the remaining He layer prior to core-collapse. Indeed, there are several known examples of massive stars that undergo vigorous outbursts just before the final explosion, leading to the ejection of large amounts of mass into the surrounding medium \cite{Smith2014}. 

Such an eruption would be consistent, for instance, with  unstable nuclear burning in the late stages of the progenitor's evolution \cite{Quataert2012, Shiode2014}, and potentially associated with Si flashes expected to occur only $\sim$ 2 weeks before the SN explosion in lower mass cores \cite{Woosley2015}. In fact, the detection of prominent high ionization He and C lines in the early spectra is consistent with a scenario where the surface layers of the star are ejected, since ultra-stripped progenitors are expected to have a surface composition rich in He and C \cite{Tauris2015}. While similar detailed calculations for He star - BH and He star - WD binaries are not available, similar physics should also apply to these ultra-stripped SN progenitors due to the similar compact nature of the companion.

\subsection*{iPTF\,14gqr and the class of Ca-rich gap transients}
iPTF\,14gqr exhibits a number of similarities to the class of Ca-rich gap transients \cite{Kasliwal2012}. Of the several proposed properties of this class, it exhibits faster photometric evolution than normal SNe, photospheric velocities comparable to normal SNe, rapid evolution to the nebular phase and a (early) nebular spectrum dominated by Calcium emission (similar to the Ca-rich transients PTF\,10iuv and SN\,2012hn). While the remote location of the transients was not used as a defining property of this class, iPTF\,14gqr was also found in the remote outskirts of its host galaxy, as were all confirmed Ca-rich gap transients \cite{Kasliwal2012,Lunnan2017}. The confirmed association of the apparent host to a small galaxy group further strengthens a possible relation to these gap transients, which are found to be preferentially located in dense group and cluster environments \cite{Lunnan2017}.

\begin{figure}
\includegraphics[width=0.49\textwidth]{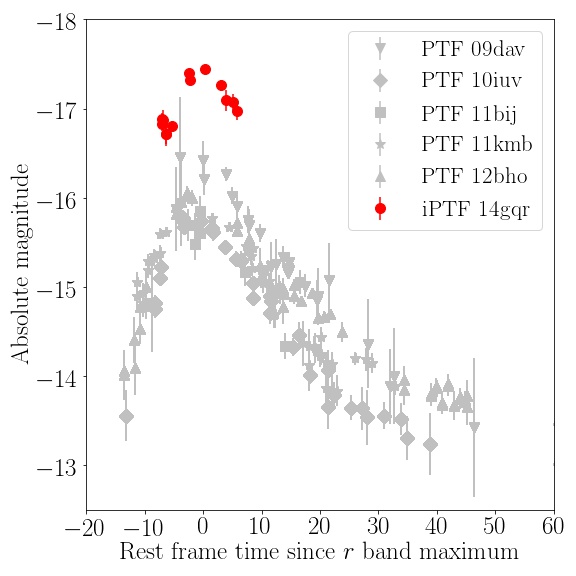}
\includegraphics[width=0.49\textwidth]{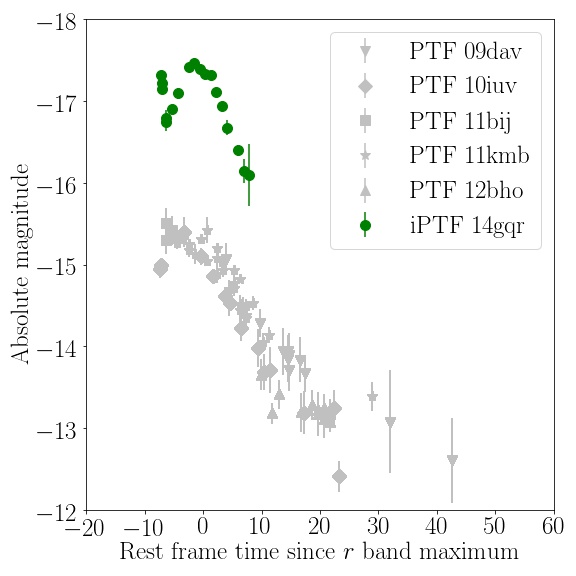}
\caption{\textbf{Comparison of the $r$ (left) and $g$ (right) band light curves of iPTF\,14gqr to those of some known Ca-rich gap transients.} These include PTF\,09dav \cite{Sullivan2011}, PTF\,10iuv and PTF\,11bij \cite{Kasliwal2012}, PTF\,11kmb and PTF\,12bho \cite{Lunnan2017}.}
\label{fig:14gqr_compareCaRich}
\end{figure}

There are several reasons why the above association is unlikely to be physical. The peak luminosity of iPTF\,14gqr is much larger than those of the Ca-rich gap transients, which occupy the luminosity `gap' between novae and SNe (peak absolute magnitude $-15.5 \geq$ $M_{peak} \geq -16.5$). Instead, the peak absolute magnitude of iPTF\,14gqr ($M_{peak} \approx -17.45$ in $r$ and $g$ band) is comparable to those of normal SNe Ib/c. The multi-color light curves of iPTF\,14gqr are also very different in shape compared to those of the Ca-rich transients, which otherwise form a fairly homogeneous class. This is demonstrated in Figure \ref{fig:14gqr_compareCaRich}, which compares the $r$ and $g$ band light curves of iPTF\,14gqr to those of Ca-rich gap transients discovered by the Palomar Transient Factory (PTF). Additionally, none of these Ca-rich gap transients exhibited double-peaked light curves as in iPTF\,14gqr, even though some of these sources had early enough photometric follow-up to be able to detect such a component if present.

Examining the final nebular spectrum of iPTF\,14gqr, we find the [Ca II]/[O I] ratio to be $\approx$ 2.5 by fitting Gaussian profiles to the line regions, which is marginally higher than that observed in Type Ib/c SNe (e.g. \cite{Milisavljevic2017}, their figure 11). In contrast, the [Ca II]/[O I] ratio in the Ca-rich gap transients are typically much higher ($\geq 3$, but typically as high as $6 - 10$ \cite{Milisavljevic2017}), and hence iPTF\,14gqr is likely to be more closely related to the Ib/c SN class (\cite{Milisavljevic2017} put the threshold for a Ca-rich classification at a ratio of 2). Overall, we conclude that while iPTF\,14gqr may be considered to be marginally `Ca-rich' as compared to other SNe, it is photometrically not a part of the class of Ca-rich gap transients. There have also been reports of other Type I SNe (e.g. iPTF\,15eqv; \cite{Milisavljevic2017}) that have been found to be Ca-rich spectroscopically but are not members of this class photometrically.

\section*{Additional Acknowledgements}

Based on observations obtained with the Apache Point Observatory 3.5-meter telescope, which is owned and operated by the Astrophysical Research Consortium. Based on observations made with the Nordic Optical Telescope, operated by the Nordic Optical Telescope Scientific Association at the Observatorio del Roque de los Muchachos, La Palma, Spain, of the Instituto de Astrofisica de Canarias. The data presented here were obtained in part with ALFOSC, which is provided by the Instituto de Astrofisica de Andalucia (IAA) under a joint agreement with the University of Copenhagen and NOTSA. Based on observations obtained at the Gemini Observatory, which is operated by the Association of Universities for Research in Astronomy, Inc., under a cooperative agreement with the NSF on behalf of the Gemini partnership: the National Science Foundation (United States), the National Research Council (Canada), CONICYT (Chile), Ministerio de Ciencia, Tecnolog\'{i}a e Innovaci\'{o}n Productiva (Argentina), and Minist\'{e}rio da Ci\^{e}ncia, Tecnologia e Inova\c{c}\~{a}o (Brazil). Some of the data presented herein were obtained at the W.M. Keck Observatory, which is operated as a scientific partnership among the California Institute of Technology, the University of California and the National Aeronautics and Space Administration. The Observatory was made possible by the generous financial support of the W.M. Keck Foundation. The authors wish to recognize and acknowledge the very significant cultural role and reverence that the summit of Mauna Kea has always had within the indigenous Hawaiian community. We are most fortunate to have the opportunity to conduct observations from this mountain. The William Herschel Telescope is operated on the island of La Palma by the Isaac Newton Group of Telescopes in the Spanish Observatorio del Roque de los Muchachos of the Instituto de Astrofisica de Canarias. The National Radio Astronomy Observatory is a facility of the National Science Foundation operated under cooperative agreement by Associated Universities, Inc. This research made use of Astropy, a community-developed core Python package for Astronomy \cite{astropy}. The data of GALEX presented in this paper were obtained from the Mikulski Archive for Space Telescopes (MAST). This publication makes use of data products from the Two Micron All Sky Survey, which is a joint project of the University of Massachusetts and the Infrared Processing and Analysis Center/California Institute of Technology, funded by the National Aeronautics and Space Administration and the National Science Foundation. This research has made use of the NASA/IPAC Extragalactic Database (NED) which is operated by the Jet Propulsion Laboratory, California Institute of
Technology, under contract with the National Aeronautics and Space Administration. IRAF is
distributed by the National Optical Astronomy Observatory, which is operated by the Association
of Universities for Research in Astronomy (AURA) under a cooperative agreement with the National Science Foundation. Funding for the Sloan Digital Sky Survey IV has been provided by the Alfred P. Sloan Foundation, the U.S. Department of Energy Office of Science, and the Participating Institutions. SDSS acknowledges support and resources from the Center for High-Performance Computing at the University of Utah. The SDSS web site is www.sdss.org. This research has made use of NASA's Astrophysics Data System.

\newpage

\noindent
Data S1: \textbf{Synthetic model spectra for the ultra-stripped SN models presented in this work.} Three model spectra, corresponding to the models presented in Figure S13 are given. The file `completemixing\_ni0.03msun.dat' is the original model in \cite{Moriya2017} for complete mixing in the ejecta (blue dashed line in Figure S13) and $^{56}$Ni mass of 0.03 M$_\odot$, the file `completemixing\_ni0.05msun.dat' corresponds to the magenta dashed line in Figure S13 with completely mixed ejecta and $^{56}$Ni mass increased to 0.05 M$_\odot$, and the file `nomixing\_ni0.05msun.dat' corresponds to the black dashed line with 0.05 M$_\odot$ of $^{56}$Ni and no mixing in the ejecta.

\begin{table*}
\centering
\caption{\textbf{Photometric follow-up of iPTF\,14gqr in the optical and UV bands.} The data have been corrected for Galactic extinctions of $0.158$ mag ($i$-band), $0.212$ mag ($r$-band), $0.307$ mag ($g$-band), $0.337$ mag ($B$-band), $0.553$ mag ($UVW1$ band) and $0.748$ mag ($UVW2$ band). Upper limits indicated are 5$\sigma$ upper limits in the respective bands.}
\begin{tabular}{lccccc}
\hline
MJD & Rest frame phase & Filter & Magnitude & Telescope\\
& (days from explosion) & & &\\
\hline
56944.41 & 0.62 & $UVW2$ & 19.14 $\pm$ 0.07 & UVOT\\
56944.83 & 1.02 & $UVW2$ & 19.86 $\pm$ 0.12 & UVOT\\
56945.08 & 1.25 & $UVW2$ & $>$ 19.62 & UVOT\\
56950.15 & 6.02 & $UVW2$ & $>$ 20.96 & UVOT\\
56954.77 & 10.37 & $UVW2$ & $>$ 21.14 & UVOT\\
56962.02 & 17.19 & $UVW2$ & $>$ 20.29 & UVOT\\
56966.30 & 21.21 & $UVW2$ & $>$ 20.99 & UVOT\\
56944.43 & 0.64 & $UVW1$ & 19.49 $\pm$ 0.10 & UVOT\\
56944.84 & 1.03 & $UVW1$ & 19.71 $\pm$ 0.16 & UVOT\\
56950.15 & 6.02 & $UVW1$ & 21.23 $\pm$ 0.30 & UVOT\\
56954.77 & 10.37 & $UVW1$ & $>$ 20.62 & UVOT\\
56962.01 & 17.18 & $UVW1$ & $>$ 19.81 & UVOT\\
56966.30 & 21.21 & $UVW1$ & $>$ 20.44 & UVOT\\
56944.34 & 0.56 & $B$ & 20.26 $\pm$ 0.08 & P60\\
56944.39 & 0.60 & $B$ & 20.22 $\pm$ 0.08 & P60\\
56944.45 & 0.66 & $B$ & 20.09 $\pm$ 0.09 & P60\\
56945.09 & 1.26 & $B$ & $>$ 19.96 & P60\\
56945.16 & 1.33 & $B$ & 20.55 $\pm$ 0.08 & P60\\
56946.09 & 2.20 & $B$ & $>$ 20.26 & P60\\
56946.15 & 2.26 & $B$ & 20.41 $\pm$ 0.10 & P60\\
56946.20 & 2.30 & $B$ & 20.42 $\pm$ 0.10 & P60\\
56946.31 & 2.41 & $B$ & $>$ 20.56 & P60\\
56949.36 & 5.28 & $B$ & 20.08 $\pm$ 0.06 & P60\\
56949.41 & 5.32 & $B$ & 20.08 $\pm$ 0.05 & P60\\
56954.75 & 10.35 & $B$ & $>$ 19.11 & UVOT\\
56962.01 & 17.18 & $B$ & $>$ 18.41 & UVOT\\
56966.28 & 21.19 & $B$ & $>$ 18.99 & UVOT\\
56940.34 & -3.21 & $g$ & $>$ 20.71 & P48\\
56941.36 & -2.25 & $g$ & $>$ 21.04 & P48\\
56942.36 & -1.31 & $g$ & $>$ 21.43 & P48\\
56943.33 & -0.40 & $g$ & $>$ 21.23 & P48\\
56944.20 & 0.42 & $g$ & 19.95 $\pm$ 0.05 & P48\\
56944.35 & 0.56 & $g$ & 20.04 $\pm$ 0.04 & P60\\
\hline
\end{tabular}
\label{tab:lc}
\end{table*}

\begin{table}
\centering
\contcaption{Photometric follow-up of iPTF\,14gqr in the optical and UV bands corrected for galactic extinction (continued).\\}
\begin{tabular}{lccccc}
\hline
MJD & Rest frame phase & Filter & Magnitude & Telescope\\
& (days from explosion) & & &\\
\hline
56944.42 & 0.63 & $g$ & 20.12 $\pm$ 0.05 & P60\\
56945.10 & 1.27 & $g$ & 20.52 $\pm$ 0.11 & P60\\
56945.17 & 1.34 & $g$ & 20.47 $\pm$ 0.10 & P60\\
56946.22 & 2.32 & $g$ & 20.36 $\pm$ 0.04 & P60\\
56947.21 & 3.25 & $g$ & 20.17 $\pm$ 0.07 & P48\\
56949.34 & 5.26 & $g$ & 19.85 $\pm$ 0.02 & P60\\
56950.26 & 6.12 & $g$ & 19.80 $\pm$ 0.03 & P48\\
56951.22 & 7.03 & $g$ & 19.87 $\pm$ 0.03 & P48\\
56952.23 & 7.98 & $g$ & 19.94 $\pm$ 0.04 & P60\\
56953.29 & 8.97 & $g$ & 19.95 $\pm$ 0.05 & P48\\
56954.27 & 9.90 & $g$ & 20.15 $\pm$ 0.05 & P48\\
56955.21 & 10.78 & $g$ & 20.32 $\pm$ 0.04 & P60\\
56956.14 & 11.66 & $g$ & 20.59 $\pm$ 0.09 & P60\\
56956.33 & 11.83 & $g$ & $>$ 19.36 & P48\\
56958.19 & 13.58 & $g$ & 20.87 $\pm$ 0.07 & P60\\
56959.26 & 14.59 & $g$ & 21.12 $\pm$ 0.15 & P48\\
56960.26 & 15.53 & $g$ & 21.17 $\pm$ 0.38 & P48\\
56960.28 & 15.55 & $g$ & $>$ 20.51 & P48\\
56965.19 & 20.17 & $g$ & $>$ 19.99 & P60\\
56966.08 & 21.01 & $g$ & $>$ 19.59 & P60\\
56954.75 & 10.35 & $V$ & $>$ 18.53 & UVOT\\
56962.01 & 17.18 & $V$ & $>$ 17.78 & UVOT\\
56966.28 & 21.19 & $V$ & $>$ 18.40 & UVOT\\
56944.34 & 0.56 & $r$ & 20.39 $\pm$ 0.07 & P60\\
56944.40 & 0.61 & $r$ & 20.45 $\pm$ 0.07 & P60\\
56944.46 & 0.67 & $r$ & 20.40 $\pm$ 0.11 & P60\\
56945.10 & 1.27 & $r$ & 20.56 $\pm$ 0.13 & P60\\
56945.17 & 1.34 & $r$ & 20.56 $\pm$ 0.11 & P60\\
56946.20 & 2.30 & $r$ & 20.47 $\pm$ 0.06 & P60\\
56949.36 & 5.28 & $r$ & 19.87 $\pm$ 0.04 & P60\\
56949.41 & 5.32 & $r$ & 19.95 $\pm$ 0.05 & P60\\
56952.15 & 7.90 & $r$ & 19.83 $\pm$ 0.04 & P60\\
56955.17 & 10.74 & $r$ & 20.01 $\pm$ 0.06 & P60\\
56956.09 & 11.61 & $r$ & 20.18 $\pm$ 0.12 & P60\\
56957.35 & 12.79 & $r$ & 20.20 $\pm$ 0.09 & P60\\
\hline
\end{tabular}
\end{table}

\begin{table}
\centering
\contcaption{Photometric follow-up of iPTF\,14gqr in the optical and UV bands corrected for galactic extinction (continued).\\}
\begin{tabular}{lccccc}
\hline
MJD & Rest frame phase & Filter & Magnitude & Telescope\\
& (days from explosion) & & &\\
\hline
56958.09 & 13.49 & $r$ & 20.30 $\pm$ 0.10 & P60\\
56965.19 & 20.17 & $r$ & $>$ 18.89 & P60\\
56966.08 & 21.01 & $r$ & $>$ 18.29 & P60\\
56944.34 & 0.56 & $i$ & 20.86 $\pm$ 0.10 & P60\\
56944.40 & 0.61 & $i$ & 20.83 $\pm$ 0.13 & P60\\
56944.45 & 0.66 & $i$ & 20.74 $\pm$ 0.17 & P60\\
56945.14 & 1.31 & $i$ & 20.66 $\pm$ 0.08 & P60\\
56946.21 & 2.31 & $i$ & 20.52 $\pm$ 0.06 & P60\\
56949.35 & 5.27 & $i$ & 20.01 $\pm$ 0.08 & P60\\
56949.41 & 5.32 & $i$ & 20.07 $\pm$ 0.09 & P60\\
56952.15 & 7.90 & $i$ & 19.99 $\pm$ 0.06 & P60\\
56955.17 & 10.74 & $i$ & 19.95 $\pm$ 0.08 & P60\\
56956.08 & 11.60 & $i$ & $>$ 19.84 & P60\\
56957.35 & 12.79 & $i$ & 20.22 $\pm$ 0.07 & P60\\
56958.08 & 13.48 & $i$ & 20.15 $\pm$ 0.11 & P60\\
56965.08 & 20.07 & $i$ & $>$ 20.44 & P60\\
56966.08 & 21.01 & $i$ & $>$ 18.64 & P60\\
56945.32 & 1.48 & $B$ & 20.68 $\pm$ 0.18 & LCO \\ 
56947.10 & 3.15 & $B$ & 20.17 $\pm$ 0.11 & LCO \\ 
56949.08 & 5.02 & $B$ & 19.99 $\pm$ 0.12 & LCO \\ 
56954.31 & 9.94 & $B$ & 20.15 $\pm$ 0.13 & LCO \\ 
56956.26 & 11.77 & $B$ & 20.63 $\pm$ 0.12 & LCO \\ 
56959.22 & 14.55 & $B$ & 21.27 $\pm$ 0.19 & LCO \\ 
56945.33 & 1.49 & $V$ & 20.36 $\pm$ 0.14 & LCO \\ 
56947.11 & 3.16 & $V$ & 20.01 $\pm$ 0.10 & LCO \\ 
56949.09 & 5.02 & $V$ & 19.85 $\pm$ 0.09 & LCO \\ 
56954.32 & 9.94 & $V$ & 19.86 $\pm$ 0.11 & LCO \\ 
56956.27 & 11.78 & $V$ & 20.25 $\pm$ 0.11 & LCO \\ 
56959.23 & 14.56 & $V$ & 20.56 $\pm$ 0.21 & LCO \\ 
56945.34 & 1.50 & $g$ & 20.33 $\pm$ 0.17 & LCO \\ 
56947.12 & 3.17 & $g$ & 20.14 $\pm$ 0.11 & LCO \\ 
56949.10 & 5.03 & $g$ & 19.81 $\pm$ 0.10 & LCO \\ 
56954.33 & 9.95 & $g$ & 20.16 $\pm$ 0.11 & LCO \\ 
56956.28 & 11.79 & $g$ & 20.38 $\pm$ 0.13 & LCO \\ 
56959.24 & 14.57 & $g$ & 21.03 $\pm$ 0.09 & LCO \\ 
\hline
\end{tabular}
\end{table}

\begin{table}
\centering
\contcaption{Photometric follow-up of iPTF\,14gqr in the optical and UV bands corrected for galactic extinction (continued).\\}
\begin{tabular}{lccccc}
\hline
MJD & Rest frame phase & Filter & Magnitude & Telescope\\
& (days from explosion) & & &\\
\hline
56945.35 & 1.50 & $r$ & 20.54 $\pm$ 0.20 & LCO \\ 
56947.13 & 3.18 & $r$ & 20.20 $\pm$ 0.10 & LCO \\ 
56949.11 & 5.04 & $r$ & 19.87 $\pm$ 0.10 & LCO \\ 
56954.34 & 9.96 & $r$ & 19.98 $\pm$ 0.12 & LCO \\ 
56956.29 & 11.79 & $r$ & 19.87 $\pm$ 0.14 & LCO \\ 
56945.36 & 1.51 & $i$ & 20.49 $\pm$ 0.26 & LCO \\ 
56947.13 & 3.18 & $i$ & 20.18 $\pm$ 0.14 & LCO \\ 
56949.12 & 5.05 & $i$ & 20.15 $\pm$ 0.15 & LCO \\ 
56954.35 & 9.97 & $i$ & 20.06 $\pm$ 0.14 & LCO \\ 
56956.29 & 11.80 & $i$ & 20.11 $\pm$ 0.13 & LCO \\ 
56959.25 & 14.59 & $i$ & 20.34 $\pm$ 0.12 & LCO \\ 
\hline
\end{tabular}
\end{table}


\begin{table*}
\centering
\caption{\textbf{Swift XRT flux limits on iPTF\,14gqr.} The upper limits indicated are 5$\sigma$. The count rates have been converted to flux upper limits for a photon index of $\Gamma =  2$ and zero galactic foreground absorption.}
\begin{tabular}{lcccc}
\hline
MJD & Rest frame phase & Exposure time & Count rate & Flux (0.3 - 10 keV)\\
 & (days since explosion) & (s) & ($10^{-3}$ counts s$^{-1})$ & $10^{-13}$ ergs cm$^{-2}$ s$^{-1}$\\
\hline
56944.41 & 0.62 & 1800 & $<$ 8.43 & $<$ 2.39\\
56944.81 & 1.00 & 990 & $<$ 15.53 & $<$ 4.41\\
56945.08 & 1.25 & 162 & $<$ 96.66 & $<$ 27.44\\
56950.14 & 6.01 & 1564 & $<$ 9.79 & $<$ 2.78\\
56954.75 & 10.35 & 3524 & $<$ 4.28 & $<$ 1.22\\
56962.01 & 17.18 & 1221 & $<$ 12.36 & $<$ 3.51\\
56966.28 & 21.19 & 2966 & $<$ 5.07 & $<$ 1.44\\
\hline
\end{tabular}
\label{tab:14gqr_xrt}.
\end{table*}

\begin{table*}
\centering
\caption{\textbf{Summary of spectroscopic observations of iPTF\,14gqr.} \dag\,Epochs where we did not get a good signal to noise ratio spectrum. *\,Slit mask observation to measure the redshifts of potential host galaxies near the transient.}
\footnotesize
\begin{tabular}{lccccc}
\hline
Observation Date & MJD & Rest frame phase & Telescope + Instrument & Range & Resolution\\
& & (days from explosion) & & (Observed {\AA}) & (FWHM \AA) \\
\hline
2014 Oct 14.36 & 56944.36 & 0.58 & APO + DIS & 3315 -- 9880 & 6\\
2014 Oct 14.86 & 56944.86 & 1.05 & WHT + ACAM & 3400 -- 9500 & 13 \\
2014 Oct 15.10 & 56945.10 & 1.27 & NOT + ALFOSC & 3450 -- 9140 & 17\\
2014 Oct 15.21 & 56945.21 & 1.38 & Gemini N + GMOS & 3790 -- 9350 & 7 \\
2014 Oct 20.36 & 56950.36 & 6.22 & Keck I + LRIS & 3100 -- 10300 & 6 \\
2014 Oct 21.41 & 56951.41 & 7.20 & Keck I + LRIS & 3100 -- 10300 & 6 \\
2014 Oct 25.40 & 56955.40 & 10.96 & Keck I + LRIS & 3200 -- 10250 & 6 \\
2014 Nov 19.27 & 56980.27 & 34.36 & Keck I + LRIS & 3060 -- 10300 & 6 \\
2014 Dec 17.25\dag & 57008.25 & 60.67 & Keck I + LRIS & 3070 -- 10290 & 6 \\
2014 Dec 21.27\dag & 57012.27 & 64.46 & Keck II + DEIMOS & 4490 -- 9650 & 4 \\
2016 Nov 28.27* & 57720.27 & -- & Keck I + LRIS mask & -- & 6 \\
\hline
\end{tabular}
\label{tab:spectra}
\end{table*}

\begin{table*}
\centering
\caption{\textbf{Redshifts and locations of galaxies near iPTF\,14gqr, as identified in the spectroscopic mask (Figure \ref{fig:lateLRISMask_14gqr}).} The last column indicates the offset of the galaxy from the location of the transient. \dag\,Apparent host galaxy of iPTF\,14gqr.}
\begin{tabular}{lcccc}
\hline
Object name & $\alpha$ (J2000) & $\delta$ (J2000) & Redshift & Offset ($''$)\\
\hline
Obj1\dag & \ra{23}{33}{29.85} & \dec{33}{38}{48.22} & 0.063 & 23.8\\
Obj2 & \ra{23}{33}{30.39} & \dec{33}{39}{0.21} & 0.063 & 33.5\\
Obj3 & \ra{23}{33}{30.70} & \dec{33}{39}{16.87} & 0.063 & 46.1\\
Obj4 & \ra{23}{33}{28.89} & \dec{33}{37}{52.44} & 0.743 & 55.0\\
Obj5 & \ra{23}{33}{32.24} & \dec{33}{38}{18.32} & 0.222 & 60.4\\
Obj6 & \ra{23}{33}{31.18} & \dec{33}{39}{39.55} & 0.292 & 66.9\\
Obj7 & \ra{23}{33}{24.46} & \dec{33}{40}{0.42} & 0.190 & 86.2\\
Obj8 & \ra{23}{33}{21.57} & \dec{33}{39}{23.25} & 0.190 & 87.9\\
Obj9 & \ra{23}{33}{25.44} & \dec{33}{40}{13.58} & 0.258 & 92.9\\
Obj10 & \ra{23}{33}{24.52} & \dec{33}{37}{10.07} & 0.480 & 105.2\\
Obj11 & \ra{23}{33}{35.20} & \dec{33}{37}{41.58} & 0.250 & 111.1\\
Obj12 & \ra{23}{33}{36.68} & \dec{33}{40}{2.76} & 0.185 & 133.2\\
Obj13 & \ra{23}{33}{38.04} & \dec{33}{36}{32.74} & 0.183 & 183.5\\
Obj14 & \ra{23}{33}{26.85} & \dec{33}{35}{31.12} & 0.197 & 195.5\\
Obj15 & \ra{23}{33}{37.42} & \dec{33}{36}{4.38} & 0.295 & 200.3\\
Obj16 & \ra{23}{33}{23.59} & \dec{33}{34}{57.65} & 0.197 & 234.9\\
Obj17 & \ra{23}{33}{26.52} & \dec{33}{34}{24.85} & 0.197 & 261.9\\
Obj18 & \ra{23}{33}{30.57} &  \dec{33}{34}{6.88} & 0.288 & 281.2\\
Obj19 & \ra{23}{33}{23.41} & \dec{33}{33}{30.78} & 0.182 & 320.4\\
\hline
\end{tabular}
\label{tab:14gqr_hostSpec}.
\end{table*}

\begin{table*}
\centering
\caption{\textbf{Basic properties of the light curve of iPTF\,14gqr in the $g$, $r$ and $i$ bands.} The apparent magnitudes have been corrected for galactic extinction, while the light curve decay rates are calculated in the rest frame of the transient. We estimate the decay rate in each filter by fitting the light curve of the transient between +10 and +16 days from explosion with a straight line. The $i$ band decline is not well constrained owing to the lack of post-peak photometry in that filter.}
\begin{tabular}{lcccccc}
\hline
Filter & MJD$_{\textrm{peak}}$ & m$_{\textrm{app,max}}$ & M$_{\textrm{abs,max}}$ & $\tau_{\textrm{rise}}$  &  Decay rate\\
& & (mag) & (mag) & (days) & (mag day$^{-1}$)\\
\hline
$g$ & 56950.39 $\pm$ 0.15 & 19.82 $\pm$ 0.01 & -17.45 $\pm$ 0.01 & 6.25 $\pm$ 0.46 & 0.21 $\pm$ 0.02\\
$r$ & 56951.36 $\pm$ 0.37 & 19.82 $\pm$ 0.03 & -17.45 $\pm$ 0.03 & 7.16 $\pm$ 0.53 & 0.11 $\pm$ 0.04\\
$i$ & 56952.51 $\pm$ 0.78 & 19.92 $\pm$ 0.09 & -17.35 $\pm$ 0.09 & 8.24 $\pm$ 0.89 & --\\
\hline
\end{tabular}
\label{tab:photoTable}
\end{table*}

\begin{table*}
\centering
\caption{\textbf{Emission line fluxes of the nucleus of the host galaxy IV Zw 155, along with broadband integrated photometry measurements.} The photometry measurements were taken from the GALEX source catalogs \cite{Bianchi2014}, SDSS \cite{Abolfathi2017} and 2MASS \cite{Skrutskie2006}. All fluxes and magnitude measurements have been corrected for galactic extinction using A$_V$=0.255 mag and a \cite{Cardelli1989} extinction law.}
\begin{tabular}{cccc}
\hline
Spectral line / Photometric band & Wavelength & Flux & AB magnitude \\
 & (\AA) & (10$^{-15}$ ergs s$^{-1}$ cm$^{-2}$) & mag\\
\hline
$[$O II$]$ & 3727 & 5.57 $\pm$ 0.79 & --\\
H$\beta$ & 4861 & 1.95 $\pm$ 1.09 & --\\
$[$O III$]$ & 5007 & 2.09 $\pm$ 0.58 & --\\
H$\alpha$ & 6563 & 8.92 $\pm$ 0.53 & --\\
$[$N II$]$ & 6583 & 2.97 $\pm$ 0.48 & --\\
$[$S II$]$ & 6717 & 1.90 $\pm$ 0.13 & --\\
$[$S II$]$ & 6731 & 1.63 $\pm$ 0.35 & --\\
GALEX FUV & 1540 & -- & 17.60 $\pm$ 0.09\\
GALEX NUV & 2314 & -- & 17.20 $\pm$ 0.05\\
SDSS $u$ & 3562 & -- & 16.740 $\pm$ 0.030\\
SDSS $g$ & 4719 & -- & 15.292 $\pm$ 0.004\\
SDSS $r$ & 6185 & -- & 14.762 $\pm$ 0.003\\
SDSS $i$ & 7500 & -- & 14.451 $\pm$ 0.003\\
SDSS $z$ & 8961 & -- & 14.314 $\pm$ 0.007\\
2MASS $J$ & 12376 & -- & 14.44 $\pm$ 0.06\\
2MASS $H$ & 16476 & -- & 14.31 $\pm$ 0.08\\
2MASS $K$ & 21621 & -- & 14.41 $\pm$ 0.12\\
\hline
\end{tabular}
\label{tab:lineFlux}
\end{table*}

\begin{table*}
\centering
\caption{\textbf{Elemental abundances at the photosphere for the ultra-stripped SN model found to be consistent with the observed data.}}
\begin{tabular}{ccc}
\hline
 Z & Element & Mass fraction (X)\\
\hline
6  & C  &  0.082\\
8  & O  &  0.464\\
10 & Ne &  0.382\\
11 & Na &  0.003\\
12 & Mg &  0.059\\
13 & Al &  0.003\\
14 & Si &  0.004\\
   & Fe group & 0.0017\\
\hline
\end{tabular}
\label{tab:USAbundance}
\end{table*}

\end{document}